%% file: fourier_lowres.tex
\documentclass[usenatbib]{mn2e}

\input{defs.tex}

\usepackage{epsf}

\begin{document}

\title[2dFGRS power spectrum]{The 2dF Galaxy Redshift Survey:
Power-spectrum analysis of the final dataset and cosmological
implications \vspace*{-0.4 truecm} }

\author[Cole et al.] {
\parbox{\textwidth}{Shaun Cole$^1$,  Will J. Percival$^2$,
John A. Peacock$^2$, Peder Norberg$^{3}$, Carlton M. Baugh$^1$,
Carlos S. Frenk$^1$, Ivan Baldry$^{4}$, Joss Bland-Hawthorn$^{5}$,
Terry Bridges$^{6}$, Russell Cannon$^{5}$, Matthew Colless$^{5}$,
Chris Collins$^7$, Warrick Couch$^8$, Nicholas J.G. Cross$^{4,2}$,
Gavin Dalton$^{9,10}$, Vincent R. Eke$^1$, Roberto De Propris$^{11}$, Simon P. Driver$^{12}$,
George Efstathiou$^{13}$, Richard S. Ellis$^{14}$, Karl Glazebrook$^{4}$,
Carole Jackson$^{15}$, Adrian Jenkins$^1$,
Ofer Lahav$^{16}$, Ian Lewis$^9$, 
Stuart Lumsden$^{17}$, Steve Maddox$^{18}$, Darren Madgwick$^{13}$, Bruce A.
Peterson$^{12}$, Will Sutherland$^{13}$, Keith Taylor$^{14}$ (The
2dFGRS Team)}
\vspace*{4pt} \\
$^1$Department of Physics, University of Durham, Science Laboratories, South
Road, Durham DH1 3LE, UK \\
$^{2}$Institute for Astronomy, University of Edinburgh, Royal
Observatory, Edinburgh EH9 3HJ, UK \\
$^{3}$Institut f\"ur Astronomie, Departement Physik, ETH Zurich,
HPF G3.1, CH-8093 Zurich, Switzerland \\
$^{4}$Department of Physics \& Astronomy, Johns Hopkins University, 3400
North Charles Street Baltimore, MD 21218--2686, USA \\
$^{5}$Anglo-Australian Observatory, P.O. Box 296, Epping, NSW 2121, Australia \\
$^{6}$Department of Physics, Queen's University, Kingston, ON, K7L 3N6, Canada\\
$^7$Astrophysics Research Institute, Liverpool John Moores University, Twelve
Quays House, Egerton Wharf, Birkenhead, L14 1LD, UK \\
$^8$School of Physics, University of New South Wales, Sydney,
NSW2052, Australia \\
$^9$Department of Physics, Keble Road, Oxford OX1 3RH, UK \\
$^{10}$Rutherford Appleton Laboratory, Chiltern, Didcot, OX11 OQX, UK \\
$^{11}$H.H. Wills Physics Laboratory, University of Bristol, 
Royal Fort, Tyndall Avenue, Bristol, BS8 1TL, UK \\
$^{12}$Research School of Astronomy \& Astrophysics, The Australian
National University, Weston Creek, ACT 2611, Australia \\
$^{13}$Institute of Astronomy, University of Cambridge, Madingley Road,
Cambridge CB3 0HA, UK \\
$^{14}$Department of Astronomy, California Institute of Technology, Pasadena,
CA 91125, USA \\
$^{15}$Australia Telescope National Facility, PO Box 76, Epping NSW 1710, Australia \\
$^{16}$Department of Physics and Astronomy, University College London, 
Gower Street, London, WC1E 6BT, UK \\
$^{17}$Department of Physics \& Astronomy, E C Stoner Building, Leeds LS2 9JT,
UK \\
$^{18}$School of Physics and Astronomy, University of Nottingham, University
Park, Nottingham, NG7 2RD, UK \\
\vspace*{-1.0 truecm}
}

\maketitle

\begin{abstract}
We present a power spectrum analysis of the final 2dF Galaxy
Redshift Survey, employing a direct Fourier method.
The sample used comprises 221\,414 galaxies with measured 
redshifts. We investigate in detail the modelling of
the sample selection, improving on previous treatments in a number
of respects. A new angular mask is derived, based on revisions to
the photometric calibration. The redshift selection function is
determined by dividing the survey according to rest-frame colour,
and deducing a self-consistent treatment of $k$-corrections and
evolution for each population. The covariance matrix for the
power-spectrum estimates is determined using two different
approaches to the construction of mock surveys, which are used to
demonstrate that the input cosmological model can be correctly
recovered. We discuss in detail the possible differences between the
galaxy and mass power spectra, and treat these using simulations,
analytic models, and a hybrid empirical approach. Based on
these investigations, we are confident that the 2dFGRS power
spectrum can be used to infer the matter content of the universe.
On large scales, our estimated power spectrum 
shows evidence for the `baryon oscillations' 
that are predicted in CDM models. Fitting to a CDM model,
assuming a primordial $\ns=1$ spectrum, $h=0.72$ and negligible neutrino mass,
the preferred parameters are $\Om  h = 0.168 \pm 0.016$
and a baryon fraction $\Ob /\Om  = 0.185\pm0.046$ (1$\sigma$ errors). 
The value of $\Om  h$ is  $1\sigma$ lower than the
$0.20 \pm 0.03$ in our 2001 analysis of the partially complete 2dFGRS.
This shift is largely due to the signal from the newly-sampled
regions of space, rather than the refinements in the treatment of
observational selection.
This analysis therefore implies a density significantly below 
the standard $\Om =0.3$: in combination with CMB data from WMAP,
we infer $\Om =0.231\pm 0.021$.

\end{abstract}
\begin{keywords}
large-scale structure of universe -- cosmological parameters
\vspace*{-0.5 truecm}
\end{keywords}

\newpage
\newpage

\section{Introduction}
\label{sec:intro}

Early investigations of density fluctuations in an expanding
universe showed that gravity-driven evolution imprints
characteristic scales that depend on the average matter density
\citep[e.g.][]{Silk68,PeeblesYu70,SunyaevZeldovich70}. Following the
development of models dominated by Cold Dark Matter
\citep{Peebles82,BondSzalay83}, it became clear that measurements of
the shape of the clustering power spectrum had the potential to
measure the matter density parameter -- albeit in the degenerate
combination $\Om  h$ ($h \equiv H_0 /  100 \kmsmpc$).

At first, the preferred CDM model was the $\Om =1$
Einstein--de~Sitter universe, together with a relatively low baryon
density from nucleosynthesis. Baryons were thus apparently almost
negligible in structure formation. However, cluster X-ray data
showed that the true baryon fraction must be at least 10--15\%, and
this was an important observation in driving acceptance of the
current $\Om  \simeq 0.3$ paradigm \citep{White93}. This
higher baryon fraction yields a richer phenomenology for the matter
power spectrum, so that non-negligible `baryon oscillations' are
expected as acoustic oscillations in the coupled matter-radiation
fluid affect the gravitational collapse of the CDM
\citep[e.g.][]{EisensteinHu98}. The most immediate effect of a large
baryon fraction is to suppress small-scale power, so that the
universe resembles a pure CDM model of lower density
\citep{PeacockDodds94,Sugiyama95}, but there should also be
oscillatory features that modify the power by of order 5--10\%, in a
manner analogous to the acoustic oscillations in the power spectrum
of the Cosmic Microwave Background.

In order to test these predictions, accurate surveys of large
cosmological volumes are required. A number of power-spectrum
investigations in the 1990s (e.g. 
Efstathiou, Sutherland \& Maddox \citeyear{efstathiou90};
Ballinger, Heavens \& Taylor 
\citeyear{ballinger95}; Tadros et al. \citeyear{tadros99}) 
confronted  the data with a simple prescription
of pure CDM using an effective value of $\Om  h$ (the
$\Gamma$ prescription of Efstathiou et al. \citeyear{efstathiou92}). The first survey
with the statistical power to make a full treatment of the power
spectrum worthwhile was the 2dF Galaxy Redshift Survey (Colless et
al. \citeyear{colless01}; \citeyear{colless03}). Observations for
this survey were carried out between 1997 and 2002, and by 2001 the
survey had amassed approximately 160\,000 galaxy redshifts. This
sample was the basis of a power-spectrum analysis by Percival et al.
(\citeyear{P01}; P01), which yielded several important conclusions.
P01 used mock survey data generated from the Hubble volume
simulation \citep{evrard02} to show that the power spectrum at
wavenumbers $k<0.15 \hompc$ should be consistent with linear
perturbation theory. Comparison with the data favoured a low-density
model with $\Om  h=0.20\pm0.03$, and also evidence, at
about the $2\sigma$ level, for a non-zero baryon fraction (the
preferred figure being around 20\%). In reaching these conclusions,
it was essential to make proper allowance for the window function of
the survey, since the raw power spectrum of the survey has an
expectation value that is the true cosmic power spectrum convolved
with the power spectrum of the survey geometry. 
The effect of this convolution is a significant distortion of the overall
shape of the spectrum, and a reduction in visibility of the baryonic
oscillations. The signal-to-noise ratio of features in the power
spectrum is thus adversely affected twice by the finite survey
volume: the cosmic-variance noise increases for small $V$, and the
signal is diluted by convolution. Both these elements need to be
well understood in order to achieve a detection.

The intention of this paper is to revisit the analysis of P01, both
to incorporate the substantial expansion in size of the final
dataset, and also to investigate the robustness of the results in
the light of our improved understanding of the survey selection.
Section~\ref{sec:survey} 
discusses the dataset and completeness masks. Section~\ref{sec:lf}
derives a self-consistent treatment of $k$-corrections and evolution
in order to model the radial selection function. Section~\ref{sec:power} 
outlines
the methods used for power-spectrum estimation, including allowance
for luminosity-dependent clustering; the actual data are
analysed in Section~\ref{sec:first_results} with the power spectrum estimate being
presented in Fig.~\ref{fig:pk_2dfgrs} and Table~\ref{tab:pk}. 
Section~\ref{sec:system} presents a
comprehensive set of tests for systematics in the analysis,
concluding that the galaxy power spectrum is robust. Section~\ref{sec:nonlin} 
then considers the critical issue of possible differences in shape
between galaxy and mass power spectra. The data are used to fit
CDM models in Section~\ref{sec:params} and Section~\ref{sec:summary} sums up.

\section{The 2\lowercase{d}F Galaxy Redshift Survey}
\label{sec:survey}

The 2dF Galaxy Redshift Survey (2dFGRS) covers approximately 1800
square degrees distributed between two broad strips, one across the
SGP and the other close to the NGP, plus a set of 99 random 2~degree
fields (which we denote by RAN) spread over the full southern galactic cap. 
The final catalogue contains reliable redshifts for
221\,414 galaxies selected to an extinction-corrected magnitude
limit of approximately $\bj=19.45$ (Colless \etal
\citeyear{colless03}; \citeyear{colless01}). In order to use these
galaxy positions to measure galaxy clustering one first needs an
accurate, quantitative description of the redshift catalogue. Here
we briefly review the properties of the survey and detail how we
quantify the complete survey selection function. Then, in 
Section~\ref{sec:lf}, we combine this with estimates of the galaxy luminosity
function to generate unclustered catalogues that will be used in the
subsequent clustering analysis.

\subsection{Photometry}
\label{sec:calib}

\begin{figure*}
\vbox{
\epsfxsize = 17.5truecm
\epsfbox{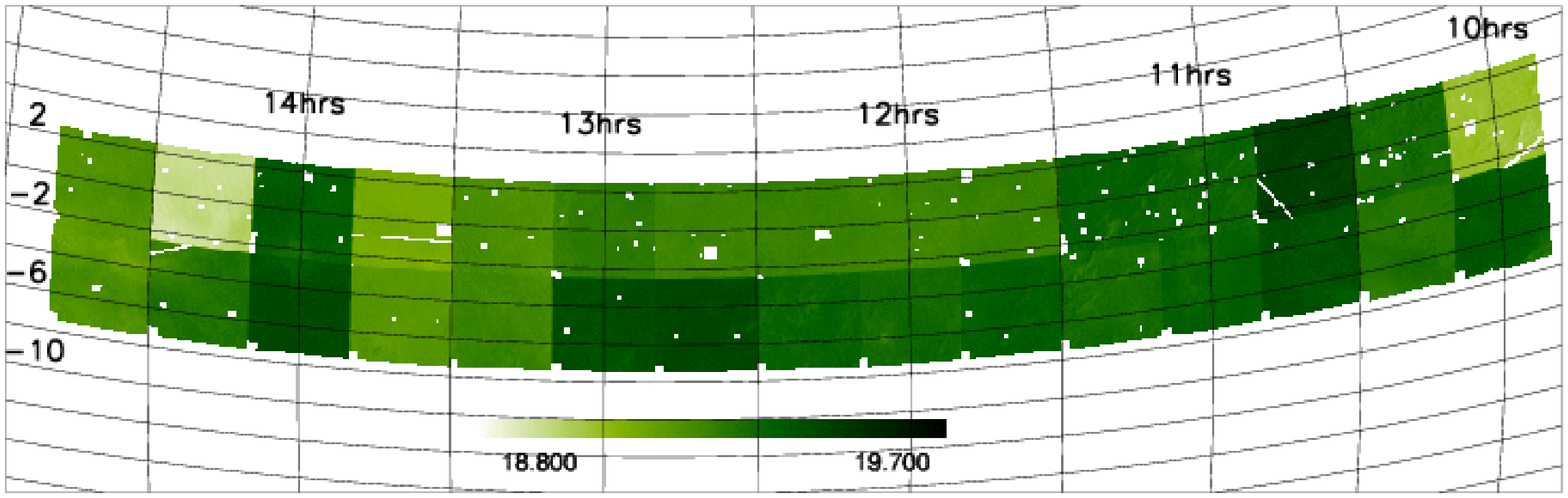}
\epsfxsize = 17.5truecm
\epsfbox{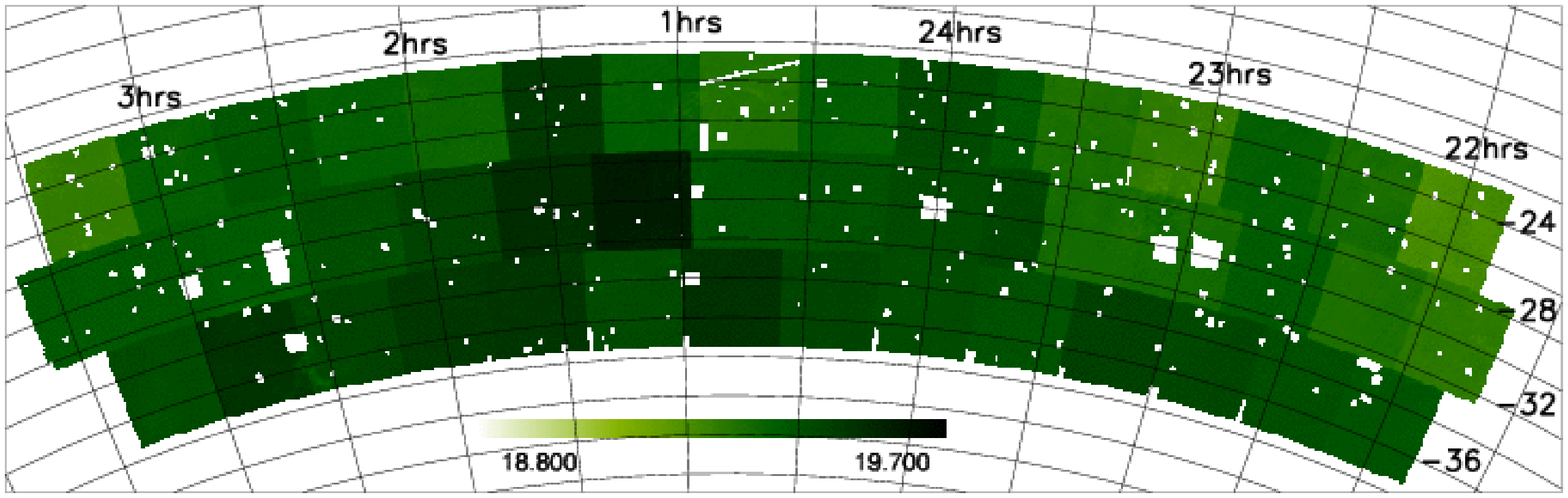} 
} \caption{Maps of the
extinction-corrected $\bj$ survey magnitude limit in the NGP (upper)
and SGP (lower) strips. The original target was a constant limit at
$\bj=19.45$; the variations from this reflect revisions to the
photometric calibration and alterations in corrections for galactic
extinction.} \label{fig:maglim}
\end{figure*}

The 2dFGRS input catalogue was intended to reach a uniform
extinction-corrected APM magnitude limit of $\bj=19.45$. However,
since the original definition of the catalogue, our understanding of
the calibration of APM photometry has improved. In the preliminary
100k release \citep{colless01}, the APM magnitudes were directly
recalibrated using CCD data from the EIS
\citep{prandoni99,arnouts01} and 2MASS \citep{jarrett00}
\citep[see][]{norberg_selfun}.  In the final data release it has
been possible to improve the calibration still further (see Colless
\etal \citeyear{colless03} and below). In addition, the Schlegel et
al. (\citeyear{schlegel98}) extinction maps were finalised after the catalogue was
selected; thus, the final survey magnitude limit varies with
position. As described in \citet{norberg_selfun} an accurate map of
the resulting magnitude limit can be constructed.
Fig.~\ref{fig:maglim} shows these maps for the final NGP and SGP
strips of the survey. Note that the maps also serve to delineate the
boundary of the survey and the regions cut out around bright stars
and satellite tracks.

The improvements in the photometry derived from the UK Schmidt
plates comes in part because these have been scanned using
the SuperCOSMOS measuring machine \citep{hambly01}.
SuperCOSMOS has some advantages in precision with respect to the
APM, yielding improved linearity and smaller random errors.
In a similar way to the APM
survey, the SuperCOSMOS recalibration matches plate overlaps (Colless \etal
\citeyear{colless03}). The magnitudes have been placed on an
absolute scale using the SDSS EDR (Stoughton et al. \citeyear{EDR}) in 33 
plates, the ESO imaging survey (e.g. Arnouts et al. \citeyear{arnouts01}) in 7,
plus the ESO-Sculptor survey (Arnouts et al. \citeyear{arnouts97} ).

When the SuperCOSMOS $\bjsc$ data are compared to the 2dFGRS APM
photometry, there is evidence for a small non-linear term, which we
eliminate by applying the correction
\begin{equation}
\bj^\prime = \bj + 0.033( [\bj-18]^2 - 1 ) \quad {\rm for } \quad
\bj
> 15.5
\end{equation}
and a fixed offset for $\bj < 15.5$. We then determine quasilinear
fits of the form
\begin{equation}
\bj^{\prime\prime} = A\, \bj^\prime + B ,
\end{equation}
where $A$ and $B$ are determined separately for each plate to
minimize the rms difference $\bj^{\prime\prime} -\bjsc$.
The final 2dFGRS magnitudes, $\bj^{\prime\prime}$, are given in
the release database. For many purposes (\eg defining the colour of
a galaxy) the SuperCOSMOS magnitudes are the preferred choice, but
for defining the survey selection function we use the final APM
magnitudes as it is for these that the survey has a well defined
magnitude limit.

\subsection{Colour data}

SuperCOSMOS has also scanned the UKST $\rf$ plates \citep{hambly01},
and these have been calibrated in the same manner as the $\bj$ plates.
The $\rf$ plates are of similar depth and quality to the $\bj$ plates, 
giving the important ability to divide
galaxies by colour.

The systematic calibration uncertainties are at the level of $0.04$~mag.
rms in each band. This uncertainty is significantly smaller
than the rms differences between the SuperCOSMOS and SDSS
photometry ($0.09$~mag $3\sigma$ clipped rms in each band, as
compared with $0.15$~mag when APM magnitudes are used). However,
some of this dispersion is not a true error in SuperCOSMOS:
SDSS photometry is not perfect, nor are the passbands and apertures used
identical. A fairer estimate of the random errors can probably be deduced from
the histogram of rest-frame colours given below in Fig.~\ref{fig:colours}.
This shows a narrow peak for the early-type population with a FWHM of about
$0.2$~mag. If the intrinsic width of this peak is extremely narrow
such that the measured width is dominated by the measurement errors
this gives us an upper limit on the errors in
photographic $\bj-\rf$ colour of
$0.2/\sqrt{8\ln(2)} \approx 0.085$~mag, 
or an uncertainty
of only $0.06$~mag in each band (including calibration systematics). 

In our power spectrum analysis, we will wish to split the sample by rest
frame colour so as to compare the clustering of intrinsically red
and blue subsamples. To achieve this we need to be able to $k$-correct
the observed colours.

\subsubsection{$k$-corrections}
\label{sec:kcorr}

The problem we face is this: given a redshift and an observed
$\bj-\rf$ colour, how do we deduce a consistent $k$-correction for
each band?  The simplest solution is to match the colours to a
single parameter, which could be taken to be the age of a
single-metallicity starburst.  This approach was implemented using the models
of \citet{bc03}.  Their Single Stellar Populations (SSPs) 
vary age and metallicity, and these variations
will be nearly degenerate. In practice, we assumed 0.4 Solar metallicity
($Z=0.008$) and found the age that matches the data. For very red
galaxies, this can imply a current age  $>13$ Gyr; in such cases
an age of 13~Gyr was assumed, and the metallicity was raised until
the correct colour was predicted.

In most cases, this exercise matched the
results of the \citet{Blanton03} KCORRECT package
(version 3.1b), which fits the
magnitude data using a superposition of realistic galaxy spectral
templates. The results of Blanton \etal 
are to be preferred in the region where
the majority of the data lie; this can be verified by taking the
full DR1 $ugriz$ data and fitting $k$-corrections, then comparing
with the result of fitting $gr$ only. The differences are small, but
are smaller than the difference between the KCORRECT results and
fitting burst models.  The main case for which this matters is for the red
$k$-correction for blue galaxies.  However, some galaxies can be
redder than the reddest template used by Blanton et al.; for such
cases, the burst models are to be preferred. In fact, the two match
almost perfectly at the join.

\begin{figure}
\vbox{
\epsfxsize=8.0truecm \centering
\epsfbox{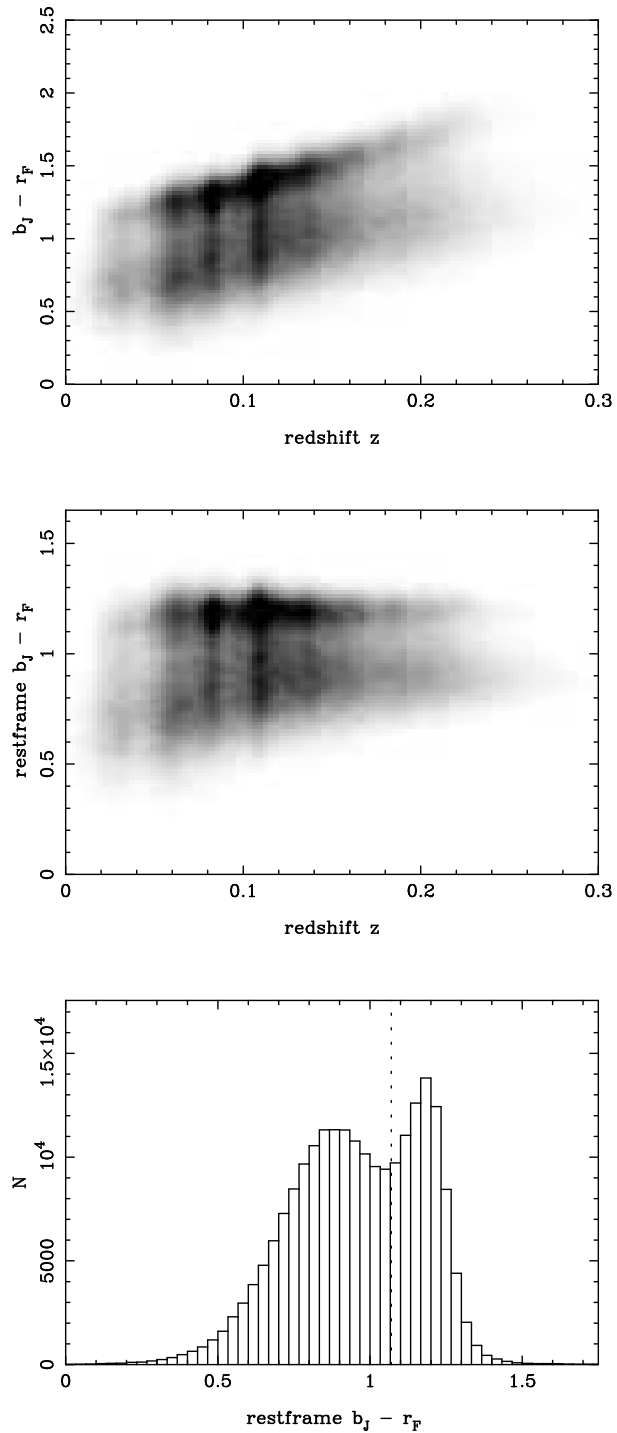} 
} 
\caption{
Photographic $\bj-\rf$
colour versus redshift for the 2dFGRS, as observed (top) and in the rest
frame (middle). The separation between `early-type' (red) and `late-type'
(blue) galaxies is very clear. The third panel shows the histogram
of $k$-corrected restframe colours, which is very clearly bimodal. This
is strongly reminiscent of the distribution of spectral type, $\eta$, and
dividing the sample at a rest frame colour of $(\bj-\rf)_{z=0}=1.07$ 
(dotted line) achieves a very similar
separation of early-type `class 1' galaxies
from classes 2--4, as was done using spectra by \citet{madgwick02}.
} 
\label{fig:colours}
\end{figure}

\begin{figure*}
\vbox{
\epsfxsize = 17.5truecm
\epsfbox{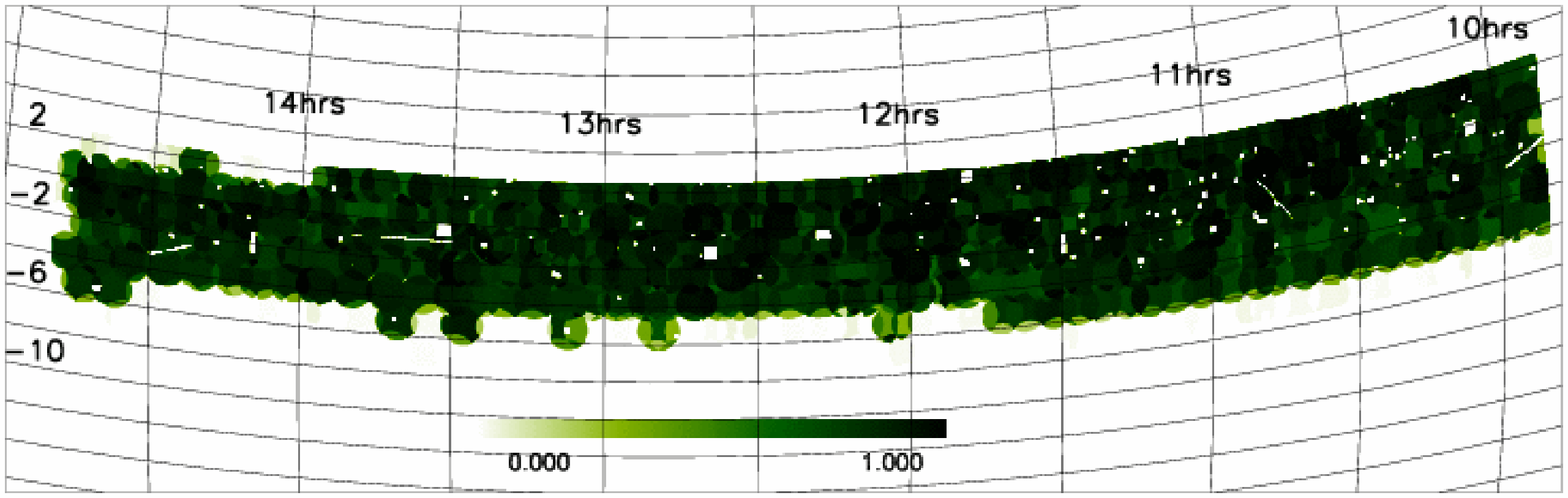}
\epsfxsize = 17.5truecm
\epsfbox{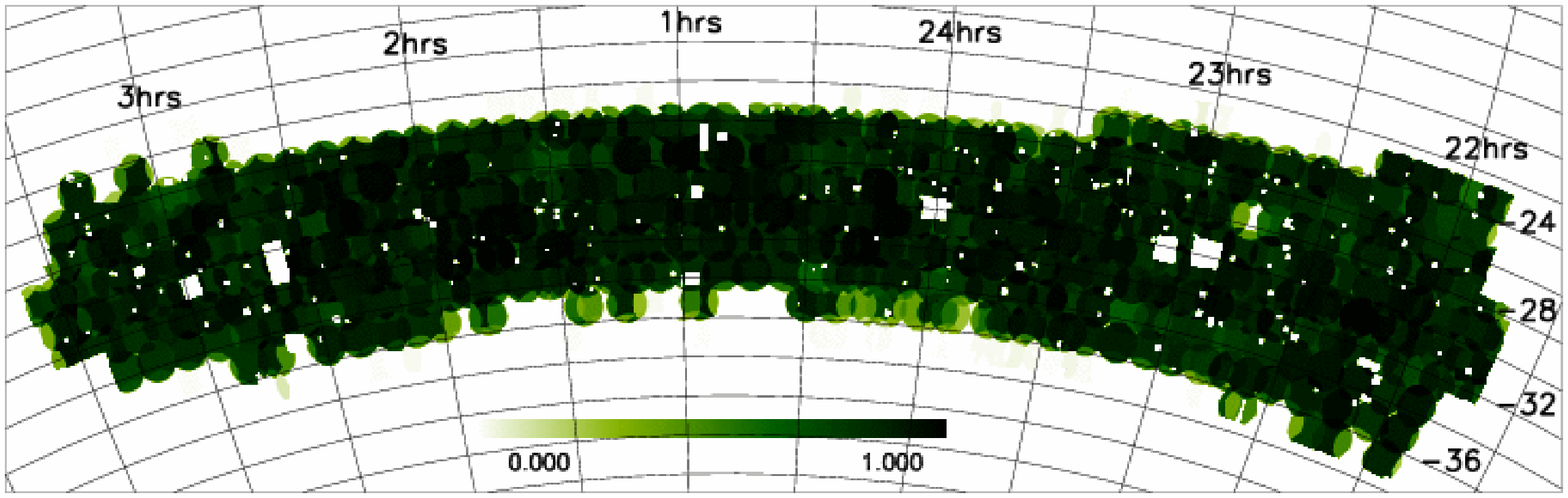}
}
\caption{Maps of the overall redshift completeness, $R(\theta)$,
averaged over apparent magnitude, in the NGP and SGP strips.
}
\label{fig:zcomp}
\end{figure*}

The following fitting formula, which we adopt, summarizes the results of this
procedure, and is good to 0.01 mag almost everywhere in the range of interest:
\begin{equation}
\eqalign{
k_{\bj} = &(-1.63+4.53C)y + (-4.03-2.01C)y^2 \cr
        &-z/(1+(10z)^4) \cr
k_{\rf} = &(-0.08+1.45C)y + (-2.88-0.48C)y^2, \cr
}
\end{equation}
where $y \equiv z/(1+z)$ and $C \equiv \bj-\rf$. In most cases, the
deviations from the fit are probably only of the order of the accuracy of
the whole exercise, so they are ignored in the interests of clarity.
The distributions of observed and $k$-corrected rest frame
colours are shown in Fig.~\ref{fig:colours}.

The histogram of rest frame colours exhibits the well known
bimodal distribution \citep{strateva01,baldry04}. 
Related spectral quantities
such as H$\delta$ absorption and the $4000\,$Angstrom break 
show similar bimodal distributions \citep{kauffmann03}.
In particular, colour is strongly correlated with
the 2dFGRS spectral type $\eta$ 
(see figure~2 of Wild \etal \citeyear{wild05}).
Thus dividing the sample at a rest frame colour of $(\bj-\rf)_{z=0}=1.07$
achieves a very similar
separation of early-type `class 1' galaxies
from classes 2--4, as was done using spectra by \citet{madgwick02}.

\subsection{Spectroscopic completeness}
\label{sec:compl}

\begin{figure*}
\vbox{
\epsfxsize = 17.5truecm
\epsfbox{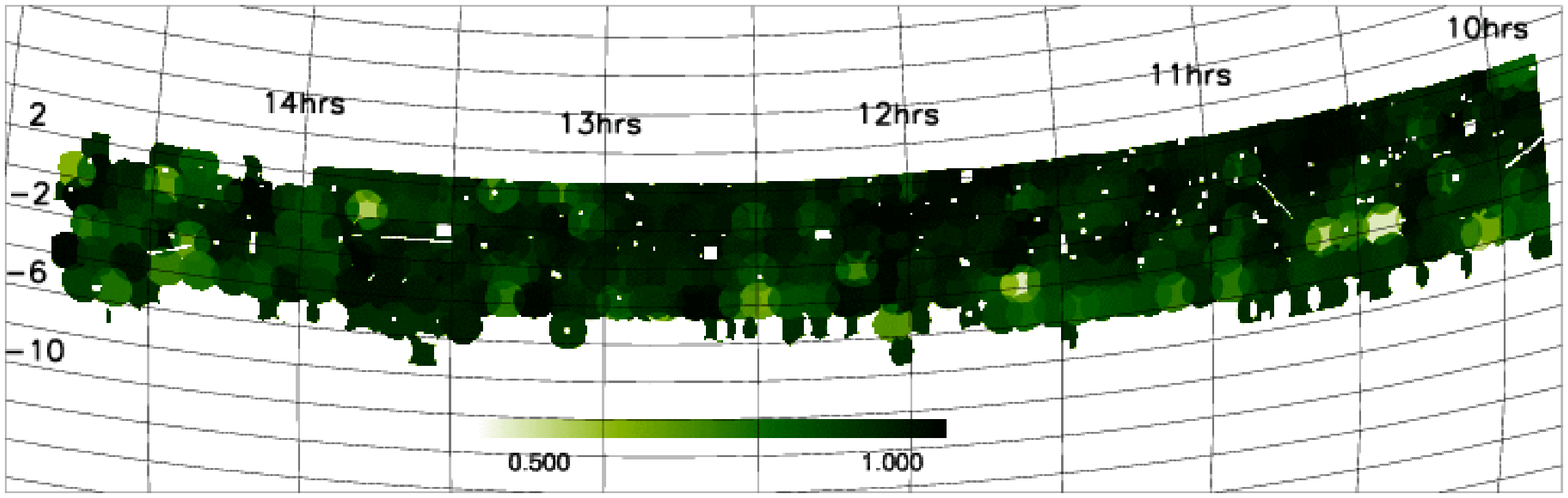}
\epsfxsize = 17.5truecm
\epsfbox{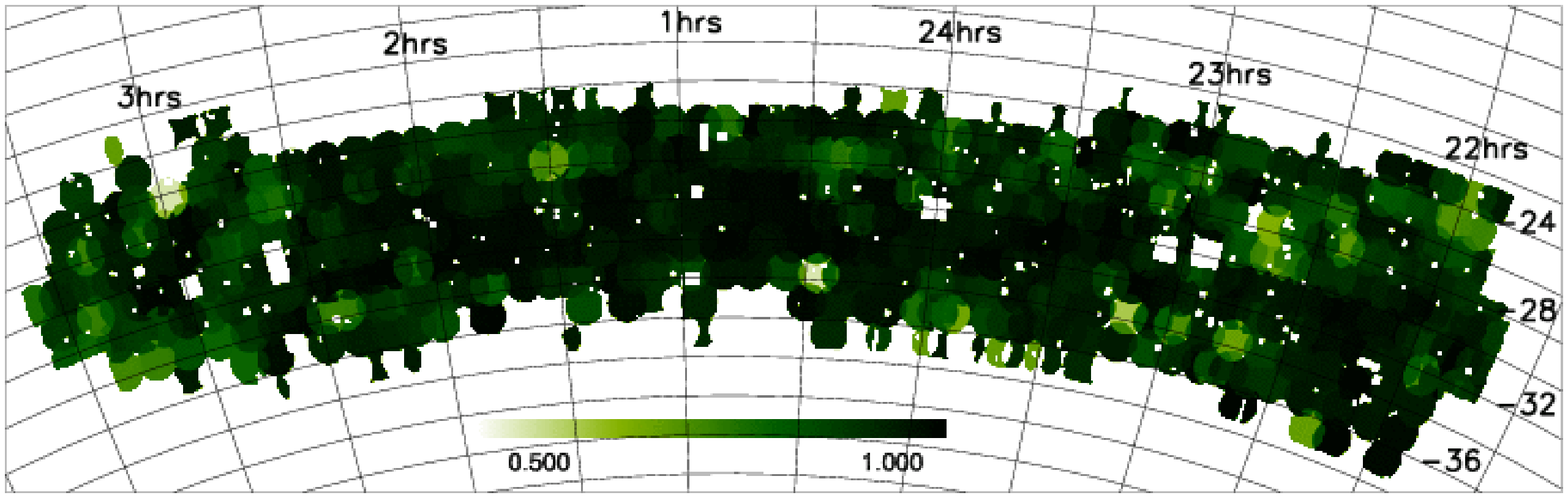}
}
\caption{Maps illustrating the redshift completeness at $\bj=19.5$
relative to that at bright magnitudes.
The magnitude dependence of this redshift completeness is
assumed to be proportional to  $1-\exp(\bj-\mu)$ and the parameter
$\mu$ is estimated for each sector in the survey mask. Here, we
plot the factor  $1-\exp(\bj-\mu)$ for a fiducial magnitude
of $\bj=19.5$.
}
\label{fig:mumask}
\end{figure*}

The spectroscopic completeness, the fraction of 2dFGRS galaxies with
reliably measured redshifts, varies across the survey. This can be due
to a failure to measure redshifts from the observed spectra
or to whole fields missing, either because they were never
observed, or because they were rejected when they had
unacceptably low spectroscopic completeness. In addition, 
there is a small level of incompleteness arising from galaxies
that were never targeted due to restrictions in fibre positioning.
In the samples we analyse we reject all fields (single observations) 
that have a spectroscopic completeness less than 70\%. As the observed 2dF
fields overlap in a complex pattern, the completeness varies from
sector to sector, where a sector is defined by a unique set of
overlapping fields. Maps of the redshift completeness,  $R(\theta)$, 
of the final
survey, constructed as detailed in \citet{norberg_selfun}, are shown
in Fig.~\ref{fig:zcomp}. Here $\theta$ denotes the angular position 
on the sky. For the two main survey strips, 80\% of the
area has a completeness greater than 80\%.

In observed fields, the fraction of galaxies
for which useful (quality $\ge 3$) redshifts have not been obtained
increases significantly with apparent magnitude. In
\citet{norberg_selfun} \citep[see also][]{colless01},
we define an empirical
model of this magnitude dependent incompleteness. In this model,
the fraction of observed galaxies yielding useful redshifts is
proportional to $1-\exp(\bj-\mu)$ and, by averaging over fields,
the parameter $\mu$ is defined for each sector in the survey.
Fig.~\ref{fig:mumask}, shows a map of the factor $1-\exp(\bj-\mu)$
for a fiducial apparent magnitude of $\bj=19.5$.
For a given apparent magnitude and position the overall redshift
completeness is given by the product
\begin{equation}
  C(\theta,\bj) = A(\theta) R(\theta) [1-\exp(\bj-\mu)] ,
\end{equation}
where $R(\theta)$ and $[1-\exp(\bj-\mu)]$ are the quantities
illustrated in Figs~\ref{fig:zcomp} and~\ref{fig:mumask}. 
In each sector, we define the normalizing constant
$A(\theta) = \langle [1-\exp(\bj-\mu(\theta))] \rangle^{-1}$
averaged over the expected apparent magnitude
distribution of survey galaxies, so that
$\langle  C(\theta,\bj) \rangle= R(\theta) $. 
In general, this magnitude dependent incompleteness is
not a large effect. At the magnitude limit of the survey, 50\%(80\%)
of the survey's area has completeness factor, $[1-\exp(\bj-\mu)]$,
greater than 88\%(80\%).

Since it is easier to measure the redshift of blue emission line galaxies
than of red galaxies, we expect the level of incompleteness to be
different for our red and blue subsamples. Since we are unable to classify
a galaxy by rest frame colour without knowing its redshift, it is
not trivial to estimate the level of incompleteness in each subsample.
However, to a first approximation, we can quantify the incompleteness
as a function of the observed colour. In fact, we can do better than this
by noting that our red and blue subsamples are quite well separated
on a plot of observed colour versus apparent magnitude. We can split
the galaxies in this plane into two disjoint samples.
Quantifying the incompleteness for red and blue subsamples split in this way
we find they are again reasonably well fit by the model $1-\exp(\bj-\mu)$,
but with $\mu_{\rm blue} = \mu + 0.65$
and $\mu_{\rm red} = \mu - 0.25 $. These are values we use
in Section~\ref{sec:redblue}, where we compare the power spectra
of the red and blue galaxies.

\section{Luminosity function and evolution} \label{sec:lf}

\begin{table*}
\caption{The parameters of the Schechter luminosity functions and 
$k+e$ corrections (see equation~\ref{eq:ke})
that define the standard model of the survey selection function.
Two Schechter functions are combined to describe the luminosity
function of red galaxies.}
\begin{center}
\begin{tabular}{lllllllll} 
\hline
\multicolumn{1}{l} {} &
\multicolumn{1}{c} {$\Phi^* / {\rm h^3 Mpc}^{-3}$} & 
\multicolumn{1}{c} {$M^{* z=0.1}_\bj - 5\logh$} &
\multicolumn{1}{c} {$\alpha$} & 
\multicolumn{1}{c} {$a$} & 
\multicolumn{1}{c} {$b$} & 
\multicolumn{1}{c} {$c$} & \\
\hline 
Combined & 0.0156  & $-19.52$  & $-1.18$  & 0.327   & 6.18    & 10.3  \\ 
Blue    & 0.00896 & $-19.55$  & $-1.3\hphantom{0}$   & 0.282   & 5.67    & 31.1   \\
Red     & 0.00909 & $-19.19$  & $-0.5\hphantom{0}$   & 1.541   & 6.78    & 7.95   \\
        & 0.00037 & $-19.87$  & $-0.5\hphantom{0}$   & 1.541   & 6.78    & 7.95   \\
\hline
\end{tabular}
\end{center}
\label{tab:lf}
\end{table*}

For a complete understanding of how the 2dFGRS probes the universe,
we need to supplement the selection masks described above with a
model for the galaxy luminosity function. It will also be necessary
to understand how the luminosity function depends on galaxy type and
how it evolves with redshift. 

In \citet{norberg_selfun}, we demonstrated that a Schechter function
was a good\footnote{In the sense that the deviations from the
Schechter form are sufficiently small that they have no important
effects on our modelling of the radial selection function. However, with
the high statistical power of the 2dFGRS even these very
small deviations are detected. As a result, the best fit
Schechter function parameters can vary by more than their formal
statistical errors when different redshift or absolute magnitude
cuts are applied to the data. } description of the overall 2dFGRS
luminosity function and we estimated a mean $k+e$ correction by
fitting Bruzual \& Charlot (\citeyear{bc93}) population synthesis models to a
subset of the 2dFGRS galaxies for which SDSS $g-r$ colours were
available. Repeating this procedure for the recalibrated final
2dFGRS magnitudes yields a Schechter function with $\alpha=-1.18$,
$\Phi^*=1.50\times 10^{-2}\, h^3 {\rm Mpc^{-3}}$, 
$M^{* z=0.1}_{\bj}-5 \log_{10} h=-19.57$, where we
have quoted the characteristic absolute magnitude at the median
redshift of the survey, $z=0.1$, $M^{*
z=0.1}_{\bj} \equiv M^{* z=0}_{\bj} +k(z=0.1)+ e(z=0.1)$ rather than
the redshift $z=0$ value. Since our purpose is to model only those
galaxies that are in the 2dFGRS, we have ignored the 9\% boost to
$\Phi^*$ that was applied in \citet{norberg_selfun} to compensate
for incompleteness in the 2dFGRS input catalogue. Thus, the
corresponding values from \citet{norberg_selfun} are
$\alpha=-1.21\pm 0.03$, 
$\Phi^*=(1.47 \pm 0.08)\times 10^{-2}\, h^3 {\rm Mpc^{-3}}$, 
$M^{* z=0.1}_{\bj}-5 \log_{10} h=-19.50 \pm 0.07$. 
The $1$-$\sigma$ shifts in $\alpha$
and $M^*_{\bj}$ are systematic changes resulting from the
photometric recalibration. The uncertainties on each of these
parameters remain essentially unchanged.\footnote{In terms of constraints
on the local galaxy population these new luminosity function estimates
do not add significantly to the results from \citet{norberg_selfun}
and \citet{madgwick02}
as the uncertainties remain dominated by systematic uncertainties in
the photometric zero-point, survey completeness and
evolutionary corrections. However, for the purposes of quantifying
the survey selection function it is important to derive estimates 
consistent with the new calibration.}
The luminosity functions
determined separately in the NGP, SGP and RAN field regions agree
extremely well in shape, but are slightly offset in $M^*_{\bj}$. In the
standard calibration used in this paper we apply a shift of
$-0.0125$ in the SGP and $0.022$ in the NGP to the galaxy magnitudes
and magnitude limits to make all the regions consistent with the
luminosity function estimated from the full survey, 
but as we shall see, these shifts are so
small that they make very little difference.

\begin{figure}
\epsfxsize=8.4truecm   \centering
\epsfbox[0 55 540 750]{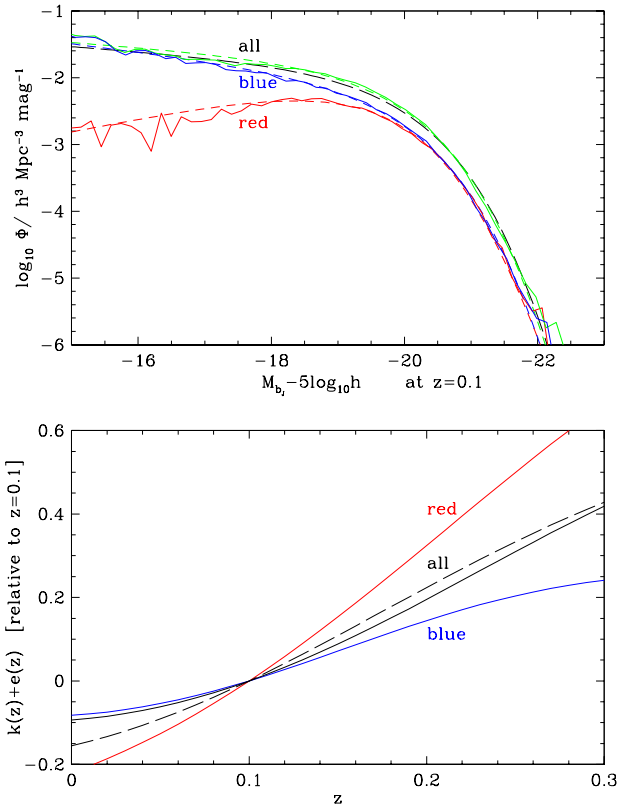}
\caption{The solid curves in the upper panel show stepwise estimates
of the overall 2dFGRS luminosity function and estimates for red and
blue subsets, split at a restframe colour of $\bj-\rf=1.07$. They
are plotted as a function of absolute magnitude at $z=0.1$,
which we define in terms of $z=0$ absolute magnitude as $M^{*
z=0.1}_{\bj}  \equiv M^{* z=0}_{\bj}  +k(z=0.1)+ e(z=0.1)$. The
smooth dashed curves are 
Schechter functions convolved with the model of the magnitude
measurement errors. It is these luminosity functions that are used
to construct random unclustered galaxy catalogues. The
corresponding maximum likelihood estimates of the $k+e$ corrections
(relative to their values at $z=0.1$) are shown by the solid curves
in the lower panel. The long-dashed line in the upper panel is an
STY estimate of the overall luminosity function when the $k+e$
correction shown as the long-dashed line in the lower panel is
adopted.} \label{fig:lf}
\end{figure}

The main problem with the previous procedure was that the evolution
is assumed to be known. Here, we take the safer approach of
estimating empirical $k+e$ corrections directly from the data. If we
model the luminosity function by a function $\phi(L)$ of the $k+e$
corrected luminosity $L$, and the radial density field by a function
of redshift $\rho(z)$, then following \citet{saunders90} we can
define the joint likelihood as
\begin{equation}
{\cal L}_1  = \frac{\Pi_i \, \rho(z_i) \phi(L_i)} {\int \!\!\! \int
\rho(z) \phi(L) \frac{dV}{dz}\, dL\, dz} ,
\end{equation}
where the product is over the galaxies in the sample. Note for
convenience one can select from the 2dFGRS a simple magnitude
limited subsample. At each redshift, the range of the luminosity
integral in the denominator is determined by the apparent magnitude
limits and the model $k+e$ correction. One could then parameterize
$\phi(L)$, $\rho(z)$  and $k(z)+e(z)$ and seek their maximum
likelihood values.  In practice, this does not work well for our data
as, without a constraint on $\rho(z)$, there is a near degeneracy
between $k+e$ and the faint end slope of the luminosity function.
This problem can be removed by introducing an additional factor into
the likelihood to represent the probability of observing a given
$\rho(z)$. Estimating this probability using the randomly
distributed clusters model of \citet{ns52} \citep[see
also][]{peebles_lss}, the likelihood becomes
\begin{equation}
{\cal L}  = {\cal L}_1 \ \times \ \Pi_r \,
\exp \left( - \frac{1}{2} \frac{(\rho(z_r)/\bar\rho-1)^2 N_r}
{(1+4\pi J_3 N_r/V_r)} \right).
\end{equation}
Here $\rho(z_r)$, $N_r$ and $V_r$ are respectively the galaxy
density, number of galaxies and comoving volume of the $r^{\rm th}$
radial bin. The overall mean galaxy density is $\bar\rho$ and $J_3$
is the usual integral over the two-point correlation function. We
adopt $J_3=400\,(\mpcoh)^3$, consistent with the measured 2dFGRS
correlation function \citep{hawkins03}.

Note that when splitting the sample into the two colour classes we
ignore any evolutionary correction to their colours. This cannot be
exactly correct, but at the quite red dividing colour of
$\bj-\rf=1.07$, galaxies are not star forming and the evolutionary
colour correction is expected to be small. This approximation is
supported empirically by the central panel of
Fig.~\ref{fig:colours}, which shows that the rest-frame colour
corresponding to the division between the red and blue populations
appears to be independent of redshift.

The luminosity functions and corresponding $k+e$ corrections that
result from applying this method are shown in Fig.~\ref{fig:lf}. 
 Note that to model the selection function all that we require
is the combined $k+e$ correction 
for the red and blue components of the luminosity function.
Thus for modelling the selection function we do not make
use of the colour dependent $k$-corrections derived in 
Section~\ref{sec:kcorr} .
To utilise these would require a bivariate model of the
galaxy luminosity function so that $\bj-\rf$ colours could be assigned 
to each model galaxy. 
We
have used a stepwise parameterization of the luminosity function and
assumed $k+e$ corrections of the form 
\begin{equation}
k + e = \frac{a z+b z^2}{1+c z^3}, 
\label{eq:ke}
\end{equation}
where
$a$, $b$ and~$c$ are constants.  The solid lines in the upper panel
show the luminosity function estimates for the full sample and for the
red and blue subsets.  The solid curves in the lower panel show the
corresponding maximum likelihood $k+e$ corrections. For the purpose
of constructing unclustered galaxy catalogues it is useful to fit
these estimates using Schechter functions convolved with the
measured distribution of magnitude errors from
\citet{norberg_selfun}. The smooth dashed curves that closely
match each of the maximum likelihood estimates are these convolved
Schechter functions.  In the case of the red galaxies we have used
the sum of two Schechter functions to produce a sufficiently good
fit. The parameters of these Schechter functions and the corresponding
$k+e$ correction parameters are listed in Table~\ref{tab:lf}.

These luminosity functions can be compared with those of
\citet{madgwick02}, who estimated the luminosity functions of 2dFGRS
galaxies classified by spectral type using a principal component
analysis.  Although there is not a one-to-one correspondence between
colour and spectral class, our red sample corresponds closely to
their class 1 and our blue sample to the combination of the
remaining classes 2, 3 and~4.  The shape and normalization of our
luminosity functions agree well: the only difference occurs fainter than
$M_{\bj}>-16$, where for the earliest spectral type,
\citet{madgwick02} find an excess over a Schechter function which is
not apparent in our red sample. At first sight, the values of $M^*$
and hence the positions of the bright end of the luminosity
functions appear to differ. \citet{madgwick02} find that late type
galaxies have significantly fainter $M^*$ than early types, while
the bright ends of our blue and red luminosity functions are very
close. This apparent difference is because Madgwick \etal correct
their luminosity functions to $z=0$ while our estimates are for a
fiducial redshift of $z=0.1$. In addition, Madgwick \etal apply
only $k$-corrections while our modelling also includes mean
$e$-corrections for each class. Because the $k+e$ corrections for
the red (early) galaxies are much greater than for the blue (late)
galaxies this brings the $M^*$ values at $z=0.1$ much closer
together. In fact, we find a very good match with the
\citet{madgwick02} results once the difference in $k+e$ corrections
is accounted for and the results translated to $z=0.1$.

We also compare the overall luminosity function and mean $k+e$
correction with the result of applying the method we used previously
in \citet{norberg_selfun}. For this purpose, we adopt  $k+e= (z+6 z^2)/(1+ 8.9
z^{2.5})$, which is shown by the long-dashed line in the lower panel
of Fig.~\ref{fig:lf}. This is essentially identical to the fit used
in \citet{norberg_selfun}. The luminosity function estimated using
the STY method, again convolved with the same model for the magnitude
errors, is shown by the long dashed line in the upper panel. We note
that apart from $z<0.05$, where there are relatively few 2dFGRS
galaxies, this $k+e$ correction is in good agreement with our new
maximum likelihood estimate. Similarly, the luminosity function is in
quite close agreement with our new estimate for the combined red and
blue sample.

\subsection{Random unclustered catalogues} \label{sec:random}

Armed with realistic luminosity functions and evolution corrections,
plus an accurate characterization of survey masks, we can now generate
corresponding random catalogues of unclustered galaxies (not to be
confused with the RAN data from the randomly-placed 2dFGRS survey
fields). The procedure we adopt to do this is as described in Section~5 of
\citet{norberg_selfun}, except that we now have the option of
treating the red and blue subsamples separately. In
this procedure, we perturb the magnitude and redshifts in accordance
with the known measurement errors. The mock catalogues include
a number of properties in addition to the angular position, apparent
magnitude and redshift of each galaxy:
\begin{itemize}
\item The overall redshift completeness, $c_i$, in the direction of
      each galaxy, as given by the completeness measure described in Section~\ref{sec:compl}.
\item The mean expected galaxy number density $n_i$ at each galaxy's position,
      taking account of the survey magnitude limit in this direction,
      and the dependence of redshift completeness on apparent magnitude as
      characterized by the parameter $\mu$.
\item The expected bias parameter $b_i$ of a galaxy of a given
      luminosity and colour as defined by the simple model
      in Section~\ref{sec:estimator}.
\end{itemize}
Note that when the  red and blue subsamples are analysed separately,
$n_i$  refers only to galaxies of the same colour class,
but if the random
catalogue is to be used in conjunction with the full 2dFGRS catalogue,
then $n_i$ is defined in terms of a sum over
contributions from both the red and blue subsamples.
The value of this parameter will also vary if one places additional
cuts on the catalogue such as varying the faint magnitude limit.
As we shall see in Section~\ref{sec:estimator}, these quantities are
useful when estimating the galaxy power spectrum.

\begin{figure}
\epsfxsize=8.5 truecm   \centering
\epsfbox[15 270 540 750]{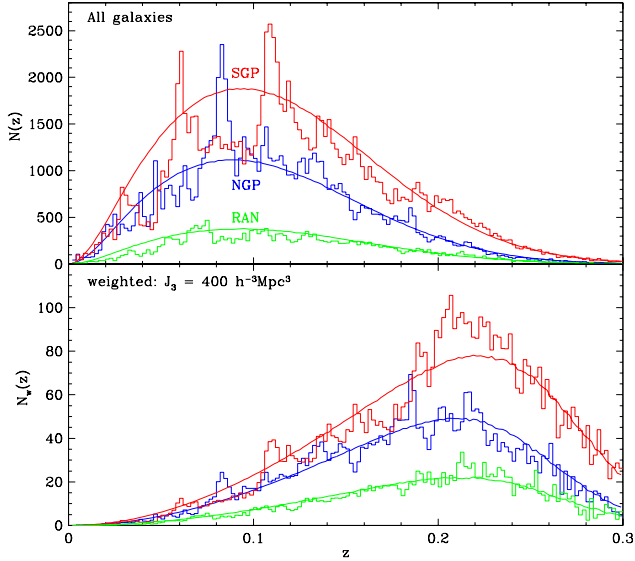} 
\caption{The histograms in the top panel
show the redshift distribution of the 2dFGRS data in the SGP, NGP and
RAN field regions. The curves show the distribution in the
corresponding random unclustered catalogues. The lower panel shows
the same distributions, but weighted with a redshift dependent function
as in the power
spectrum analysis, using $J_3=400$~h$^{-3}$Mpc$^3$. In all cases, the
histograms for the random catalogues are normalized so that the sum
of the weights matches that of the corresponding data. }
\label{fig:dndz_all}
\end{figure}

\begin{figure}
\epsfxsize=8.5 truecm  \centering
\epsfbox[15 270 540 750]{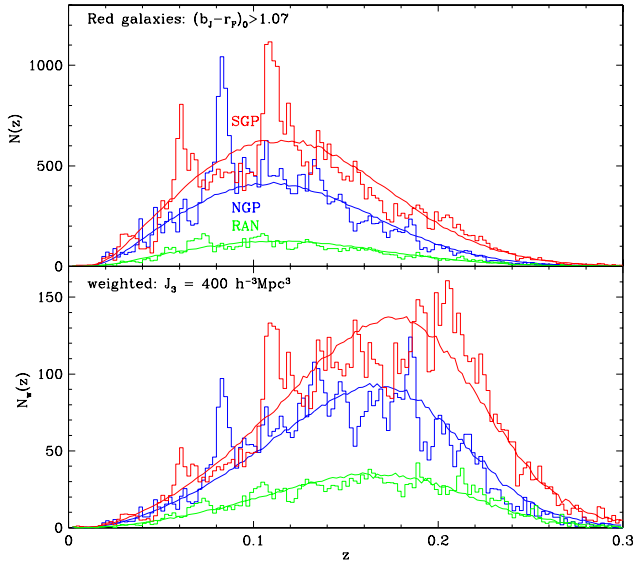} 
\caption{As Fig.~\ref{fig:dndz_all} but
for the red subset with rest frame colours redder than
$(\bj-\rf)_{z=0}>1.07$. } \label{fig:dndz_red}
\end{figure}

\begin{figure}
\epsfxsize=8.5 truecm \centering
\epsfbox[15 270 540 750]{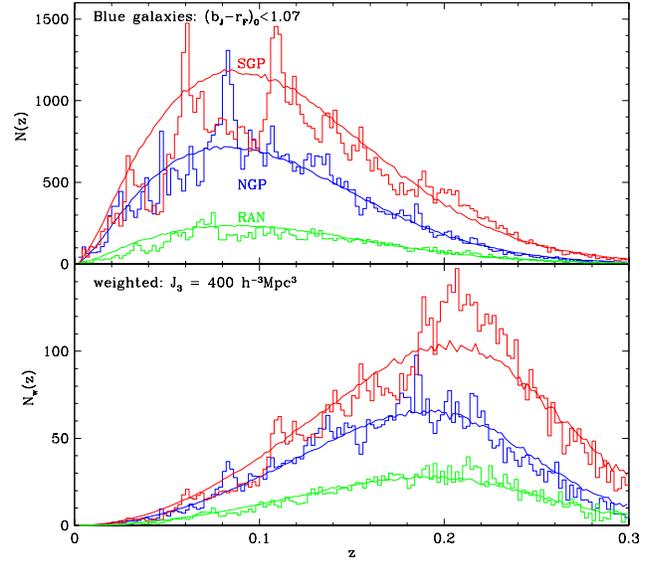} 
\caption{As Fig.~\ref{fig:dndz_all} but
for the blue subset with rest frame colours bluer than
$(\bj-\rf)_{z=0}<1.07$. } \label{fig:dndz_blue}
\end{figure}

Fig.~\ref{fig:dndz_all} compares the redshift distribution of the
genuine 2dFGRS data with that of the random catalogues. The upper
panel shows the number of galaxies, binned by redshift, that pass
the selection criteria defining the samples used in the power
spectrum analysis. The lower panel shows the same distributions but
weighted by the radial weight that is used in the power spectrum
analysis. Figs~\ref{fig:dndz_red} and~\ref{fig:dndz_blue} repeat
this comparison for the red and blue subsets. For the NGP and RAN
fields the smooth redshift distributions of the random catalogues
match quite accurately the mean values for the full dataset and for
the red and blue subsamples all the way to the maximum  redshift of our
samples ($z=0.3$). The SGP exhibits greater variation and in
particular is underdense compared to the random catalogue at
$z\la0.06$. This local underdensity in the SGP has been noted and
discussed many times before
\citep[e.g.][]{sgphole90,sgphole95,sgphole03,sgphole04}. In discussing the
2dFGRS 100k data, \citet{norberg_selfun} demonstrated, in their
figures~13 and~14, that similar redshift distributions were not
unexpected in $\Lambda$CDM mock catalogues. Moreover, the lower
panels in Figs~\ref{fig:dndz_all}, ~\ref{fig:dndz_red}
and~\ref{fig:dndz_blue}, which weight the galaxies in the same way
as in the power spectrum analysis, indicate that the contribution
from this local region is negligible. With this weighting, it is the
excess in the SGP at $0.2<z<0.24$ that appears more prominent. This
excess appears to be due to two large structures around ${\rm RA}=
23^{\rm h}$ and~$2^{\rm h}$. It is therefore likely that these
excursions are due to genuine large scale structure. Nevertheless, in
Section~\ref{model_dndz} we assess the sensitivity of our power spectrum
estimate to the assumed redshift distribution by also considering an
empirical redshift distribution.

\section{Power spectrum estimation and error assignment} \label{sec:power}

\subsection{The power spectrum estimator}
\label{sec:estimator}

We employ the Fourier based method of Percival, Verde \& Peacock
\defcitealias{PVP}{PVP} (\citeyear{PVP};  \citetalias{PVP})
which is a generalization of the minimum variance
method of Feldman, Kaiser \& Peacock
\defcitealias{FKP}{FKP} (\citeyear{FKP};  \citetalias{FKP})
to the case where the galaxies have a known luminosity and/or
colour dependent bias.
This procedure requires an assumed cosmological geometry
in order to convert redshifts and positions on the sky to comoving
distances in redshift space. Strictly, this geometry
should vary with the cosmological model being tested. However, the
effect of such a change is very small (this was extensively tested in P01). We have
therefore simply assumed a cosmological model with $\Om=0.3$ 
and $\Omega_\Lambda=0.7$ for this transition.

In our implementation, we carry out a summation over galaxies
from the data and
random catalogues to evaluate the weighted density field
\begin{equation}
  \label{fr}
  F(\r) = \frac{1}{N} \int \frac{w(\r,L)}{b(L)}
\, \left[ \ngl(\r,L) - \alpha \nrl(\r,L) \right] \, dL
\label{eqn:field}
\end{equation}
on a cubic grid. Here $\ngl(\r,L)$ and $\nrl(\r,L)$
are the number density of galaxies of luminosity $L$
to $L+dL$ at position $\r$ in the data and random catalogues
respectively. It is straightforward to generalize
this to include a summation
over galaxies of different types or colours.
Early analysis of the 2dFGRS found a bias parameter of $b \approx 1$     
\citep{verde02,lahav02} for $L_*$ galaxies averaged over all
types. Subsequently we have determined that the bias parameter    
depends both on luminosity and galaxy type or colour. Here we adopt
a scale independent bias parameter 
\begin{equation}
b(L) =  0.85 + 0.15(L/L_*),
\label{eq:bias}
\end{equation}
as found by \citet{norberg_xil}.
For red and blue subsets, split by  a rest frame colour of
$(\bj-\rf)_{z=0}=1.07$, the bias is significantly different and
we adopt
\begin{equation}
b_{\rm red} = 1.3\, [0.85 + 0.15(L/L_*)]
\label{eqn:red}
\end{equation}
and
\begin{equation}
b_{\rm blue}= 0.9\,  [0.85 + 0.15(L/L_*)] ,
\label{eqn:blue}
\end{equation}
which, as we find in Section~\ref{sec:redblue},
 empirically describes the difference in amplitudes
of the power spectra of red and blue galaxies around
$k=0.1 \hompc$.
Note that in all these formulae the $L_*$ refers to the
Schechter function fit to the overall 2dFGRS luminosity function.

The minimum variance weighting function is given by \citepalias{PVP}
\begin{equation}
\label{eq:rweight}
w(\r,L) = \frac{b^2(L)\, \wa(\r) }{1 + 4 \pi (J_3/b^2_{\rm T}) \int
b^2(L^\prime) \, \bngl(\r,L^\prime)
\, dL^\prime} .
\end{equation}
Here $\bngl(\r,L^\prime) $ ($\equiv \alpha \bnrl(\r,L^\prime)$) is
the expected mean density of galaxies of luminosity $L^\prime$ at
position $\r$ in the survey. Our standard choice for $J_3$ is
$400\,h^{-3}$Mpc$^3$ and refers to the value for typical galaxies
in the weighted 2dFGRS, for which the typical bias factor relative
to that of $L_*$ galaxies is $b_{\rm T}=1.26$. To revert to the
standard weighting function for the \citetalias{FKP} estimator, we
replace $b(L)$ by $b_{\rm T}$. The weighting function, $w(\r,L)$,
takes account of the galaxy luminosity function, varying survey
magnitude limits, varying completeness on the sky and its dependence
on apparent magnitude. The angular weight $\wa(\r)$ has a mean of
unity and gives a statistical correction for missing close
pairs of galaxies caused by fibre placing constraints (see Section~\ref{w_pairs}
for details).

The factor $\alpha$ in (\ref{fr}) is related to the ratio of the number
of galaxies in the random catalogues to that in 
the real galaxy catalogue. It is defined as
\begin{equation}
\alpha = \frac{ \int\!\!\!\int \frac{w(\r,L)}{b(L)}\, \ngl(\r,L)\, dL\, d^3\r}
{ \int\!\!\!\int \frac{w(\r,L)}{b(L)}\,  \nrl(\r,L)\, dL\, d^3\r},
\end{equation}
which reduces to a sum over the real and random galaxies
\begin{equation}
\alpha = \sum_{\rm data} \frac{w_i}{b_i} \; \Big / \, \sum_{\rm random} \frac{w_i}{b_i}  .
\end{equation}
Similarly, the constant $N$ in (\ref{fr}), which normalizes the survey
window function, is defined as
\begin{equation}
N^2 \equiv \int \left[ \int \bngl(\r,L) w(\r,L) dL \right]^2 d^3\r ,
\end{equation}
which can be written as a sum over the random galaxies
\begin{equation}
N^2 = \alpha \sum_{\rm random}  \bng_i  w_i^2 ,
\end{equation}
where $\bng_i$ is the expected mean galaxy density at the position
of the $i$th galaxy in the random catalogue.
This quantity is evaluated and tabulated at the position of each
galaxy so that we can use this simple summation to evaluate $N^2$.

To evaluate $F(\r)$ we loop over real and random galaxies, calculate
their spatial positions assuming   a flat $\Om =0.3$ cosmology, and use
cloud-in-cell assignment \citep[e.g.][]{efstathiou85}
 to  accumulate the difference
in $(w_i/b_i)_{\rm data} - \alpha (w_i/b_i)_{\rm random} $ on a
grid. We first do this with a $256^3$ grid in a cubic box of
$L^0_{\rm box} = 3125 \mpcoh$. Periodic boundary conditions are
applied to map galaxies whose positions lie outside the box. To
obtain estimates at smaller scales we repeat this with $256^3$ grids
of size $L_{\rm box} = L^0_{\rm box}/4$, $L^0_{\rm box}/16$.  We
then use an FFT to Fourier transform these fields and explicitly
correct for the smoothing effect of the cloud-in-cell assignment
\citep[e.g.][chap.~5]{he81}. From each grid we retain only estimates
for $k<0.63\, k_{\rm Nyquist}$, where the correction for the effects
of the grid are highly accurate. Thus, from the largest box we sample
the power spectrum well on a 3D grid of spacing $dk\simeq 0.002 \hompc$
covering $0.002<k<0.16 \hompc$.
The smaller boxes give a coarse sampling of the power spectrum with
resolutions of $dk = 0.008$ and $0.032 \hompc$ well
into the non-linear regime $0.16<k<2.5 \hompc$, where
our estimates become shot noise limited.

The shot noise corrected power spectrum estimator is
\begin{equation}
\hat P(\k) = \langle \vert F(\k) \vert^2 \rangle - P_{\rm shot},
\end{equation}
where  $P_{\rm shot}= S/N^2$ with
\begin{equation}
S \equiv \sum_{\rm data} \frac{w_i^2}{b_i^2} + \alpha^2
\sum_{\rm random} \frac{w_i^2}{b_i^2} .
\end{equation}
Finally, we average the power over direction, in shells of
fixed $|{\bf k}|$ in redshift space.

\begin{figure}
\epsfxsize=0.9\columnwidth \centering  \epsfbox{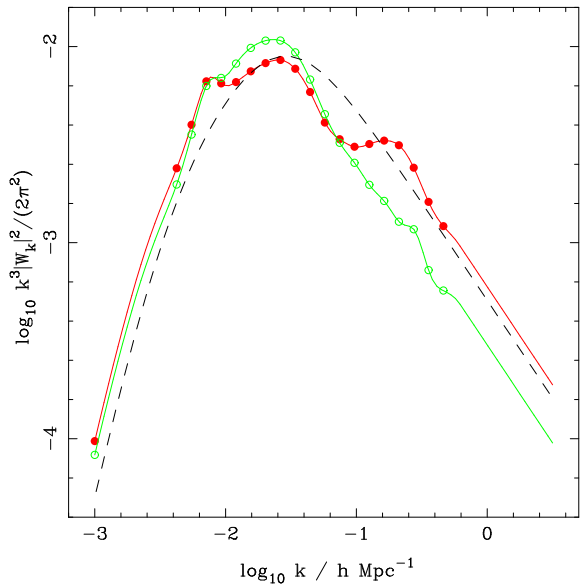}
\caption{ The amplitude of the spherically averaged 2dFGRS window
function of our standard weighted dataset in Fourier space (filled
circles). The solid line passing through these symbols gives a
cubic spline \citep{press92} fit to these data and was used to perform
the spherical convolution of model power spectra. 
For comparison, we plot the window function and spline fit
that results from excluding the random fields (open circles). 
Also the dashed line shows 
the fit used in \citeauthor{P01} (\citeyear{P01}; P01),
which ignores the structure in the window for $k>0.02\hompc$.
\label{fig:win}}
\end{figure}

The power spectrum, $\hat P(k)$, is an estimate of the true
underlying galaxy power spectrum convolved with the 
power spectrum of the survey window function,
\begin{equation}
W(\r) =   \frac{\alpha}{N} \int w(\r,L)
\, \nrl(\r,L) \, dL.
\end{equation}
Thus, to model our results we also need
an accurate estimate of the window function. This we obtain
using the same techniques as above.
The normalization is such that $\int W^2(\k)\; d^3\k=1$.
Under the approximation that the underlying power spectrum
is isotropic, i.e. ignoring redshift distortions, then
the operations of spherical averaging and convolving commute.
In Section~\ref{sec:mock_analysis}
we test the effect of redshift distortions via direct
Monte Carlo simulations using the Hubble volume mock
catalogues described in Section~\ref{sec:HV_mocks}.
Thus, all that we require is a model of the spherically
averaged window function. 
The curve going through the filled circles
in Fig.~\ref{fig:win} shows the window function 
that results for our standard choice of data cuts and weights. 
The window function marked by the open circles results from removing
the random fields. This comparison shows that the secondary
peak in the window function of our standard dataset is due to
the discrete random 2~degree fields.
Also shown, as the dashed line, is the window function computed for the
smaller 100k sample in P01.

We use a 2-step process to compute the effect of this window on the
recovered power spectrum. Firstly we interpolate the measured
window function using a
cubic spline \citep{press92}; examples of these interpolated window functions
are shown by the solid lines in Fig.~\ref{fig:win}. Secondly, we use a
modified Newton-Cotes integration scheme to perform a
spherical integration numerically using this fit and determine the
$k$-distribution of power required for each data point. This
integration is performed once, with the result stored in a `window matrix' 
giving the contribution from each of 1000 bins linearly
spaced in $0<k<2 \hompc$ to each measured $P(k)$ data point. We have
performed numerical integrations with fixed convergence limits for a
number of power spectra, and find results similar to those calculated
using this matrix. The effect of the 2dFGRS window on the recovered
power is demonstrated in the next Section using mock catalogues.

\subsection{Mock catalogues}
\label{sec:mocks}

To determine the statistical error in our power spectrum estimates
and also to test our codes thoroughly, we employ two sets of mock
catalogues.

\subsubsection{Hubble volume mocks}  \label{sec:HV_mocks}

The first set of mock catalogues are based on the $\Lambda$CDM
Hubble Volume cosmological N-body simulation \citep{evrard02}. The
Hubble Volume simulation contained $10^9$ particles in a box of
comoving size $L_{\rm box}=3000 \mpcoh$ with cosmological parameters
$h=0.7$, $\Om  =0.3$, $\Omega_{\Lambda} =0.7$,
$\Ob  =0.04$ and $\sigma_8=0.9$ . The galaxies are biased
with respect to the mass. This is achieved by computing the local
density, $\delta_{\rm s}$, smoothed with a Gaussian of width $r_{\rm s}=2
\mpcoh$, around each particle in the simulation and selecting the
particle to be a galaxy with a probability 
\begin{equation}
P(\delta_{\rm_s}) \propto \cases { 
\exp(0.45\, \delta_{\rm s} -0.14\, \delta_{\rm s}^{3/2}) & 
$\delta_{\rm_s}\ge 0$ 
\cr
\exp(0.45\, \delta_{\rm s}) & $\delta_{\rm_s}<0$ 
}
\label{eq:HV_bias}
\end{equation}
\citep{cole98}. 
The constants in this expression were chosen to produce a galaxy
correlation function matching that of typical galaxies in the
2dFGRS. This can be seen in figure~6 of \citet{hawkins03} as in this
analysis of the 2dFGRS correlation function we used the same set of
22 mock catalogues. They were also used in the analysis
of the dependence of the correlation function on galaxy type and luminosity
\citep{norberg_xil,norberg_xieta} and when analysing higher order
counts in cells statistics \citep{baugh04,croton04}.

The attractive features of the \HV\ mocks are that
their clustering properties are a good match to that of 2dFGRS and that 
they are fully non-linear: their density field is
appropriately non-Gaussian and they have realistic levels of
redshift space distortion. The limitations are that they lack
luminosity or colour dependent clustering and that the 22 simulations
are too few to determine the power covariance matrix accurately. 
We could generate
more catalogues, but given the finite volume of the Hubble Volume
simulation this would be of little value as they are not strictly
independent.

\subsubsection{Log-normal mocks} \label{sec:LN_mocks}

For an accurate determination of the covariance matrix of our power spectrum
estimates, we need sets of mock catalogues with of order 1000
realizations. In P01, we achieved this by generating realizations of
Gaussian random fields. Here, we slightly improve on this method by
generating fields of a specified power spectrum with a log-normal
1-point distribution function.  The log-normal model \citep{coles91}
is known to match both the results of large scale structure
simulations \citep{kayo01} and agree empirically with 1-point
distribution function of the 2dFGRS galaxy density field on large
scales \citep{wild05}. 
The power spectrum we adopted for these
mocks was generated using the Eisenstein \& Hu (\citeyear{EisensteinHu98}) 
algorithm with
cosmological parameters $\Om  h = 0.168$ and $\Ob /\Om  = 0.17$. 
The normalization we chose corresponds
to $\sigma_8^{\rm gal}=0.89$ for $L_*$ galaxies 
and $\sigma_8^{\rm gal}=1.125$ for the
typical galaxy in our weighted 2dFGRS catalogue. The method for
constructing the log-normal field and random galaxy catalogue is
similar to that described by \citetalias{PVP}.

We generate a log-normal field with the required power spectrum in a
cuboid aligned with the principal axes of the 2dFGRS. We have chosen
to use a cuboid of dimensions $3125\times  1565.25\times  3125
\hmpc$ covered by a grid of $512\times 256 \times 512$ cubic cells.
To convert this field into a mock catalogue we simply loop over all
the galaxies in our random catalogue, determine which cell they 
occupy (applying periodic boundary conditions if necessary) and select
the galaxy according to a Poisson probability distribution. The mean of the
Poisson distribution is modulated by the amplitude of the log-normal field and
normalized to achieve the right overall number of galaxies in the
mock catalogue.

These catalogues are computationally cheap, so we can generate
sufficient realizations to determine the power covariance 
matrix accurately. Also, by modulating the rms
amplitude of the log-normal field we can build in luminosity and
colour dependent clustering. Their limitations are that they are
restricted to quite large scales, the level of non-Gaussianity is
not necessarily realistic and they have no redshift space
distortion. We assess these shortcomings by comparison to the Hubble
Volume mocks.

\begin{figure}
\epsfxsize = 8.4 truecm \centering
\epsfbox{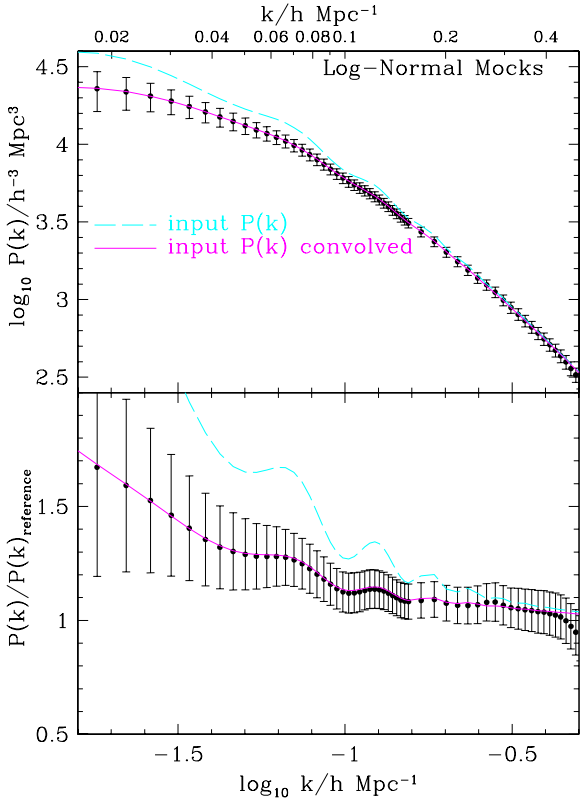}
\caption{Comparison of the recovered and input power spectra
for a set of log-normal mocks. The two curves in the
upper panel show the model input power spectrum (dashed)
and its convolution with the window function
of the catalogue (solid). The mean recovered power spectrum
from a set of 1000 mocks and the rms scatter about this
mean are shown by the points and error bars.
In the lower panel, instead of using a logarithmic scale
we plot, on a linear scale, the ratio of the three
power spectra of the top panel to a reference model with $\Om  h =0.2$
and $\Ob =0$. The line types and symbols have the same
meaning as in the upper panel.
}
\label{fig:pk_ln1}
\end{figure}

\begin{figure} 
\epsfxsize = 8.4 truecm \centering
\epsfbox{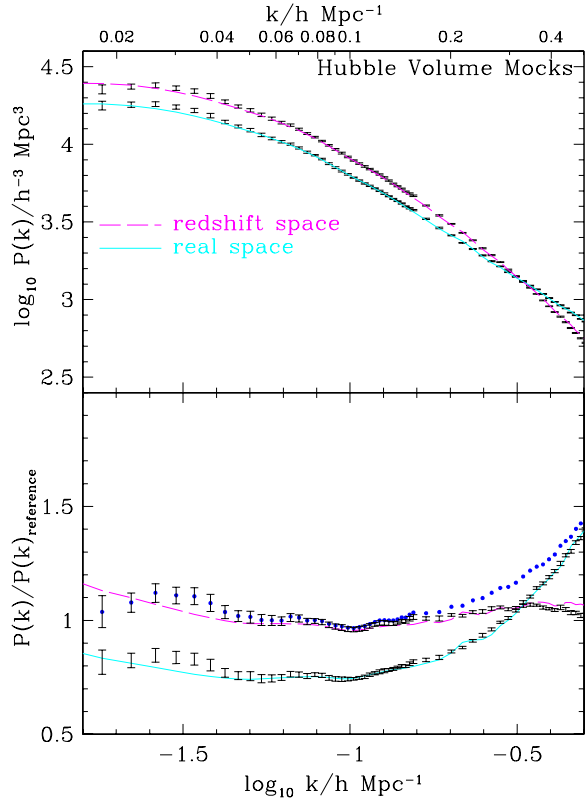}
\caption{Comparison of the expected and recovered power spectra for
the Hubble Volume mock catalogues. For mocks constructed using both
real-space and redshift-space galaxy positions, we compare the input
non-linear power spectrum (curves) with the mean recovered power
spectrum (error bars).  Unlike Fig.~\ref{fig:pk_ln1}, the error bars
here indicate the error in the mean recovered power, computed assuming
the 22 mocks to be independent. The lower panel shows, on a linear
scale, these same two power spectra but divided
by the same reference model as in Fig.~\ref{fig:pk_ln1} with
$\Om  h =0.2$ and $\Ob =0$.
Also shown as filled circles in the lower panel is the estimated power
from the 22 mock catalogues in redshift space but after applying
the cluster collapsing algorithm.
These match the redshift-space estimates on large
scales but have more power on small scales.
\label{fig:pk_hub1} }
\end{figure}

\subsubsection{Analysis of mock catalogues}
\label{sec:mock_analysis}

We now apply the method for estimating the power spectrum, described
in Section~\ref{sec:estimator}, to our two sets of mock catalogues. This
exercise allows us to test our code, illustrate the effect of the
window function and assess the level of systematic error that
results from ignoring the anisotropy of the redshift space power
spectrum.

Fig.~\ref{fig:pk_ln1} compares power spectrum estimates from the
log-normal mocks with the input power spectrum.  These mocks have
clustering that depends on luminosity according to equation~\ref{eq:bias} and
are analysed using the \citetalias{PVP} method assuming the same dependence
of bias parameter on luminosity.  The dashed curve shows the intrinsic input
power spectrum and the solid curve the result of convolving it with the
survey window function.  In the lower panel one sees that the
baryon oscillations present in the input power spectrum are greatly
suppressed by the convolution with the window function. The points
with error bars show the mean recovered power spectrum and the rms
scatter about the mean for a set of 1000 mocks.  We see that for 
$k<0.4 \hompc$ the mean recovered power is in excellent agreement with the
convolved input spectrum.  In particular, there is no perceptible glitch
in the recovered power at $k=0.16 \hompc$ where we switch between
the $3125$ and $781.25 \mpcoh$ boxes used for the FFTs.
The \citetalias{PVP} method has correctly recovered the
input power spectrum with no biases due to the luminosity dependence of
the clustering.  At the edge of the plots, as we approach the
Nyquist frequency, $k_{\rm ny}=0.51 \hompc$, of the grid on
which the log-normal field was
generated, the recovered power begins to deviate significantly from
input power. For $k>0.4 \hompc$ our log-normal mocks are of
limited value.

Fig.~\ref{fig:pk_hub1} compares the recovered power spectra with the
expected values for three sets of Hubble Volume mocks.  As the bias
is independent of luminosity for these samples, the power spectrum
estimator we use is equivalent to the \citetalias{FKP} method. In
the first set of mocks, redshift space distortions were eliminated
by placing the galaxies at the their real space positions.  Here, the
power spectrum is isotropic (as in the log-normal mocks) and
again we expect, and find, that the recovered power spectrum
accurately matches the exact non-linear spectrum from the full \HV,
convolved with the survey
window function. The error bars
shown on this plot are the errors in the mean power. The rms error
for an individual catalogue will be $\sqrt{21}$ times larger,
comparable to the error bars in Fig.~\ref{fig:pk_ln1}. The second
set of points in Fig.~\ref{fig:pk_hub1} are the Hubble Volume mocks
constructed using the galaxy redshift space positions. These are
compared to the expectation computed by taking the spherically
averaged redshift space power spectrum from the full simulation cube
and convolving with the window function.  We see that over the range
of scales plotted, the recovered power agrees well with this
expectation. This indicates that ignoring the anisotropy when
fitting models will not introduce a significant bias.

In the third set of Hubble Volume mocks, groups
and clusters were identified in the redshift space mock catalogues
using the same friends-of-friends algorithm and parameters that
\citet{eke04} used to define the 2PIGG catalogue of 2dFGRS groups
and clusters. Each group member was then shifted to the mean group
redshift perturbed according to a Gaussian random distribution with width
corresponding to the projected group size. This has the effect of
collapsing the clusters along the redshift space direction, removing
the `fingers of god' and making the small scale clustering much
less anisotropic. In the lower panel of Fig.~\ref{fig:pk_hub1} we
see that, on large scales, this procedure has no effect on the
recovered power. In contrast, on small scales the smoothing effect
of the random velocities of galaxies in groups and clusters is
removed and the recovered power spectrum has a shape much closer to that
of the real space mocks. In Section~\ref{sec:params} we will compare the
results of analyzing the genuine 2dFGRS data in redshift space with
and without this cluster collapsing algorithm.

\begin{figure*}
\epsfxsize = 16.8 truecm \centering
\epsfbox[0 40 575 775]{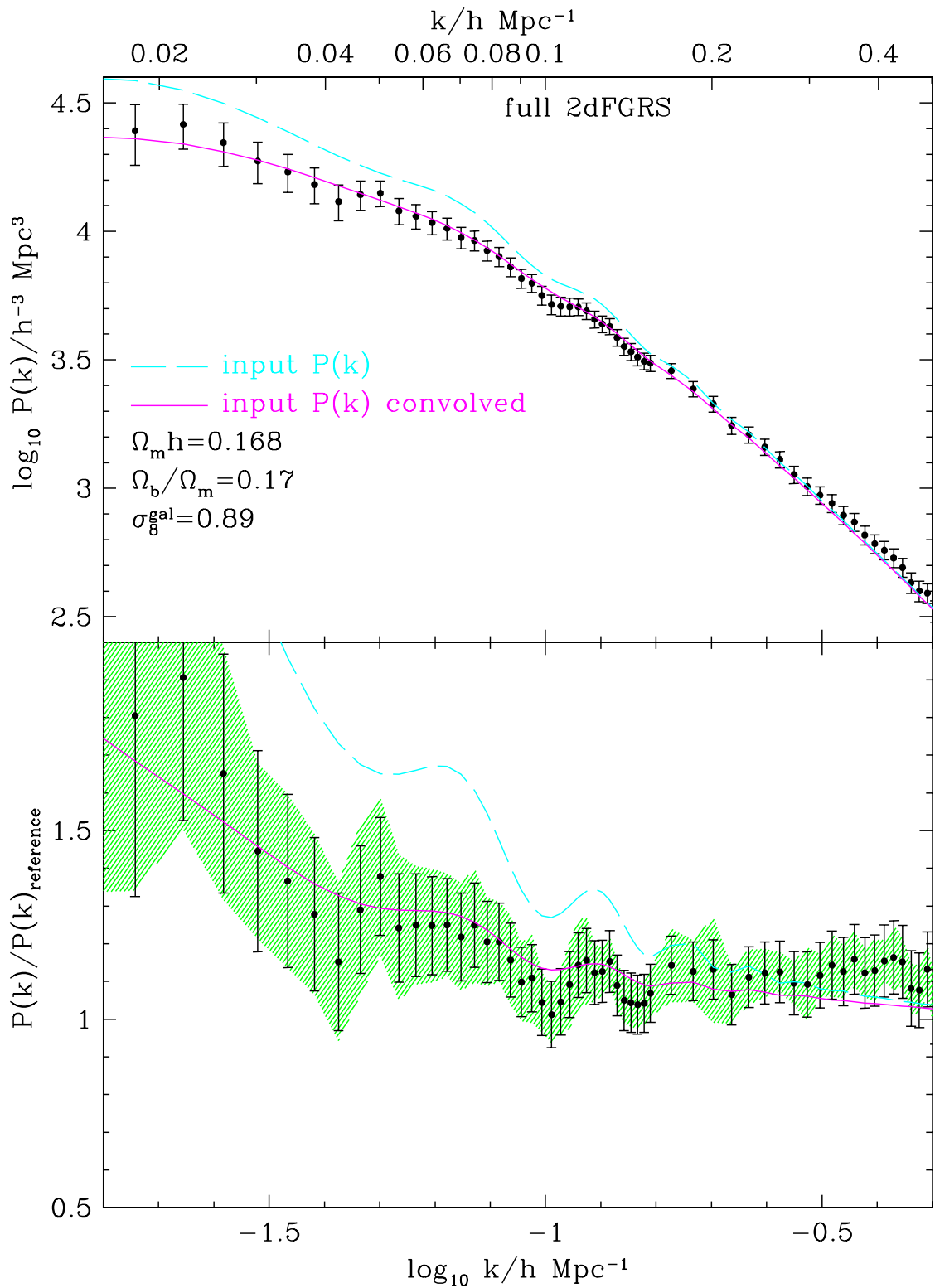}
\caption{ The data points show the
recovered 2dFGRS redshift space galaxy power spectrum
for our default set of cuts and weights.  The curves show the same
realistic model as in Fig.~\ref{fig:pk_ln1}, both before and after
convolving with the survey window function. In the lower panel, where
we have again divided through by an unrealistic reference model with 
$\Om  h =0.2$ and $\Ob =0$, we show both the log-normal estimate of
the errors (error bars) and an alternative error estimate based on
jack-knife resampling of the 2dFGRS data (shaded region). Note that
the window function, shown in Fig.~\ref{fig:win}, causes the data points 
to be correlated.}
\label{fig:pk_2dfgrs}
\end{figure*}

\begin{table}
\caption{The 2dFGRS redshift-space power spectrum.
The 3rd column gives the square root of the
diagonal elements of the covariance matrix calculated from 1000 
realizations of model log-normal density fields. 
The 4th column gives an alternative empirical estimate of the error 
based on 20 jack-knife samples.
The 5th column gives the value of $P(k)$ 
convolved with the survey window function
for a fiducial linear theory model 
with $\sigma_8^{\rm gal}=0.89$, 
$h=0.7$, $\Omega_{\rm m} h=0.168$, 
$\Omega_{\rm b}/\Omega_{\rm m}=0.17$. 
}
\begin{center}
\begin{tabular}{ccrrrrrrrrrrrr} 
\hline
\multicolumn{1}{c} {$k/\hompc$} &
\multicolumn{1}{l} {$\;\;P_{\rm 2dFGRS}(k)$} & 
\multicolumn{1}{l} {$\;\;\sigma_{\rm LN}$} &
\multicolumn{1}{l} {$\;\;\sigma_{\rm jack}$} &
\multicolumn{1}{l} {$P_{\rm ref}(k)$} \\
\hline 
 0.010 &  43\,791.0 &  19\,640.0 &  15\,571.9 &  22\,062.9 \\
 0.014 &  27\,021.7 &   9\,569.3 &   9\,538.0 &  23\,280.4 \\
 0.018 &  24\,631.7 &   7\,058.4 &   6\,291.8 &  22\,818.3 \\
 0.022 &  26\,076.4 &   6\,201.8 &   5\,442.2 &  21\,783.8 \\
 0.026 &  22\,163.8 &   4\,603.7 &   4\,441.0 &  20\,477.8 \\
 0.030 &  18\,784.6 &   3\,430.5 &   3\,006.7 &  18\,991.5 \\
 0.034 &  17\,050.0 &   2\,785.1 &   2\,850.8 &  17\,524.0 \\
 0.038 &  15\,233.3 &   2\,283.4 &   2\,521.6 &  16\,153.5 \\
 0.042 &  13\,069.6 &   1\,801.0 &   2\,349.1 &  14\,985.6 \\
 0.046 &  13\,904.3 &   1\,808.1 &   2\,420.1 &  14\,040.4 \\
 0.050 &  14\,085.4 &   1\,703.1 &   2\,110.7 &  13\,183.9 \\
 0.054 &  12\,021.6 &   1\,348.5 &   1\,840.4 &  12\,405.9 \\
 0.058 &  11\,452.8 &   1\,221.2 &   1\,414.9 &  11\,738.7 \\
 0.062 &  10\,829.3 &   1\,099.9 &   1\,283.4 &  11\,114.0 \\
 0.066 &  10\,269.5 &    985.9 &   1\,115.3 &  10\,490.4 \\
 0.070 &  \phantom{0}9\,477.6 &    870.1 &   1\,088.4 &   9\,849.1 \\
 0.074 &  \phantom{0}9\,209.2 &    822.0 &   1\,107.1 &   9\,205.2 \\
 0.078 &  \phantom{0}8\,418.5 &    737.4 &    807.5 &   8\,571.7 \\
 0.082 &   \phantom{0}7\,985.5 &    682.6 &    697.7 &   7\,967.9 \\
 0.086 &   \phantom{0}7\,275.4 &    603.2 &    737.4 &   7\,426.2 \\
 0.090 &   \phantom{0}6\,557.0 &    521.3 &    607.7 &   6\,916.7 \\
 0.094 &   \phantom{0}6\,290.2 &    491.3 &    658.6 &   6\,462.1 \\
 0.099 &   \phantom{0}5\,636.1 &    440.8 &    421.1 &   6\,070.1 \\
 0.103 &   \phantom{0}5\,196.2 &    407.6 &    385.8 &   5\,748.2 \\
 0.107 &   \phantom{0}5\,113.0 &    401.2 &    406.1 &   5\,479.2 \\
 0.111 &   \phantom{0}5\,086.4 &    393.4 &    536.8 &   5\,242.0 \\
 0.115 &   \phantom{0}5\,080.4 &    384.2 &    515.6 &   5\,028.1 \\
 0.119 &   \phantom{0}4\,902.5 &    366.8 &    482.1 &   4\,820.7 \\
 0.123 &   \phantom{0}4\,549.7 &    338.3 &    298.1 &   4\,606.2 \\
 0.127 &   \phantom{0}4\,362.7 &    317.4 &    244.0 &   4\,392.3 \\
 0.131 &   \phantom{0}4\,269.7 &    310.0 &    241.1 &   4\,181.5 \\
 0.135 &   \phantom{0}3\,862.7 &    278.2 &    220.7 &   3\,969.9 \\
 0.139 &   \phantom{0}3\,563.6 &    257.3 &    209.2 &   3\,767.2 \\
 0.143 &   \phantom{0}3\,396.6 &    244.7 &    205.1 &   3\,577.2 \\
 0.147 &   \phantom{0}3\,242.2 &    231.9 &    202.2 &   3\,401.9 \\
 0.151 &   \phantom{0}3\,121.7 &    222.4 &    162.3 &   3\,248.7 \\
 0.155 &   \phantom{0}3\,074.0 &    218.7 &    175.6 &   3\,112.9 \\
 0.169 &   \phantom{0}2\,867.9 &    203.8 &    239.5 &   2\,728.6 \\
 0.185 &   \phantom{0}2\,438.2 &    173.9 &    170.9 &   2\,362.8 \\
\hline
\end{tabular}
\end{center}
\label{tab:pk}
\end{table}

\begin{figure*}
\epsfxsize=17 cm \epsfbox[35 520 540 770]{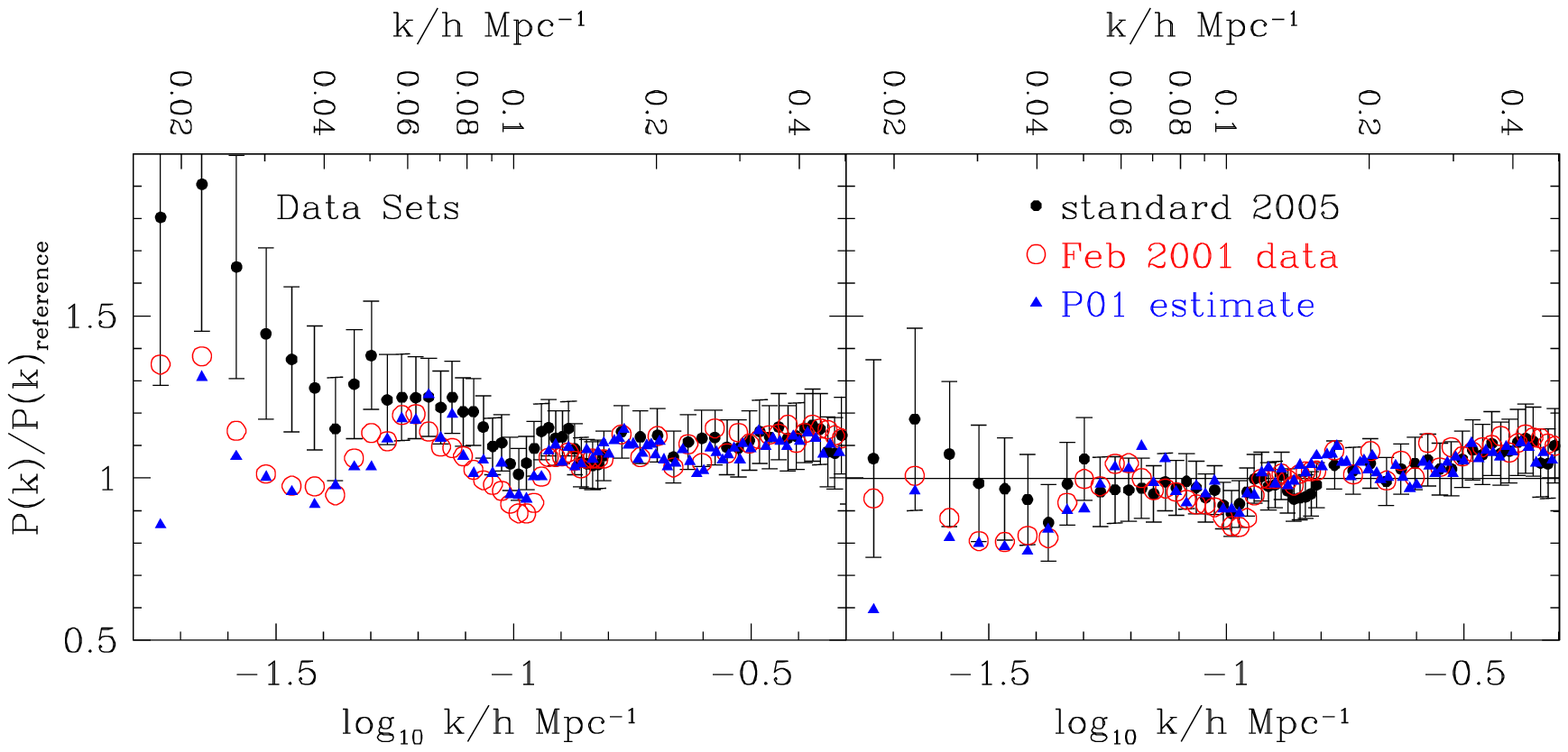} 
\caption{
Comparison of the power spectrum estimate from P01 with our current
estimates. 
To compare the amplitude of the new PVP and old FKP
estimates, we have scaled the FKP estimate by a factor
$\langle b^{-2} \rangle$ (the weighted average value of the 
bias factor appearing in equation~\ref{eqn:field}).
The left hand panel shows each power spectrum
estimate divided by a reference power spectrum with parameters
$\Om h=0.2$ and $\Ob /\Om =0$. In the
right hand panel the reference power spectrum has 
$\Om h=0.168, \Ob /\Om =0.17$,  and is convolved
with the window function of the final or February 2001 data as
appropriate. The solid circles with error bars show our standard
estimate from the final 2dFGRS catalogue. The triangles show the P01
estimate and the open circles show an estimate using only the
pre-February 2001 data but our current calibration and modelling of
the survey selection function. } \label{fig:pk_p01}
\end{figure*}

\section{Final 2\lowercase{d}FGRS results} \label{sec:first_results}

Fig.~\ref{fig:pk_2dfgrs} shows the application of the above machinery to
the 2dFGRS data for our default choice of selection cuts, weights and
model of the selection function. The error bars on this plot come
from a set of \LN\ mocks selected, weighted and analysed in the same
way. The model power spectrum of these mocks, shown by the curve,
has $\Om  h=0.168$, $\Ob /\Om = 0.17$
and $\sigma_8^{\rm gal}=0.89$ and closely matches what we recover from the
2dFGRS. The shaded region shows
as an alternative a jack-knife estimate of the power spectrum
errors. For this, we divided the 2dFGRS data into 20 samples split by
RA such that each sample contained the same number of galaxies. We
then made twenty estimates of the power, excluding one of the 20
regions in each case. The error bars are $\sqrt{20}$ times the rms
dispersion in these estimates. We see that the log-normal and
empirical jack-knife error estimate agree remarkably well. 

The survey window function causes the power estimates
to be correlated and so the plotted error bars alone do not allow one to 
properly assess the viability of any given model. If the
correlations were ignored then the model plotted in
Fig.~\ref{fig:pk_2dfgrs} would have an improbably low value of
$\chi^2$, whereas when the covariance matrix is used one
finds a very reasonable $\chi^2/{\rm d.f.}=37/33$ 
for $k<0.2\hompc$. At $k>0.3 \hompc$
the estimated power begins to significantly exceed that of the
linear theory model. This is due to non-linearity which we discuss
in Section~\ref{sec:nonlin}.
These power spectra and error estimates are tabulated in
Table~\ref{tab:pk}
\footnote{The power spectra estimates in Table~\ref{tab:pk}
 along with the full error covariance matrix are available 
in electronic form at
{\tt http://www.mso.anu.edu.au/2dFGRS/Public/Release/PowSpec/} }.
We show in Section~\ref{sec:system} that this power-spectrum 
estimate is robust with respect to
variations in how the dataset is treated and we fit models
to these data in Section~\ref{sec:params}.

\subsection{Comparison with Percival et al. (\citeyear{P01})}  
\label{sec:cmpr_P01}

In Fig.~\ref{fig:pk_p01} we compare our new power spectrum estimate
with that from P01. There are significant differences in the shape of
the recovered power spectrum on scales larger than $k<0.1 \hompc$, but
this is largely due to the difference in the window function. In the
right hand panel, where this has been factored out, the old and new
estimates only begin to differ significantly for $k<0.04 \hompc$.  
The main reason for this difference is sample variance. The
estimate shown by the open circles is based on the same dataset as
the P01 estimate, but uses the updated calibration, modelling of the
selection function and PVP estimator described in this paper. 
Our current model of the survey selection
function differs in many details from that used in P01, but in general
these differences make very little difference to the recovered power.
The two differences that cause a non-negligible change are the improvement
in photometric calibration and the empirical fitted model of the 
redshift distribution. The perturbation these changes cause 
are small, restricted to $k<0.04 \hompc$ and largely cancel one
another out. For the case of the final data set this is discussed
in Sections \ref{sec:sys_calib} and~\ref{model_dndz} and 
shown in Figs~\ref{fig:pk_var}d and~n.

\begin{figure}
\epsfxsize=8.5 truecm \centering  \epsfbox{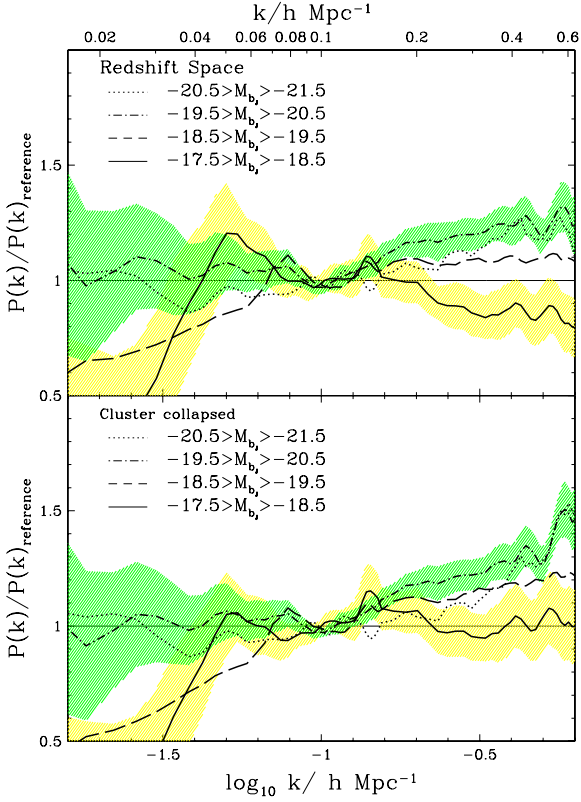}
\caption{The various lines show recovered power spectra from 2dFGRS
galaxies split into different bins of absolute
magnitude (in redshift space -- top panel and cluster-collapsed --
bottom panel). The power spectra have been divided by a reference
model with $\Om h=0.168$ and $\Ob /\Om =0.17$, convolved with
the window function corresponding to each data cut. We have
illustrated statistical errors estimated from 
log-normal mocks by showing the $\pm1\sigma$ range for two of the
samples by the corresponding shaded regions.
\label{fig:lumdep_pk}}
\end{figure}

\subsection{Dependence on luminosity}

The power spectrum we measure comes from combining galaxies of 
different types, whose clustering properties may be different.
We now complete the presentation of the basic results
from the survey by dissecting the power spectrum according
to galaxy luminosity and colour.

Fig.~\ref{fig:lumdep_pk} shows the power spectrum estimated as described in
Section~\ref{sec:estimator}, but for galaxies in fixed bins of absolute magnitude.
Because the 2dFGRS catalogue is limited in apparent
magnitude, each of these power spectrum measurements will have a
different window function; however, we can consider the
effect of the window on each power spectrum approximately by dividing
the recovered $P(k)$ by the appropriately convolved
version of a CDM model that fits the large-scale combined $P(k)$. The power
spectra have been renormalized to a common large-scale
($0.02<k<0.08\hompc$) amplitude.

The luminosity-dependent spectra show differences at large and small
scales. The variations at $k\ls 0.1 \hompc$ are cosmic variance:
the different redshift distributions corresponding to different
luminosity slices implies that the samples are close to independent.
Using the separate covariance matrices for these samples, a $\chi^2$
comparison shows that the large-scale variations are as expected.
The differences at high $k$, however,  reflect genuine differences in
the non-linear clustering and/or pairwise velocity dispersions
as a function of luminosity. We discuss below in Section~\ref{sec:nonlin}
how these systematic differences affect our ability to extract cosmological
information from the 2dFGRS.

\begin{figure}
\epsfxsize=8.5 truecm \epsfbox{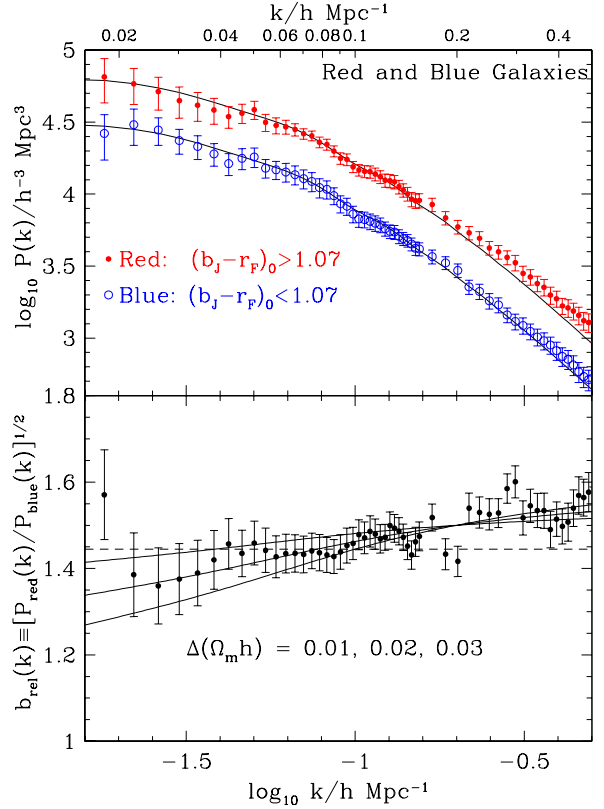} \centering
\caption{Power spectrum for matched red and blue galaxy subsamples. The
symbols and error bars in the upper panel show our estimates with
errors derived from the log-normal mocks. For reference, the solid
curves show the linear power spectrum used for the log-normal mocks,
which has $\Om h=0.168$ and $\Ob/\Om =0.17$. In each
case, the model power spectra are normalized according to the bias
parameters defined in equations (\ref{eqn:red}) and~(\ref{eqn:blue})
and convolved with the window function of the sample. The lower
panel shows the relative bias, the square root of the ratio of these
power spectra. The error bars, determined from our mock catalogues,
take account of the correlation induced by the fact the red and blue
subsamples sample the same volume. The horizontal line in
the lower panel shows the expectation for scale independent bias given
by the ratio of $b(L_*)$ for the adopted red and blue bias
factors from equations (\ref{eqn:red}) and~(\ref{eqn:blue}). The
solid curves show the ratio that would result if the red and blue
galaxies had power spectra that were well described by linear theory
models whose values of $\Om h$ differed by $0.01$, $0.02$
or $0.03$ from top to bottom on large scales. } \label{fig:pk_redblue}
\end{figure}

\subsection{Dependence on colour}  \label{sec:redblue}

In Fig.~\ref{fig:pk_redblue}, we show estimates of the galaxy power
spectrum for the two samples defined by splitting the catalogue at a
rest frame colour of $(\bj-\rf)_{z=0}=1.07$. As the redshift
distribution of the blue  sample is more extended than that of the
red, the optimal PVP weighting for the blue sample weights the
volume at high redshift more strongly. Since we wish to
compare the shapes of the red and blue power spectra it would be
preferable if they sampled the same volume. Hence, when
analyzing the blue sample, we have chosen to apply an additional
redshift dependent weight, so as to force the mean weight per unit
redshift to be the same for both samples. The estimates were made
using the PVP estimator and the bias parameters defined in equations
(\ref{eqn:red}) and~(\ref{eqn:blue}). However, to illustrate 
that at  fixed luminosity the red galaxies are more clustered than the
blue galaxies we have multiplied each estimate by their respective
values of $b(L_*)^2$, where $L_*$ is the characteristic luminosity
of the full galaxy sample. To first order, we see that the two power
spectra have very similar shapes, with both becoming more clustered
than the linear theory model on small scales. 

\begin{figure*}
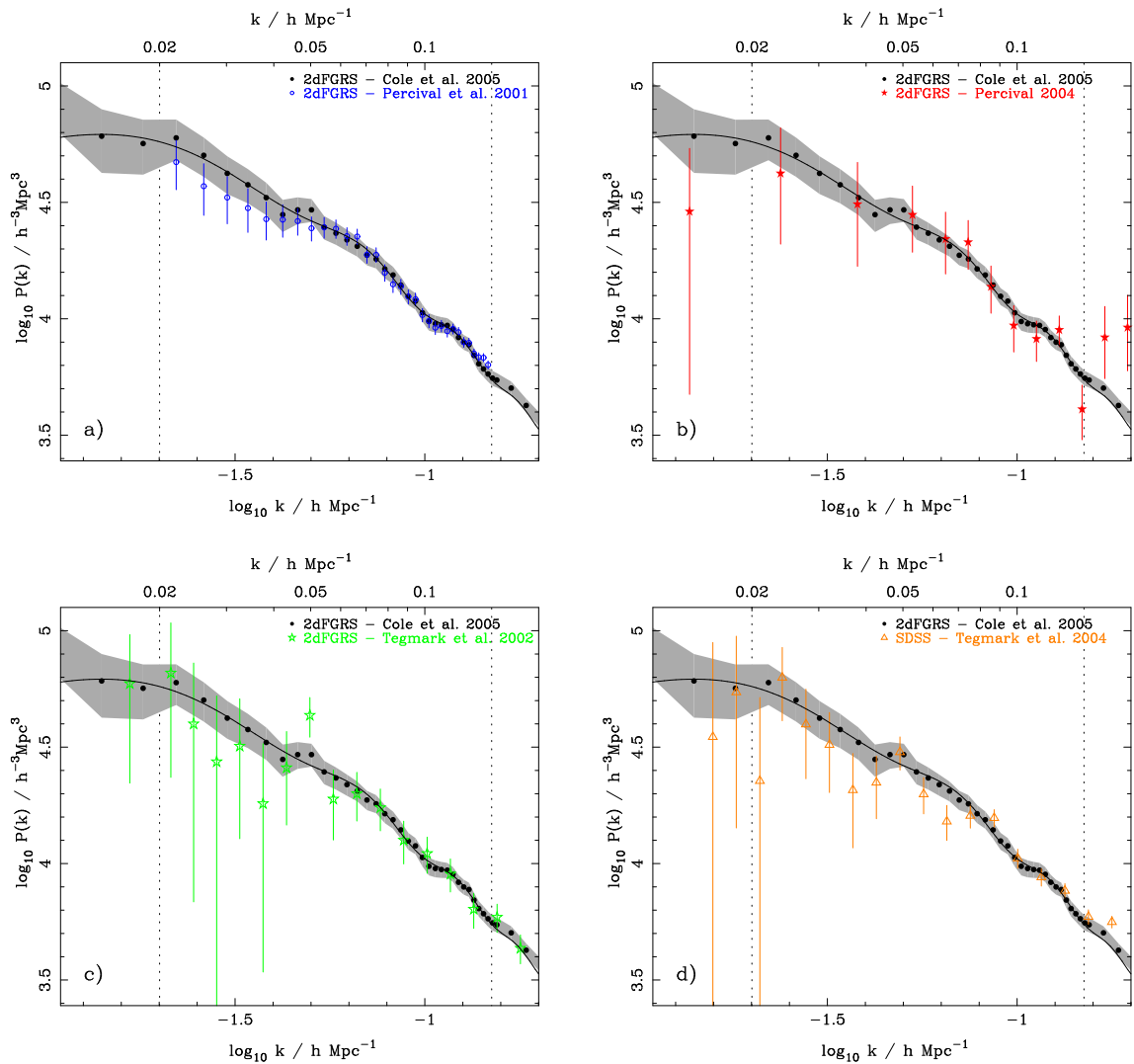

\hbox{
\epsfxsize=0.45\textwidth 
\epsfbox[65 110 600 570]{figs/pk_panel1.ps}
\epsfxsize=0.45\textwidth 
\epsfbox[65 110 600 570]{figs/pk_panel2.ps}}
\hbox{
\epsfxsize=0.45\textwidth 
\epsfbox[65 110 600 600]{figs/pk_panel3.ps}
\epsfxsize=0.45\textwidth 
\epsfbox[65 110 600 600]{figs/pk_panel4.ps}}
\caption{ The redshift-space power spectrum calculated in this paper
(solid circles with 1-$\sigma$ errors shown by the shaded region)
compared with other measurements of the 2dFGRS power spectrum shape by
a) Percival et al. (\citeyear{P01}), b)
Percival (\citeyear{percival04}), and c) Tegmark et     
al. (\citeyear{tegmark02}). For the data with window    
functions, the effect of the window has been approximately corrected
by multiplying by the net effect of the window on a model power
spectrum with $\Om h=0.168$, $\Ob /\Om =0.0$, $h=0.72$ \& $\ns=1$. A
zero-baryon model was chosen in order to avoid adding features into  
the power spectrum. All of the data are renormalized to match the new 
measurements. Panel d) shows the uncorrelated SDSS real space
$P(k)$ estimate of Tegmark et al. (\citeyear{tegmark04}), calculated using
their `modelling method' with no FOG compression (their
Table~3). These data have been corrected for the SDSS window as  
described above for the 2dFGRS data. The solid line shows a model
linear power spectrum with $\Om h=0.168$, $\Ob /\Om =0.17$, $h=0.72$,
$\ns=1$ and normalization matched to the 2dFGRS power spectrum.
\label{fig:pk_cmpr_SDSS}}
\end{figure*}

The lower panel shows
the relative bias, $ b_{\rm rel}(k) \equiv \sqrt{P_{\rm red}(k)/ P_{\rm
blue}(k)}$, as a function of scale. On large scales, this relative
bias is consistent with a constant and is, by construction, 
close to the value given by
the ratio of the adopted bias parameters of equations
(\ref{eqn:red}) and~(\ref{eqn:blue}) shown by the horizontal dashed
line. In fact, for $k < 0.12 \hompc$, fitting a constant bias using
the full covariance matrix produces a fit with  $\chi^2=25.5$ for
25~degrees of freedom. 
 We note that this value of the bias, $b_{\rm red}/b_{\rm blue}=1.44$
is in very good agreement with the 
$b_{\rm passive}/b_{\rm active}=1.45 \pm 0.14$ found in section~3.3 of
\citet{madgwick03}, when analysing the correlation function of
spectrally classified 2dFGRS galaxies. The value also agrees well
with that found in the halo model analysis of red and blue 2dFGRS 
galaxies by \citet{collister05}.
At smaller scales, there is an increasingly
significant deviation, with the red galaxies being more clustered
than the blue \citep[in agreement with][]{madgwick03}.
Also shown in the lower panel are curves indicating
the relative bias that would result if the red and blue power
spectra were well fitted by linear theory models whose values of
$\Om h$ differed by $0.01$, $0.02$ or $0.03$. 
From this we see that a simple fit of linear theory to the red
and blue samples would yield values of $\Om h$ that differ by
$\Delta \Om h \simeq 0.015$. This small difference is
comparable to the statistical uncertainty. In any case, in
Section~\ref{sec:nonlin} we discuss systematic nonlinear
corrections to the power, and show how a robust measurement
of $\Om h$ can be achieved even in the presence of small
distortions of the spectrum.

\subsection{Comparison with other power spectra}

In Fig.~\ref{fig:pk_cmpr_SDSS}, we compare the power spectrum
measured in this paper with previous estimates of the shape of the
power spectrum on large-scales measured from the 2dFGRS and SDSS. In addition
to the data of Percival et~al. (\citeyear{P01}), with which we compared in
detail in Section~\ref{sec:cmpr_P01}, we additionally plot the data
of Percival (\citeyear{percival04}) who extracted the real-space power 
spectrum from
the 2dFGRS. In that work, Markov-chain Monte-Carlo mapping of the
likelihood surface was used to deconvolve the power spectrum from a
Spherical Harmonics decomposition presented in Percival et~al.
(\citeyear{P04}). Because of the method used, a cut-down
version of the final 2dFGRS catalogue was analysed with a radial
selection function that was independent of angular position.
Consequently, the volume analysed is smaller, and this method
provides weaker constraints on the power spectrum shape. However, we
see from Fig.~\ref{fig:pk_cmpr_SDSS}b that the general shape of the
recovered power spectrum is very similar over $0.02<k<0.15\hompc$,
the range of scales probed in Percival (\citeyear{P04}).

In Fig.~\ref{fig:pk_cmpr_SDSS}c we
plot the power spectrum measured by Tegmark et al. (\citeyear{tegmark02}) from
the 2dFGRS 100k data release. Because of the weighting scheme they used, 
these data are expected to be tilted relative to
the true power spectrum because of
luminosity-dependent bias. The plot shows evidence for such a bias 
and the Tegmark et al. (\citeyear{tegmark02}) data have a
lower amplitude on large scales than any of the other 2dFGRS $P(k)$
measurements. Given the small sample analysed, these data
provide a far weaker constraint on the power spectrum shape than
our current analysis.

In addition to the 2dFGRS power spectrum measurements described
above, we also plot in Fig.~\ref{fig:pk_cmpr_SDSS}d
the recent estimate from the SDSS by Tegmark
et~al. (\citeyear{tegmark04}). This analysis differed from 
the analysis of the 2dFGRS by
Tegmark et al. (\citeyear{tegmark02}) by including a crude correction for
luminosity-dependent bias, which corrects for an amplitude offset
for each data point, but does not allow for the changing survey
volume (Percival et al. \citeyear{PVP}). 
The SDSS work quotes a somewhat larger value of $\Omega_{\rm m}h$ than
that found here: $0.213 \pm 0.023$, which is formally a $1.6-\sigma$
deviation. However, this SDSS figure assumes a known baryon fraction,
which makes the error on $\Omega_{\rm m}h$ unrealistically low. As can be
seen from Fig.~\ref{fig:pk_cmpr_SDSS}d, the basic shapes of the 2dFGRS and
SDSS galaxy power spectra in fact agree remarkably well.

We have chosen not to compare with galaxy power spectrum estimates
obtained from surveys prior to the 2dFGRS, or calculated by
deprojecting 2D surveys because the 2dFGRS and SDSS data offer a
significant improvement over these data. However, we do note that the
general shape of our estimate of the power spectrum is very similar to
that obtained in such studies (e.g. Efstathiou \& Moody \citeyear{efstathiou01}; Padilla \& Baugh \citeyear{PB03}; 
Ballinger, Heavens \& Taylor \citeyear{ballinger95}; Tadros et
al. \citeyear{tadros99}).

\section{Tests of systematics}
\label{sec:system}

Given the cosmological significance of the 2dFGRS power spectrum estimates,
it is important to be confident that the results presented in the previous Section
are robust, and not sensitive to particular assumptions made in the
analysis. This Section presents a comprehensive investigation
into potential sources of systematic error in the final result.

\renewcommand{\theenumi}{(\roman{enumi})} 

Our default set of assumptions in modelling and analyzing the
2dFGRS data are:
\begin{enumerate}
\item Our standard choice for the photometric calibration
of the catalogue is essentially that of the final data
release \citep{colless03}
but with small shifts of $-0.0125$ and $0.022$ mag.
applied to the NGP and SGP respectively to bring their
estimated luminosity functions into precise agreement.
\item We combine data from the NGP and SGP strips and also the RAN
fields.
\item We model the galaxy population
by a single Schechter luminosity function and $k+e$ correction as
described in Section~\ref{sec:lf} and shown in Fig.~\ref{fig:lf}.
Magnitude measurement errors are then applied using the empirical
model of Norberg \etal (\citeyear{norberg_selfun} see their
figure~3f).
\item Incompleteness in the redshift survey is modelled
in the mock catalogues using a combination of the
mean completeness in each sector $R(\theta)$
(Fig.~\ref{fig:zcomp}) and its dependence on apparent
magnitude as parameterized by $\mu(\theta)$
(Fig.~\ref{fig:mumask}; see Colless \etal \citeyear{colless01} Section~8
and appendix~A of Norberg \etal \citeyear{norberg_selfun} for details).
\item We discard data from sectors with redshift completeness
$R(\theta)<0.1$.
\item We impose a maximum redshift of $z_{\rm max}=0.3$ .
\item We use the \citetalias{PVP} estimator with the bias parameter given
by equation~\ref{eq:bias}.
\item We use angular weights that attempt to correct
for missed close pairs due to fibre collisions and positioning constraints.
Their construction
is explained in Section~\ref{w_pairs}.
\item We use the radial weighting given by equation
(\ref{eq:rweight}) with $J_3=400 h^{-3}$Mpc$^3$.
\end{enumerate}

\begin{figure*}
\epsfxsize = 16.0 truecm \centering
\epsfbox[25 55 540 765]{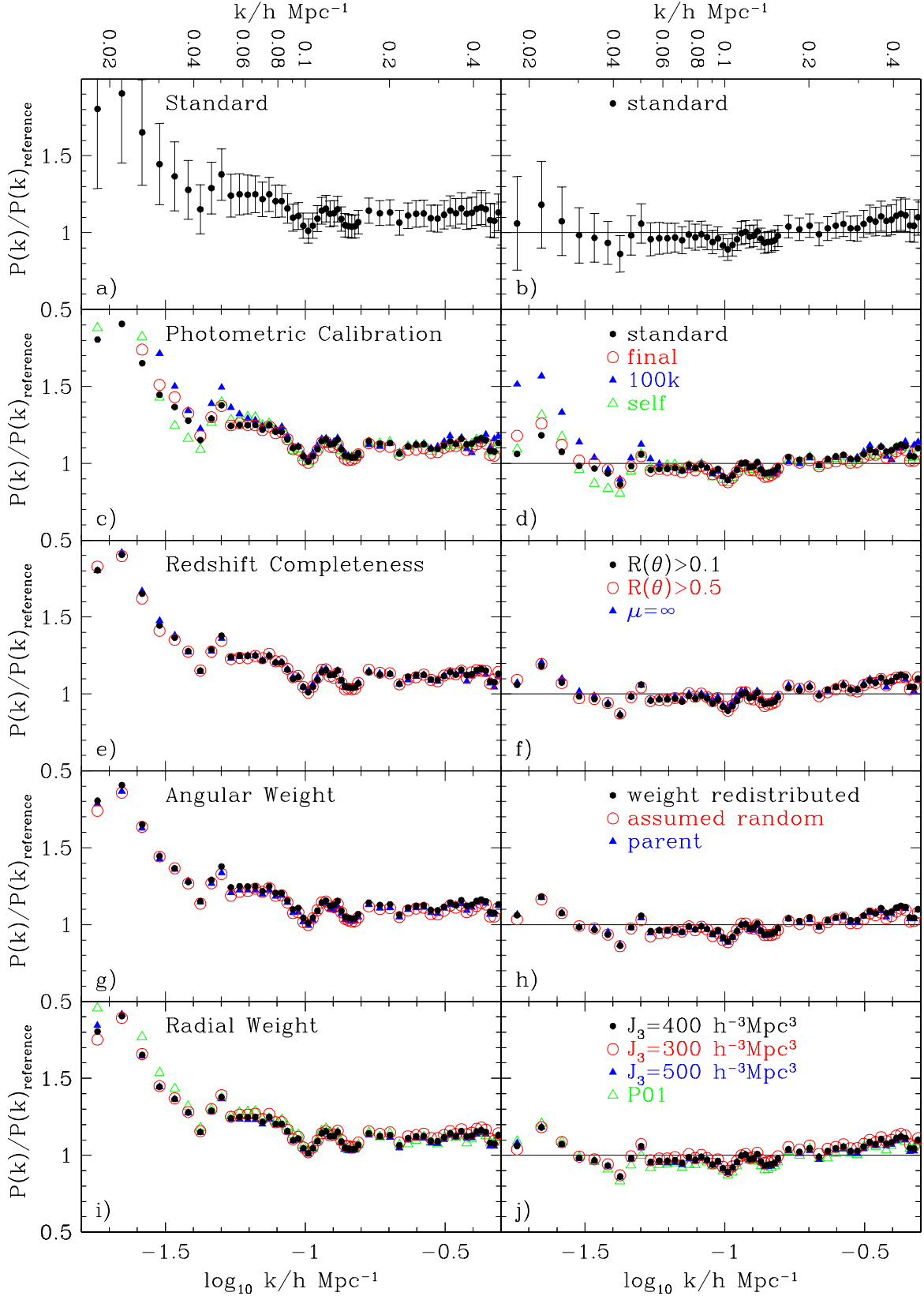}
\caption{
Test power spectra calculated for different data cuts and
assumptions. The data are divided by a reference power
spectrum. In the left hand column, the reference power spectrum has 
parameters $\Om h=0.2$ and $\Ob /\Om =0$.
In the right hand column, the reference power spectrum has
$\Om h=0.168, \Ob /\Om =0.17$
(as used for the log-normal mock catalogues), and is
convolved with the correct window function (which varies with data
cuts and weighting scheme).
The top row shows the power spectrum estimate and associated statistical
errors resulting from our standard choices of data cuts and weighting.
Subsequent rows give results for different tests, as described in
Section~\ref{sec:system}.
}
\label{fig:pk_var}
\end{figure*}

\begin{figure*}
\epsfxsize = 16.0 truecm \centering
\epsfbox[25 55 540 765]{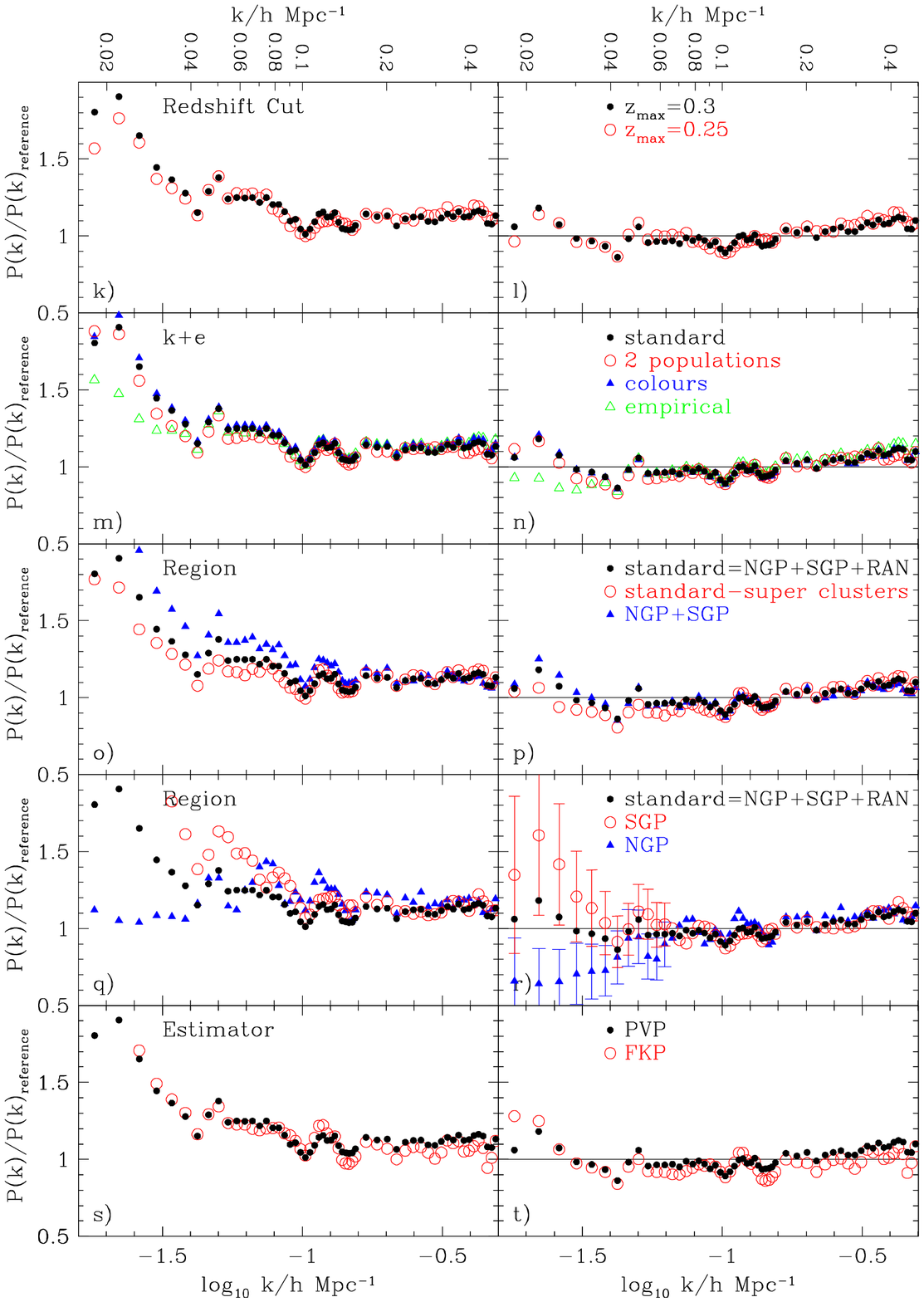}
\contcaption{}
\end{figure*}

In Fig.~\ref{fig:pk_var}, the left hand panels show the ratio of 
estimated power spectra to an (unrealistic) reference model power spectrum with
$\Om h=0.2$ and $\Ob /\Om =0$. These
panels allow one to see the effect of our modelling assumptions on
the shape and amplitude of the recovered power spectrum. However,
part of this variation will be due to how the survey window function
changes when we modify the weighting or selection cuts. Thus, the
right hand panels show the same power spectra, but divided instead 
by the realistic model with  $\Om h=0.168$ and $\Ob /\Om =0.17$ 
that was used for the log-normal mocks, but now
taking into account the correct window function for each dataset.
Unless stated otherwise, no adjustments are made to the normalization
of the power spectrum estimates.

The top panels of Fig.~\ref{fig:pk_var} show the estimated power
spectrum for the full 2dFGRS for the standard choices listed above.
The error bars are those we estimate from the \LN\ mock catalogues.
In the subsequent panels of Fig.~\ref{fig:pk_var}, we show the
effects of varying these assumptions.

\subsection{Photometric calibration}
\label{sec:sys_calib}

Over the years the earlier calibrations of the APM photographic
plates have been the source of much debate
\citep[e.g.][]{sgphole95,sgphole04}. Thus it is important to try and
quantify at what level uncertainties in the photometry have an impact on our
ability to measure the galaxy power spectrum.

Results of four different calibrations are shown 
in Figs~\ref{fig:pk_var}c and~d.
As described in Section~\ref{sec:calib},
our standard choice (standard) differs from the final 2dFGRS
calibration (final) by the small offsets that we apply to the NGP and
SGP regions so as to bring their luminosity functions into
good agreement. We see that these offsets cause very little
change in the recovered power spectrum. We also show
results for the older calibration from the preliminary
100k data release (100k) \citep{colless01}. Here, the systematic
shift in the recovered power spectrum is somewhat larger,
but as the size of the error bars in the upper
panels show the shift is never larger than
the statistical error. If we had used this calibration,
then the maximum likelihood value of $\Om h$
inferred in Section~\ref{sec:params} would have been reduced by $0.01$
and the baryon fraction $\Ob /\Om $ increased
by $0.04$. These shifts are almost equal to the  $1-\sigma$ statistical
errors in these quantities.

For the last calibration model shown
in Figs~\ref{fig:pk_var}c and~d, we take a novel approach and first
calibrate each photographic plate without the use of external photometric
data.  The magnitudes in the final released catalogue, $\bj^{\rm final}$ and
magnitudes, $\bj^{\rm self}$, resulting from this self-calibration are
assumed to be related by a quasilinear relation
\begin{equation}
  \bj^{\rm self} =  a_{\rm self}  \, \bj^{\rm final} + b_{\rm self} .
\end{equation}
The calibration coefficients $a_{\rm self}$ and $b_{\rm self}$ are
allowed to vary from plate to plate.  To set the values of these
calibration coefficients two constraints are applied. First, on each
plate we assume that the galaxy luminosity function can be represented
by a Schechter function with faint-end slope $\alpha=-1.2$ and make a
maximum likelihood estimate of $M_*$.  The value of $M_*$ is sensitive
to the difference in $\bj^{\rm self}$ and $\bj^{\rm final}$ at around
$\bj=17.5$ and the number of galaxies on each plate is such that the
typical random error on $M_*$ is $0.03$ magnitudes.  Second, we compare
the number of galaxies, $N(z>0.25)$, with redshifts greater than
$z=0.25$ with the number we expect, $N_{\rm model}(z>0.25)$, based on
our standard model of the survey selection function.  The value of
$N_{\rm model}(z>0.25)$ depends sensitively on the survey magnitude
limit and so constrains the difference in $\bj^{\rm self}$ and
$\bj^{\rm final}$ at $\bj \simeq 19.5$. By demanding that on each plate
both $N(z>0.25) =N_{\rm model}(z>0.25)$ and $M_*-5 \log h\equiv -19.73$, we
determine $a_{\rm self}$ and $b_{\rm self}$.
Note that this method of calibrating the catalogue is extreme.
It ignores the information available in the plate overlaps
and ignores the CCD calibrating data (apart from setting the overall
arbitrary zero point of $M_*-5 \log h= -19.73$).
Nevertheless, we see that this (self) and the default (standard)
calibration results in only a very small shift in the recovered
power spectrum. The corresponding shifts in the inferred cosmological
model parameters $\Om h$ and baryon fraction
$\Ob /\Om $
are $-0.006$ and $0.02$ respectively. These shifts are small compared
to the corresponding statistical errors.

We conclude from these comparisons that
the final 2dFGRS photometric calibration is more accurate
than the preliminary 100k calibration and the residual
systematic uncertainties are at a level that they have
negligible impact on the accuracy of the recovered galaxy power spectrum.

\subsection{Redshift incompleteness}

In Figs~\ref{fig:pk_var}e and~f, we investigate the effect of varying
the treatment of incompleteness in the redshift survey. As described
above, our default choice is to keep all sectors of the survey with a
completeness $R(\theta)>0.1$ and use the completeness maps shown in
Fig.~\ref{fig:zcomp} and Fig.~\ref{fig:mumask} to reproduce this in
the random catalogues. In Figs~\ref{fig:pk_var}e and~f, we show the
effect of using the much more stringent cut $R(\theta)>0.5$ and so
removing the tail of low completeness sectors that are visible in
Fig.~\ref{fig:zcomp}. These are mainly around the edges of
the survey where constraints on observing time meant that
overlapping fields were never observed. This has a very small, but
measurable effect on the $P(k)$ shown in Fig.~\ref{fig:pk_var}e, but
once the effect of the changed window function is accounted for, no
perceptible difference remains in Fig.~\ref{fig:pk_var}f.

Also shown in these panels is  the effect of ignoring the apparent
magnitude dependence of the incompleteness by setting $\mu=\infty$
when constructing our random catalogues. Again, there is negligible
effect, clearly demonstrating that the accuracy of the 2dFGRS galaxy
power spectrum is not affected by uncertainty in the incompleteness.

\subsection{Angular weight}
\label{w_pairs}

In Figs~\ref{fig:pk_var}g and~h, we show the effect of varying the
angular weights which compensate for redshifts that are missed due
to fibre collisions. Our default choice of the angular weights,
$w^{\rm A}$, that attempt to correct for missing close pairs, are
defined by a multi-step process.  We assign unit weight to all
objects in the 2dFGRS parent catalogue, then loop over the subset
that lack measured redshifts and redistribute their weight to
their 10 nearest neighbours with redshifts. The angular weights,
$w^A$, are then defined by multiplying these weights by $R(\theta)$
and explicitly normalizing them to have an overall mean of unity.
The inclusion of the $R(\theta)$ factor means that the overall weight
assigned to a given sector is proportional to the number of galaxies
in that sector with measured redshifts, rather than to the number in the
parent catalogue.  The estimate labelled `assumed random' instead
has $w^A\equiv 1$ and so no correction is made for missing close
pairs other than their contribution to the overall completeness of a
given sector, \ie within a sector the missing galaxies are assumed
to be a random subset. We see that, on the scale of interest,
correcting for the missing close pairs has a negligible effect.

For the estimate labelled `parent' we
omit the factor $R(\theta)$ in the construction of the angular
weights for the main NGP and SGP strips.
This has the effect of up-weighting regions with low
completeness so that each sector has a weight proportional to the
number of galaxies in the parent catalogue. Hence the
angular dependence of the window function that is due to varying
redshift incompleteness is removed.
Figs~\ref{fig:pk_var}g and~h show that
even for this very different weighting, 
the change in the recovered power spectrum is extremely small.

\subsection{Radial weight}

In Figs~\ref{fig:pk_var}i and~j, we investigate the effect
of varying the radial weighting function. We show the result
of using equation~\ref{eq:rweight} with $J_3=300$, $400$
and~$500 h^{-3}$Mpc$^3$. The choice of weighting alters
the effective window function and so there is some variation
in the left hand panel on the very largest scales,
but in the right hand panel, where this is factored out, there is
very little variation in the recovered power.
For each of these values of $J_3$, we generated a set of 1000 \LN\ mocks
and compared the statistical error
in the recovered power, measured from the rms scatter
in the individual estimates. This exercise
explicitly verified that the value $J_3 \simeq 400\, h^{-3}$Mpc$^3$,
that we adopt as a default,
is close to optimal in terms of giving a minimal variance
estimate of the power.

The weighting function, equation~\ref{eq:rweight}, depends not only 
on redshift, but also on angular position through the angular dependence
of the quantity $\bngl$. That is, it takes account of the variation in the
expected galaxy  number density due to angular variation of
redshift incompleteness and survey magnitude limit.
The estimates labelled `P01' use instead a purely redshift
dependent weight of
\begin{equation}
w =(1.0+100/ [1+(z/0.12)^3]^2 )^{-1}
\end{equation}
as was done in \citet{P01}. On average, this is
close to our $J_3=400 h^{-3}$Mpc$^3$ weighting.
It slightly modifies the window function, but once this is
factored out we see, in  Fig.~\ref{fig:pk_var}j,
that there is little effect on the recovered power.

\subsection{Redshift limit}

In Figs~\ref{fig:pk_var}k and~l, we reduce the redshift limit
from $0.3$ to $0.25$. This alters the window function and
so has an effect on the power plotted in the left hand panel,
but in the right hand panel, where this is corrected for,
the variation is minimal.  The accurate agreement here is reassuring
and indicates there are no problems in pushing the survey
and the model of its selection function to the full volume that
it probes.

\subsection{Luminosity function and evolution}
\label{model_dndz}

In Figs~\ref{fig:pk_var}m and~n, we investigate the uncertainty in the
recovered power induced by the uncertainty involved in the
radial selection function of the survey.
Our default determination of the 2dFGRS selection function
involves modelling the galaxy luminosity function  as
a single Schechter luminosity function and the evolution by a single
$k+e$ correction (magnitude measurement errors are also included).
These are derived empirically by the maximum
likelihood method presented in Section~\ref{sec:lf}.  This `standard' model is
compared with the result of using a `2 population' model with individual
luminosity functions and $k+e$ corrections for the red and blue galaxy
populations. Again, the luminosity functions and $k+e$ corrections used
are the empirically determined ones presented in Section~\ref{sec:lf}.  We
see that adding these extra degrees of freedom to the description of
the galaxy population has a negligible effect on the recovered power.

As a separate test, we show the results (labelled `colours')
of a single population model in which the mean $k+e$ correction
has been determined
using \citet{bc93} stellar population models matched to
the galaxy colours as was done in \citet{norberg_selfun}.  This model
again produces highly consistent results, which is perhaps not
surprising given that its $k+e$ correction, shown in Fig.~\ref{fig:lf}, is quite
similar to the one found by the maximum likelihood method.

The three radial selection function models discussed above produce
consistent results, but all make common assumptions such
as Schechter function forms for the luminosity  functions and smooth
$k+e$ corrections. To demonstrate that these assumptions are not
artificially distorting our estimate of the power, we present results
for an alternative empirical model of the redshift distribution.
For this, we compare the observed redshift distribution averaged over the
whole survey with that of our standard model. The two redshift
distributions are shown in the top panel of Fig.~\ref{fig:dndz_all}.
We then resample our
default random catalogues so that the redshift distribution of the
remaining galaxies exactly matches that of the data. In this process
we also correspondingly modify the tabulated galaxy number densities
in the random catalogue so that our power spectrum estimator
remains correctly normalized. Note that this procedure is not
equivalent to simply generating a random catalogue by shuffling the
data redshifts as we retain the modulation of the redshift
distribution caused by the varying survey magnitude limit that was
built into the standard random catalogue.

Fig.~\ref{fig:pk_var}n shows that the only effect of adopting this
empirical redshift distribution is, unsurprisingly 
a reduction of the power on the
very largest scales and that even here the shift is not large
compared to the statistical errors.  Adopting this estimate rather
than our standard one only shifts our estimates of the cosmological
parameters $\Om h$ and $\Ob /\Om $ by
$+0.006$ and $-0.03$ respectively. These shifts, which are smaller
than the $1\sigma$ statistical errors, should be considered extreme,
as adopting an empirical redshift distribution will undoubtedly lead
to the removal of some genuine large scale radial density
fluctuations.

\subsection{Region}

In Figs~\ref{fig:pk_var}o,p,q and~r, we show the effect of excluding 
various regions from our analysis. 
The effect of excluding the random fields is very modest. 
In particular, we note that the oscillatory features in the
estimated power spectra around $k \approx 0.15 \mpcoh$ 
are present both with and without the random fields and that
once the effects of the very different window functions 
(see Fig.~\ref{fig:win}) have been compensated for,
Fig.~\ref{fig:pk_var}p, the power spectra agree very accurately.
This clearly demonstrates that these features are not related
to the presence or absence of a secondary peak in the window
function.
Also shown in Figs~\ref{fig:pk_var}o and~p is the effect of excluding from
our data the two superclusters identified by \citet{baugh04} and
\citet{croton04} and mapped using the Wiener filtering
technique by \citet{erdogdu04}.  
The northern supercluster is the heart of the structure
that has also become known as the Sloan great wall \citep{gott03}.  Here
we have simply excised these superclusters by cutting out regions of
9x9~degrees and $\Delta z=0.1$ centred on the superclusters. These
superclusters are known to perturb higher order clustering statistics
significantly \citep{croton04}, but we see that their removal causes a
negligible reduction in the large-scale power. Using this dataset only
shifts our estimates of the cosmological parameters $\Om h$ and $\Ob/\Om$
by $+0.008$ and $-0.026$ respectively.  These shifts are much smaller than
the $1-\sigma$ statistical errors. Also, as genuine structure
is being removed one expects the large-scale power to be suppressed.
Clearly these superclusters do not significantly perturb our estimated
power spectrum.

Excluding either the NGP or SGP strips
has a large effect on the window function and this is partly
responsible for the changes in the recovered power seen in 
Fig.~\ref{fig:pk_var}q. 
But in this case cosmic variance is also important and so,
even when we compensate for the window, as is done in 
Fig.~\ref{fig:pk_var}r, 
we do not expect perfect agreement between these estimates; for
independent samples, we would expect differences comparable to the
statistical errors. The errors from our log-normal mocks shown on the
independent NGP and SGP estimates indicate that only on the very
largest scales, where the data points are highly correlated, do
the estimates differ by more than $1\sigma$. If the likelihood
analysis described in Section~\ref{sec:params} is applied separately
to these two samples we find $\Om h=0.168\pm0.035$, $\Ob /\Om
=0.163\pm0.075$ for the SGP and $\Om h=0.205\pm0.037$, $\Ob /\Om
=0.116\pm0.072$ for the NGP, which are entirely consistent within
their statistical errors.

\subsection{Estimator}

In Figs~\ref{fig:pk_var}s and~t, we compare the result of using
the \citetalias{FKP} rather than the \citetalias{PVP} estimator.
We have adjusted the normalization of the \citetalias{FKP}
estimate by a factor $\langle b^{-2} \rangle$ to account for the
normalization difference in the definition of the two estimators.
If galaxies have a luminosity dependent bias, then
the \citetalias{FKP} estimator
is biased, with the result that one recovers a power spectrum 
convolved with an effective window function that is slightly
different to the one assumed \citepalias{PVP}. Provided the model
of luminosity dependent bias is correct, then the \citetalias{PVP} estimator
removes this bias. The two recovered power spectra
shown in  Fig.~\ref{fig:pk_var}t differ only slightly in shape
indicating that the bias resulting from using the \citetalias{FKP} estimator,
as was done in P01, is small. Furthermore, even if our model of
bias dependence on luminosity and colour is not highly accurate,
the effect on the recovered power spectrum will be significantly
smaller than the difference between the \citetalias{FKP}
and \citetalias{PVP} estimates and
so entirely negligible.

\subsection{Summary}

In conclusion, we have not identified any systematic effects at a
level that is significant compared to the statistical errors. We return to
this point in Section~\ref{sec:params}, where we show explicitly how various
systematic uncertainties affect the likelihood surfaces that
quantify our constraints on cosmological parameters.

\section{Non-linearity and scale-dependent bias}  \label{sec:nonlin}

The previous Section has demonstrated that we can measure the
spherically-averaged redshift-space power spectrum of the
2dFGRS in a robust fashion. We now have to consider in detail
the critical issue of how the galaxy measurements relate to the 
power spectrum of the underlying density field.

The conventional approach is to assume that, on
large enough scales, linear theory provides an adequate description 
of the shape of the galaxy power spectrum. In reality, this agreement
can never be perfect, and we need a  model for the differences
between the galaxy power spectrum and linear theory. In this
Section, we pursue a number of approaches for estimating such corrections;
detailed simulations, analytical models, and an empirical hybrid approach
are all considered.

\subsection{Simulated galaxy catalogues} \label{sec:simcat}

\begin{figure}
\epsfxsize=8.4 truecm \centering  
\epsfbox{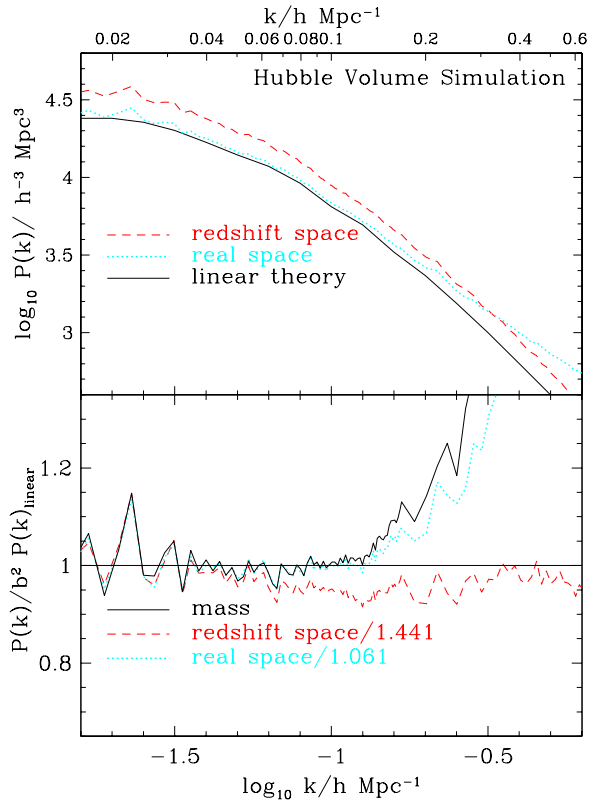}
\caption{The power spectrum of the mass and galaxies in the
Hubble Volume simulation cube. The solid curve in the upper
panel shows the input linear theory power spectrum. The
dotted and dashed curves show the power spectrum for the
galaxies in real and redshift space respectively.
In the lower panel, using the same line types, we show these
galaxy power spectra divided by the linear theory power spectrum,
scaled by the square of expected bias factor.
The solid curve shows the ratio of the mass to linear
theory power spectra.
\label{fig:pk_hv}}
\end{figure}

We start by considering the power spectrum of the \HV\ galaxies.
Fig.~\ref{fig:pk_hv} shows results from the full \HV,
both in real and redshift space. Here, we
use all $10^9$ particles in the
simulation cube weighted by the probability of each particle being
selected as a galaxy. 
On large scales ($k\ls 0.1\, \mpcoh$) both the real-space
and redshift-space galaxy power spectra are related to
linear theory by a simple scale independent
constant. The large scale linear bias factor for the galaxies in
real space is $b=1.03$.
On these large scales the 
redshift-space power is boosted by the \citet{kaiser87} factor
$b^2(1+2/3\, \beta + 1/5 \, \beta^2)$; here $\beta=\Om^{0.6}/b=0.471$, 
so the expected boost factor is 1.441, in excellent agreement with the
simulation results.

In real space, both the mass and galaxy power spectra begin to
exceed the linear theory prediction significantly for
$k\gs 0.12\hompc$.
In redshift space, the smearing effect
of the random galaxy velocities reduces the small scale power with
the result that deviations from linear theory are greatly reduced.
This cancellation of the distortions caused by non-linearity, bias
and mapping to redshift space was used in P01 to motivate fitting
the 2dFGRS with linear theory for $k< 0.15\hompc$. The \HV\ results
presented here reinforce this argument, although there is a suggestion
that the redshift-space power underestimates linear theory
by up to 10\% on small scales.

The \HV\ simulations are realistic in some respects, but they
do not treat the relation between mass and light in a particularly
physical way. According to current understanding, the location of
galaxies within the dark matter is determined largely by
the dark-matter halos and their merger history.
Full semi-analytic models of galaxy formation follow
halo merger trees within a numerical simulation, and can yield
impressively realistic results. 
An important landmark for this kind of work was the paper
by Benson et al. (\citeyear{benson00}),
who showed how a semi-analytic model could naturally
yield a correlation function that is close to a
single power law over $1000 \gs \xi \gs 1$, even though the
mass correlations show a marked curvature over this range.
We have used the most recent version of this code to
predict the galaxy power spectra, and their ratio to
linear theory. This is shown in Fig.~\ref{fig:prats1},
for a model close to our final preferred cosmology:
$\Om=0.25$, $\Ob=0.045$, $h=0.73$ and $\sigma_8=0.9$.
The simulation volume has a side of $1000\mpcoh$ and
$10^9$ simulation particles. One advantage of this more
detailed simulation is that the predicted colour distribution is
bimodal, and so we are able to identify red and blue subsets
in the same way as for the real data.

These results paint a similar picture to what was seen in the Hubble Volume,
despite the very different treatment of bias. On intermediate scales,
there is a tendency for galaxy power to lie below linear theory.
In real space, this trend reverses around $k=0.1\hompc$, and galaxy
power exceeds linear theory for $k\ga 0.2\hompc$. A small-scale
increase is also seen in redshift space, but redshift-space smearing
naturally means that the effect is reduced.

It will be convenient to consider a fitting formula for the
distortion seen in this simulation, and the following simple
form works well:
\begin{equation}
  P_{\rm gal} = {1+Qk^2 \over 1 + Ak}\; P_{\rm lin}.
  \label{eq:Qdef}
\end{equation}
The required parameters to fit the `all galaxy' data, shown
by the dashed lines in Fig.~\ref{fig:prats1},
are $A=1.7$ and $Q=9.6$ (real space) or
$A=1.4$ and $Q=4.0$ (redshift space). 
The critical question is whether this correction is
robust, both with respect to variations in the
galaxy-formation model and variations in cosmology.
It is impractical to address this directly by running a
large library of simulations, so we consider an alternative
analytic approach.

\subsection{The halo model}

The success of Benson et al.'s work stimulated the analytic `halo model',
which allows one to understand rather simply the differences in shape
between the galaxy and mass power spectra 
(Seljak \citeyear{seljak00}; Peacock \& Smith \citeyear{peacock00};
Cooray \& Sheth \citeyear{cooray02}). In this approach, the galaxy 
density field  results from a
superposition of dark-matter halos, with small-scale clustering arising
from neighbours in the same halo.

Using the halo model, it is possible to predict the relation between the galaxy 
power spectrum and linear theory. This can be done as a function of galaxy type,
by an appropriate choice of prescription for the occupation numbers of halos as 
a function of their mass. 
In effect, we can give particles in halos a weight that depends on halo mass, as
was first considered by Jing, Mo \& B\"orner (\citeyear{jing98}).
A simple but instructive model for this is
\begin{equation}
w(M) = \cases{
0\quad &($M<M_c$)\cr
(M/M_c)^{\alpha-1}\quad &($M>M_c$)\cr
}
\end{equation}
A model in which mass traces light would have $M_c \rightarrow 0$ and $\alpha=1$.
In practice, data on group $M/L$ values argues for $\alpha$ substantially less
than unity (Peacock \& Smith \citeyear{peacock00}, see also
Collister \& Lahav \citeyear{collister05}).
More elaborate occupation models can be considered 
(e.g. Tinker et al. \citeyear{tinker04};
Zheng et al. \citeyear{zheng04}) and have previously been applied
to model 2dFGRS galaxy clustering \citep{vdbosch03,maglio03},
but this simple model will suffice for the present purpose: we are trying to
estimate a small correction in any case, and are largely interested in
how it may vary with cosmology.

\begin{figure}
\epsfxsize = 8.2truecm \centering
\epsfbox{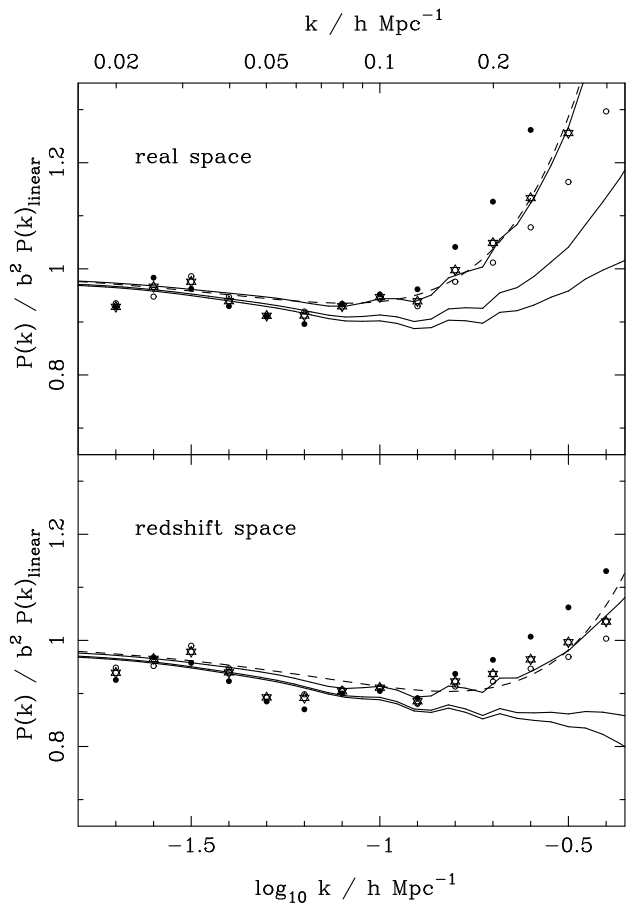}
\caption{Predictions for how the 
redshift-space power spectrum of galaxies may be expected to
deviate from linear theory. A model with
$\Om=0.25$, $\Ob=0.045$, $h=0.73$ and $\sigma_8=0.9$ is assumed.
The predictions of semi-analytic modelling are shown as points:
filled circles denote red galaxies; open circles denote blue galaxies;
stars denote all galaxies (to $M_{\bj} < -19$).
The dashed line shows the fitting formula described in the text.
The solid lines show the predictions made using the halo model for
red (upper), blue (lower) and all (intermediate) galaxies.
The occupation parameters are adjusted so
as to fit the real-space correlation functions from \citet{madgwick03}.
We attempt to make the calculation more robust by modelling
the conversion between real and redshift space using the
Ballinger et al. (\citeyear{ballinger96}) prescription. We use
observed values of $\beta=0.49$
and $\beta=0.48$ and large-separation effective pairwise dispersions of
$280\kms$ and $410\kms$ for late and early types respectively
from \citet{madgwick03} and \citet{hawkins03}
\label{fig:prats1}}
\end{figure}

\begin{figure}
\epsfxsize = 8.2truecm \centering
\epsfbox{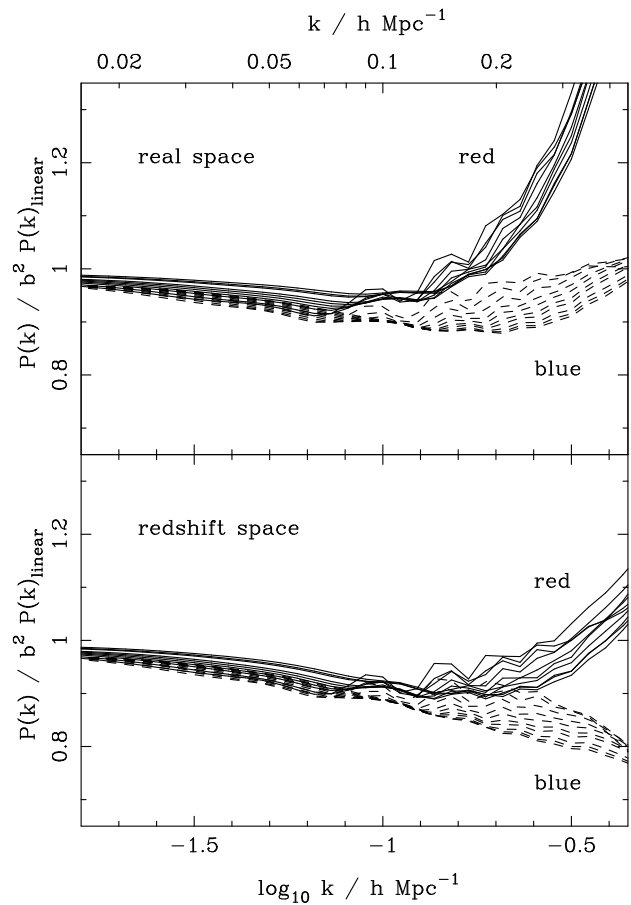}
\caption{A more extensive set of predictions for the deviation
of the galaxy power spectrum from linear theory, using the
halo model as in Fig.~\ref{fig:prats1}.
We retain $\Ob=0.045$ and $h=0.73$, but vary $\Om$ between
0.17 and 0.35. The normalization is chosen to scale as 
$\sigma_8 = 0.9 (0.25/\Om)^{0.6}$, as expected for
a normalization to redshift-space distortions or cluster abundance.
The plotted ratio is a weakly declining function of $\Omega_{\rm m}$
(i.e. the lowest $\Omega_{\rm m}$ gives the strongest kick-up at high $k$).
\label{fig:prats2}}
\end{figure}

The translation of the halo model into redshift space 
has been discussed by White (\citeyear{White01}), 
Seljak (\citeyear{seljak01}) and Cooray (\citeyear{cooray04}).
In the halo model, one thinks of the real-space power spectrum as
being a combination of two parts:
\begin{equation}
P_r = P_{\rm 2-halo} + P_{\rm 1-halo},
\end{equation}
representing the effect of correlated halo centres (the first term), plus
power owing to halo discreteness and internal structure of a single halo (the second term).
In redshift space, we expect the first term to undergo
Kaiser (\citeyear{kaiser87}) distortions, so that it gains a factor $(1+\beta \mu^2)^2$, where
$\mu$ is the cosine of the angle between the wavevector and the line of sight.
Having shifted the halo centres to redshift space, the
effect of virialized velocities is to damp the total power
for modes along the line of sight:
\begin{equation}
P_s = \left((1+\beta \mu^2)^2 P_{\rm 2-halo} + P_{\rm 1-halo}\right)\; D^2(\mu k).
\end{equation}
For a Gaussian distribution of velocities within a halo, the damping factor is
\begin{equation}
D(x) = \exp\left(-x^2/2\sigma_v^2\right);
\end{equation}
here, $\sigma_v$ denotes the one-dimensional halo velocity dispersion 
in units of length (i.e. divided by $H_0$).
This expression applies for the case where $\beta$ and $\sigma_v$
are the same for all halos. Since both vary with mass, the expression
must be appropriately averaged over halo mass, as described in the above references.

This completes in outline the method needed to calculate the
redshift-space power spectrum. However, we will not use all
this apparatus: the halo model was not designed to work at the
precision of interest here, and we will therefore use it only
in a differential way which should minimize systematics in the modelling.
The power ratio of interest can be expressed as
\begin{equation}
{P_{\rm gal}^s\over P_{\rm lin}} 
=
{P_{\rm gal}^r\over P_{\rm nl}} 
\times
{P_{\rm nl}\over P_{\rm lin}}
\times
{P_{\rm gal}^s\over P_{\rm gal}^r}.
\end{equation}
For the ratio to linear theory in real space, the last factor on the
rhs is not required. The advantage of this separation is that we
have accurate empirical methods of calculating the second and third
terms on the rhs. The halo model is thus only required to give
the real-space ratio between galaxy and nonlinear mass power
spectra. 

The ratio between non-linear and linear mass power is given by
the HALOFIT fitting formula from Smith et al. (\citeyear{halofit}).
This procedure uses the same philosophy as the halo model, but
is tuned to give an accurate fit to $N$-body data. 
For the ratio between real-space and redshift-space galaxy
data, we adopt the model used in past 2dFGRS papers: 
a combination of the Kaiser linear boost and the damping corresponding
to exponential pairwise velocities:
\begin{equation}
P_s = P_r (1+\beta \mu^2)^2 (1+k^2\mu^2\sigma_p^2/2)^{-1},
\end{equation}
where $\sigma_p$ is the pairwise velocity dispersion translated into length units,
and $P_r$ is the full real-space galaxy power spectrum
(e.g. Ballinger et al. \citeyear{ballinger96}).
This has been shown to work well in comparison with $N$-body data.
For the present purpose, the advantage is that this correction is
an observable, and therefore does not need to be modelled in a 
way that introduces a cosmology-dependent uncertainty.

We therefore used the azimuthal average of the Ballinger et al.
expression to convert to redshift space.
This allows us to use observed values: $\beta=0.49$,  $\beta=0.48$ 
and $\beta=0.46$ and large-separation effective pairwise dispersions of
$280\kms$, $410\kms$ and $340\kms$ for red, blue and all galaxies 
respectively.
These figures are derived from \citet{madgwick03}, but reduced
by a factor 1.5 to allow for the fact that pairwise
dispersions appear to fall at large separations \citep{hawkins03}.
The ratio of bias parameters we found in Section~\ref{sec:redblue} implies an
expected value of $\beta_{\rm blue}/\beta_{\rm red}\approx 1.45$.
This is larger than the ratio of best fit values found by 
\citet{madgwick03}, but within their quoted errors.

The resulting galaxy power spectra are also shown in Fig.~\ref{fig:prats1}, and are
relatively consistent with the direct simulation results. 
The observed power is expected
to fall progressively below linear theory as we move to
higher $k$, with a reduction of approximately 10\% at $k=0.1\hompc$.
Beyond this, the trend reverses as non-linearities add power -- 
although in redshift space the effect is more of a plateau until
$k\simeq 0.3\hompc$.

The remaining issue is whether the correction depends on the
cosmological model. If we were to ignore all corrections and
fit linear theory directly, as in P01, there is a relatively
well-defined apparent  model. In the spirit of perturbation
theory, there would then be a case for simply calculating the
correction for that model and applying it. However, it is
more reassuring to be able to investigate the model dependence
of the correction.
This is shown in Fig.~\ref{fig:prats2}. Here, we take the
approach of varying the most uncertain cosmological
parameter, $\Om$. 
We hold $\Ob=0.045$ and $h=0.73$ fixed, but vary $\Om$ between
0.17 and 0.35. The normalization is chosen to scale as
$\sigma_8 = 0.9 (0.25/\Om)^{0.6}$, as expected for
a normalization to redshift-space distortions or cluster abundance.
A less realistic choice ($\sigma_8 = 0.9$ independent of $\Om$)
shows similar trends: the fall in power to $k=0.1\hompc$ is virtually identical, but 
there is a dispersion in where the small-scale upturn becomes important
(a maximum range of a factor 2 in $k$).

To summarise, it seems clear that we should expect small systematic
distortions of the galaxy power spectrum with respect to linear theory.
The robust prediction is that the power ratio should fall monotonically 
between $k=0$ and $k=0.1\hompc$. Beyond that, the trend reverses, but
the calculation of the degree of reversal is not completely robust.
This motivates our final hybrid strategy. We adopt the formula 
(equation  \ref{eq:Qdef}) that was used to fit the data from the
simulation populated by the semi-analytic model
\begin{equation}
  P_{\rm gal} = {1+Qk^2 \over 1 + Ak}\; P_{\rm lin},
\end{equation}
but we do not take the parameters as fixed. Rather, the large-scale
parameter is assumed to be $A=1.4$ (redshift space) or $A=1.7$
(cluster collapsed/ real space) as in the simulation fits,
but the small-scale quadratic $Q$ parameter is allowed to vary over
a range up to approximately double the expected value
($Q_{\rm max}=12$ in real space and 8 in redshift space).
This allows the residual uncertainty in the small-scale
behaviour to be treated as a nuisance parameter to be determined
empirically and marginalized over.

As we will see in the following Section, the net result of following
this strategy is a systematic shift in the recovered cosmological
parameters of almost exactly $1\sigma$. In a sense, then, this
apparatus is unnecessarily complex (and was justifiably
neglected in P01). However, the fact that we can make a reasonable
estimate of the extent of systematics at this level should increase
confidence in the final results.

\section{Likelihood analysis and model fitting}
\label{sec:params}

\subsection{Likelihood fitting}  \label{sec:likefit}

Having measured the 2dFGRS power spectrum in a
series of bins, we now wish to model the likelihood -- i.e. the probability
density function of the data given different cosmological
models. Assuming that the power spectrum errors have Gaussian statistics 
that are independent of the model being tested, the
likelihood function is
\begin{equation}
\eqalign{
  -2\ln{\cal L} & =   \ln |C| + \cr
    &  \sum_{ij}
    [P(k_i)^{\rm TH}-P(k_i)]C^{-1}_{ij}[P(k_j)^{\rm TH}-P(k_j)], \cr
}
  \label{eq:like_Gauss}
\end{equation}
up to an irrelevant additive constant. Here $P(k)^{\rm TH}$ is the
theoretical power spectrum to be tested, $P(k)$ is the observed power
spectrum and $C$ is the covariance matrix of the data.

This form for the likelihood is only an approximation.
For a Gaussian random field where the window and shot noise are negligible, 
the exact likelihood is given by
\begin{equation}
  -2\ln{\cal L} = \sum_{i} \left[ \ln P(k_i)^{\rm TH}
    + \frac{\bar\delta^2(k_i)}{P(k_i)^{\rm TH}} \right],
  \label{eq:like_true}
\end{equation}
where $\bar\delta(k_i)$ gives the observed transformed overdensity
field. This equation is simply the standard Gaussian likelihood as in
Eq.~\ref{eq:like_Gauss}, but now with $\bar\delta(k_i)$ as the
Gaussian random variable. Equation~\ref{eq:like_true} has been
simplified because $\langle\bar\delta(k_i)\rangle=0$ independent of
model to be tested, and
$\langle\bar\delta(k_i)\bar\delta(k_i)\rangle=P(k_i)^{\rm TH}$. In
Percival et al. (\citeyear{P04}), where we presented a decomposition
of the 2dFGRS into spherical harmonics, the likelihood was calculated
assuming Gaussianity in the Fourier modes of the decomposed density
field $\delta(k_i)$, as in this equation.  However, in practice this
method is difficult: the window function causes $\bar\delta(k_i)$ and
$\bar\delta(k_j)$ to be correlated for $i\ne j$, and shot noise means
that $\langle\bar\delta^2(k_i)\rangle\ne P(k_i)^{\rm TH}$.  We
therefore prefer in the present work to use the faster approximate
likelihood, knowing that the method can be validated empirically using
mock data.

Following the assumption that the likelihood can be written in the
form of equation~\ref{eq:like_Gauss}, we need to define a covariance matrix
for each model under test. In Section~\ref{sec:LN_mocks} we used mock
2dFGRS catalogues for a single theoretical power spectrum in order to
estimate the power covariance matrix. In principle, this procedure
should be repeated for each model under test. However, in the case of
an ideal survey with no window or noise, the appropriate covariance
matrix should be diagonal and dependent on the power spectrum to be
tested through
\begin{equation}
  C_{ii} \propto [P(k_i)^{\rm TH}]^2.
\end{equation}
We use this scaling to suggest the following dependence of the
covariance matrix on the theoretical model being considered:
\begin{equation}
  C_{ij} = \frac{P(k_i)^{\rm TH}P(k_j)^{\rm TH}}
                {P(k_i)^{\rm LN}P(k_j)^{\rm LN}} C^{LN}_{ij};
    \label{eq:cov_revised}
\end{equation}
here $C^{LN}_{ij}$ is the original covariance matrix estimated using
the log-normal catalogues, and $P^{\rm LN}$ is the true power spectrum
of these catalogues. In other words, we assume that the correlation coefficient
between modes will be set by the survey window and will be independent
of the theoretical power spectrum.
The primary results on parameter estimation
are calculated following this assumption.
We show in Section~\ref{sec:results} that, in any case, the results obtained
by using equation~\ref{eq:cov_revised} are very similar to those that follow
from the assumption of a fixed covariance matrix.

\begin{figure*}
\epsfxsize=0.9\textwidth \centering
\epsfbox{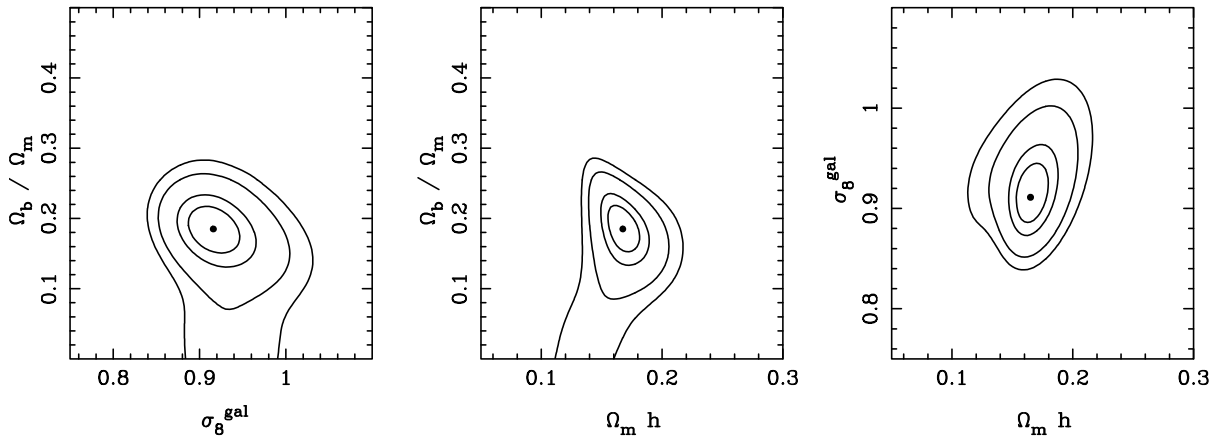}
\caption{Contour plots showing changes in the likelihood from the
maximum of $2\Delta\ln{\cal L}=1.0, 2.3, 6.0, 9.2$ for different
parameter combinations for the redshift-space 2dFGRS power spectrum,
assuming a $\Lambda$CDM cosmology with $h=0.72$ and $\ns=1.0$. These
contour intervals correspond to 1$\sigma$ 1-parameter, and
1,2,3$\sigma$ 2-parameter confidence intervals for independent
Gaussian random variables. The power spectrum was fitted for
$0.02<k<0.20\hompc$, marginalizing over $0<Q<8$. The solid circle
marks the maximum likelihood position for each 2dFGRS likelihood
surface. \label{fig:2dF_like} }
\end{figure*}

\subsection{Models, parameters and priors}  \label{sec:model}

When fitting the 2dFGRS data, the
parameter space has the five dimensions needed to describe
the matter power spectrum in the simplest CDM models:
\begin{equation}
{\bf p} =
(\Om, \Ob, h, \ns, \sigma_8),
\end{equation}
where $\ns$ is the scalar spectral index and the other
parameters are as discussed earlier. For analyses including
the CMB, one would add four further parameters: 
spatial curvature, an amplitude
and slope for the tensor spectrum, plus $\tau$, the optical depth
to last scattering. These do not affect the matter spectrum,
which we calculate using the formulae of Eisenstein \& Hu (\citeyear{EisensteinHu98}).

In practice, this dimensionality can be reduced. 
For a given $\ns$, the shape of the matter power spectrum depends
mainly on two parameter combinations:
(1) the matter density times the Hubble parameter $\Om h$;
(2) the baryon fraction $\Ob /\Om$. There is a
weak residual dependence on $h$, but we neglect this because
$h$ is very well constrained by any analysis that includes
CMB data. We therefore adopt the fixed value $h=0.72$.
A similar argument is not so readily made for $\ns$; even though
this too is accurately determined in joint analyses with CMB
data, there is strong direct degeneracy between the value of
$\ns$ and our main parameters. Fortunately, this is easy enough
to treat directly: raising $\ns$ increases small-scale power
and thus requires a lower density compared to the figure
deduced when fixing $\ns=1$, for which an adequate approximation is
\begin{equation}
(\Om h)_{\rm true} = (\Om h)_{\rm apparent} + 0.3(1-\ns).
\end{equation}
Similarly, we choose to neglect possible effects of a neutrino rest
mass. It is known from oscillation experiments that this is justified
provided that the mass eigenstates are non-degenerate. Again, it
is straightforward to allow directly for a violation of this
assumption:
\begin{equation}
(\Om h)_{\rm true} = (\Om h)_{\rm apparent} + 1.2(\Omega_\nu/\Om)
\end{equation}
(Elgaroy et al. \citeyear{elgaroy}).  Finally, as discussed above in
Section~\ref{sec:nonlin}, we assume a simple quadratic model
(equation~\ref{eq:Qdef}) with a single free parameter, $Q$, 
for the small-scale deviations
from linear theory caused by non-linear effects and redshift-space
distortions. The parameter $A$ in equation~\ref{eq:Qdef} describing
large-scale quasilinear effects is held constant at $A=1.4$.

The calculation of the likelihood of a cosmological model given just
the 2dFGRS data is computationally inexpensive, and we can therefore
use grids to explore the parameter space of interest. When each
likelihood calculation becomes more computationally expensive, or the
parameter space becomes larger, then a different technique such as
Markov-Chain Monte-Carlo (MCMC) would be expedient. In
Section~\ref{sec:implications} we use the MCMC technique when
combining large-scale structure and CMB data. For our exploration of
the cosmological implications of the 2dFGRS data alone, grids of
likelihoods were calculated using the method described in
Section~\ref{sec:likefit}, uniformly distributed in parameter space
over $0.05<\Om h<0.3$, $0<\Ob /\Om <0.5$ and $0.6<\sigma_8^{\rm
gal}<1.1$ and $0<Q<8$ for standard redshift-space catalogues, and
$4<Q<12$ for cluster-collapsed catalogues, which we treat as if they
were in real space. These grids were used to marginalize over
parameters assuming uniform priors with these limits.

Compared to the shape parameters $\Om h$ and $\Ob /\Om$, the
normalization of the model power spectrum is a relatively
uninteresting parameter, over which we will normally
marginalize. However, it has some interesting degeneracies with the
shape parameters, which are worth displaying. It should be emphasised
that the meaning of the normalization is not straightforward, owing to
the depth of the survey. We thus measure an amplitude at some mean
redshift greater than zero (the fitted parameters $\Om$ and $\Ob$,
however, do correspond strictly to $z=0$). We normalize the power
spectrum using the rms density contrast averaged over spheres of
$8\mpcoh$ radius. If we define $\sigma_8$ to correspond to the linear
mass overdensity field at redshift zero, then the normalization of the
measured power spectrum corresponds to an `apparent' value
$\sigma_8^{\rm gal}$, which should not be confused with an estimate
for the true value of $\sigma_8$:
\begin{equation}
  \sigma_8^{\rm gal}=b(L_*,z_s)\,D(z_s)\,K^{1/2}(\beta[L_*,z_s])\ \sigma_8,
\end{equation}
where $z_s$ is the mean redshift of the
weighted overdensity field,
$D(z_s)$ is the linear growth factor between redshift $0$ and
$z_s$, $K(\beta[L_*,z_s])=1+2/3\ \beta+1/5\ \beta^2$ 
is the spherically averaged Kaiser linear boost factor that
corrects for linear redshift distortions of $L_*$ galaxies 
at redshift $z_s$, and $b(L_*,z_s)$ is the bias of $L_*$ galaxies 
between the real space galaxy overdensity field and the linear mass
overdensity field at redshift $z_s$ \citep[see][]{lahav02}.

\subsection{2dFGRS results}  \label{sec:results}

Fig.~\ref{fig:2dF_like} shows our default set of likelihood contours
for $\Om h$, $\Ob /\Om $ and the normalization $\sigma_8^{\rm gal}$,
calculated using the redshift-space power spectrum data with
$0.02<k<0.20\hompc$, marginalizing over the distortion parameter $Q$.
There is a weak degeneracy between $\Om h$ and $\Ob /\Om$ as found in
P01, corresponding to power spectra with approximately the same
shape. However this degeneracy is broken more strongly than in P01 and
we find maximum-likelihood values of $\Om h=0.168\pm0.016$ and $\Ob
/\Om =0.185\pm0.046$.  Here, the errors quoted are the rms of the
marginalized probability distribution for the parameter under
study. For a Gaussian random variable, this corresponds to the 68\%
confidence interval for 1 parameter. The normalization is measured to
be $\sigma_8^{\rm gal}=0.924\pm0.032$, and we find a marginalized
value of $Q=4.6\pm1.5$, when fitting to $0.02<k<0.30\hompc$, well
within the range of $Q$ considered.
The improvement in the accuracy of the parameter constraints compared to
those of P01 is the result of three factors. For example the error
on $\Om h$ is reduced by $0.006$, $0.005$ and $0.003$ by  
the increased angular coverage, increasing $z_{\rm max}$ to $0.3$ and
adopting the more optimal PVP weighting respectively.

\begin{figure}
\epsfxsize=0.97\columnwidth \centering
\epsfbox{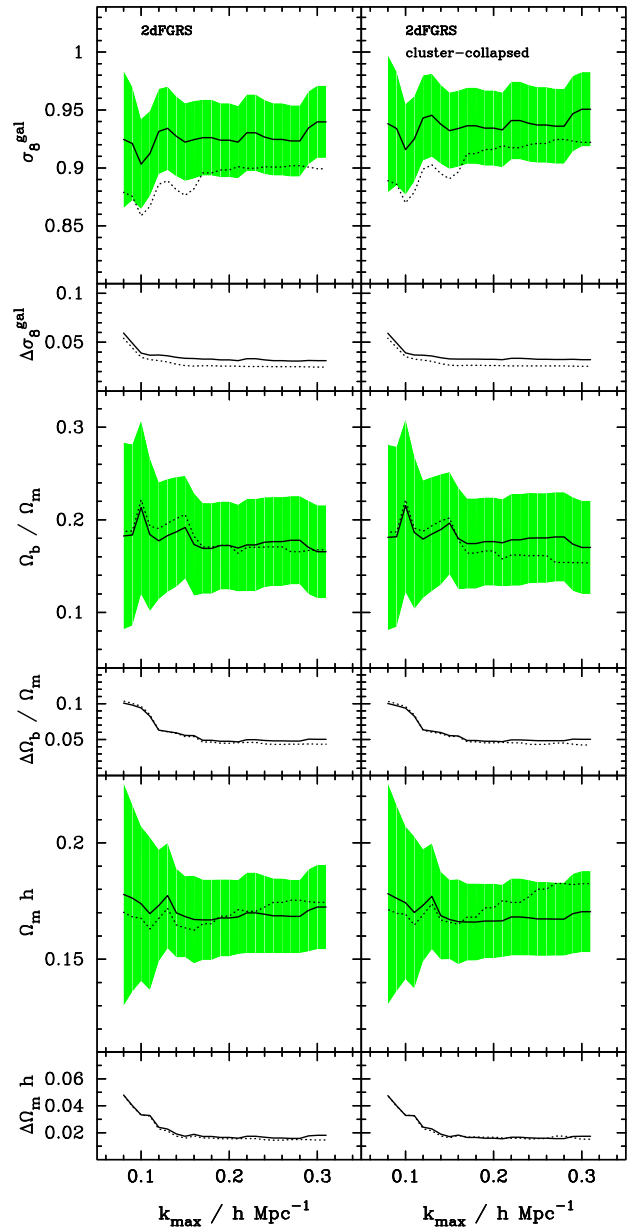}
\caption{Marginalized parameters as a function of the maximum fitted
$k$ for the 2dFGRS redshift-space catalogue (left column), and after
collapsing the clusters (right column). The
rows are for different parameters and the recovered errors calculated
by marginalizing over the region of parameter space considered (see
text for details). Solid lines (both for marginalized parameter and
error) include a possible correction for non-linear and small-scale
redshift space distortion effects parameterized by $Q$, dotted lines
make no corrections to linear theory. The shaded region shows the
$\pm 1\sigma$ confidence region, indicating that systematic corrections are
at most comparable to the random errors.
\label{fig:param_2dFGRS_kmax} }
\end{figure}

\subsubsection{Dependence on scale}

\begin{figure*}

\hspace{0.125\columnwidth}
\epsfxsize=0.75\columnwidth
  \epsfbox{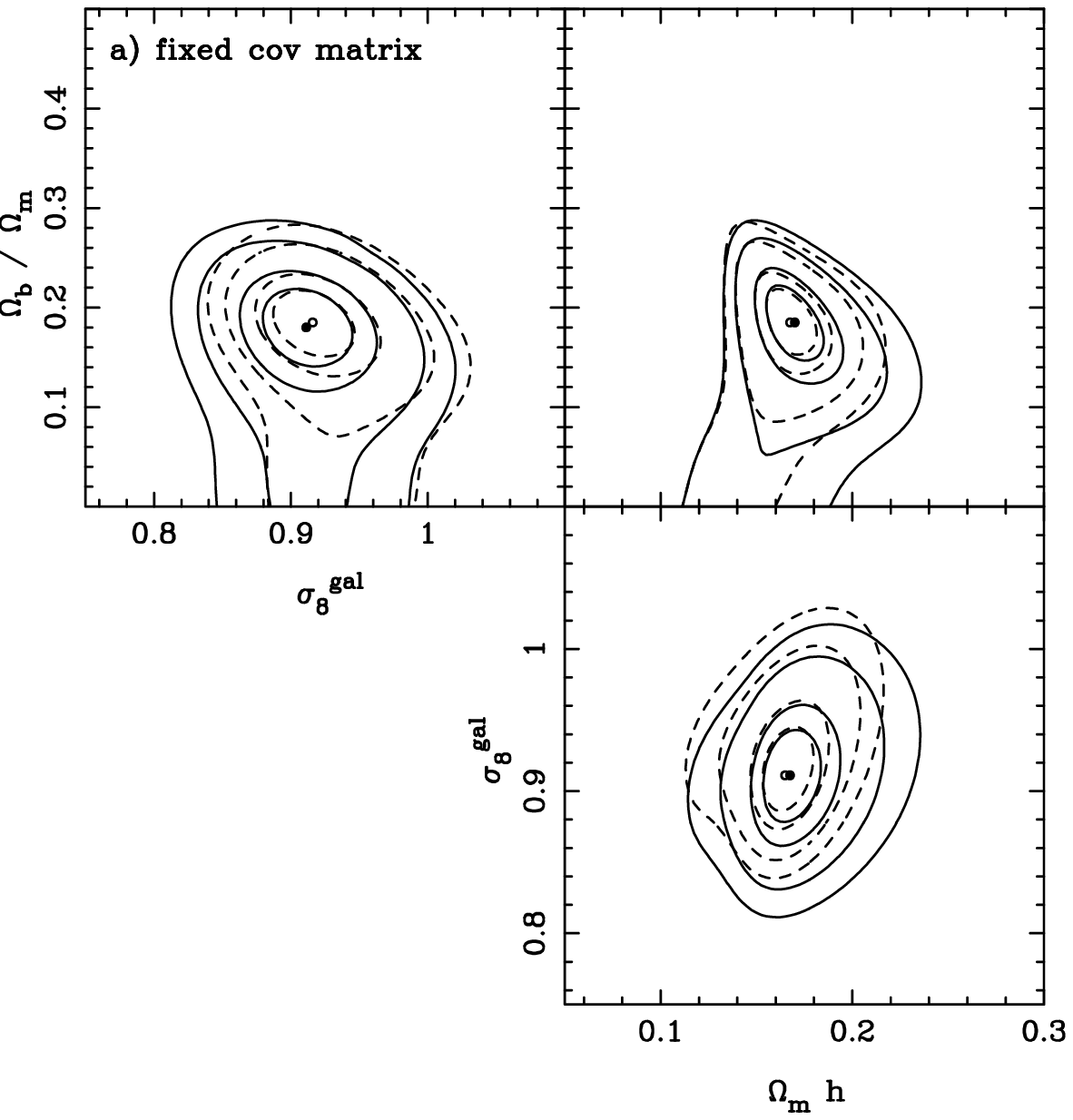}
\hfill
\epsfxsize=0.75\columnwidth
  \epsfbox{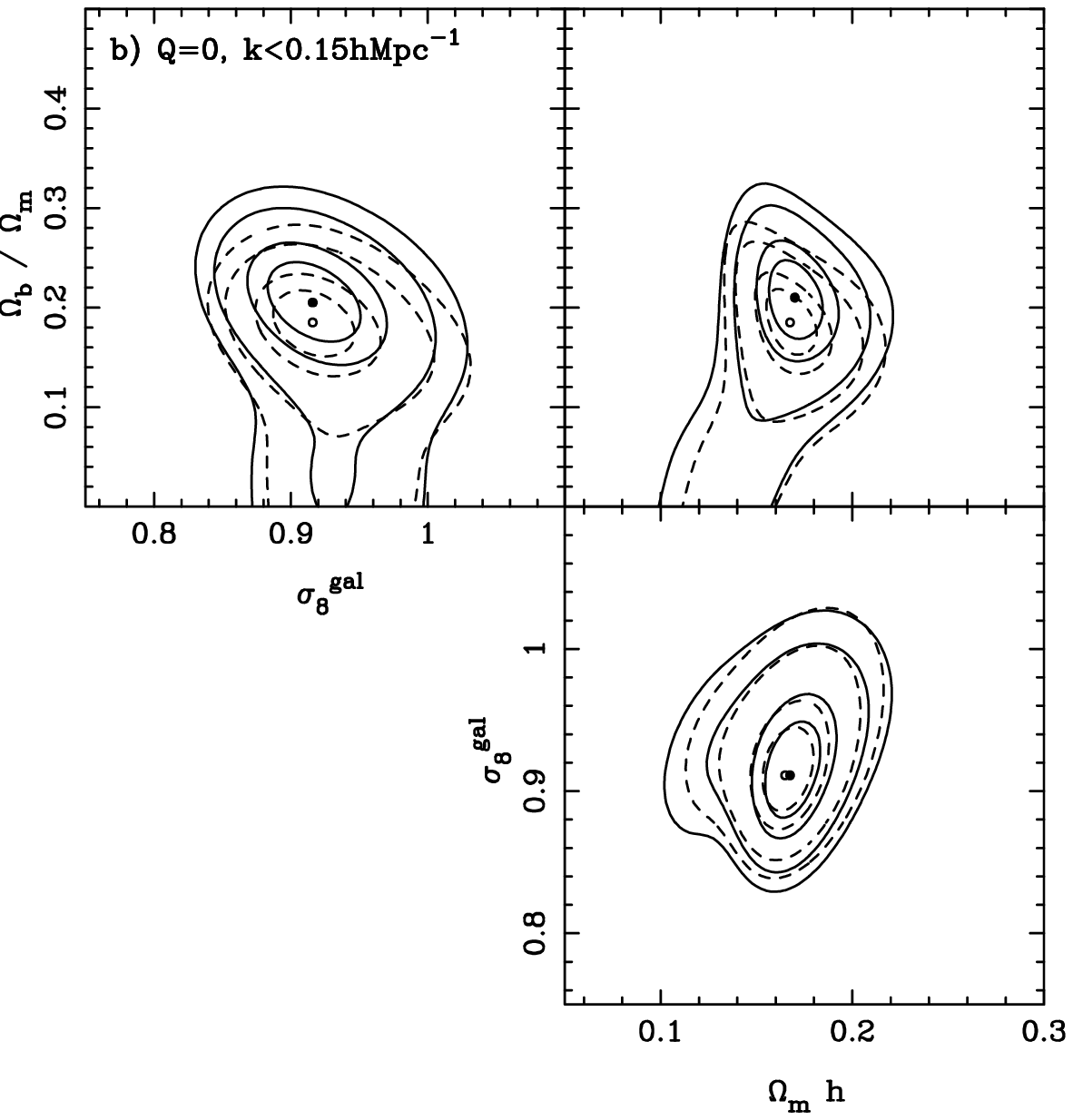}
\hspace{0.125\columnwidth}

\hspace{0.125\columnwidth}
\epsfxsize=0.75\columnwidth
  \epsfbox{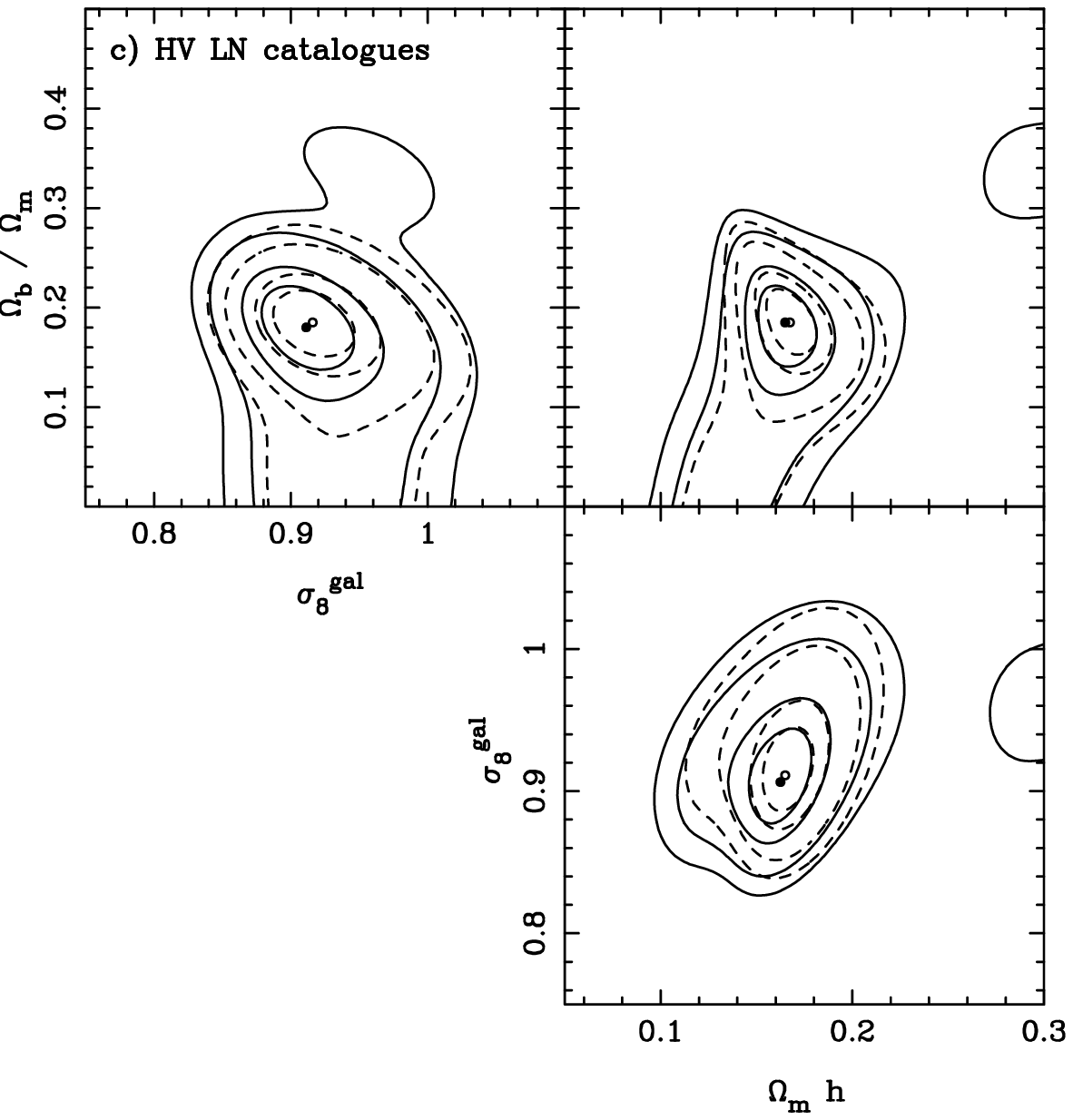}
\hfill
\epsfxsize=0.75\columnwidth
  \epsfbox{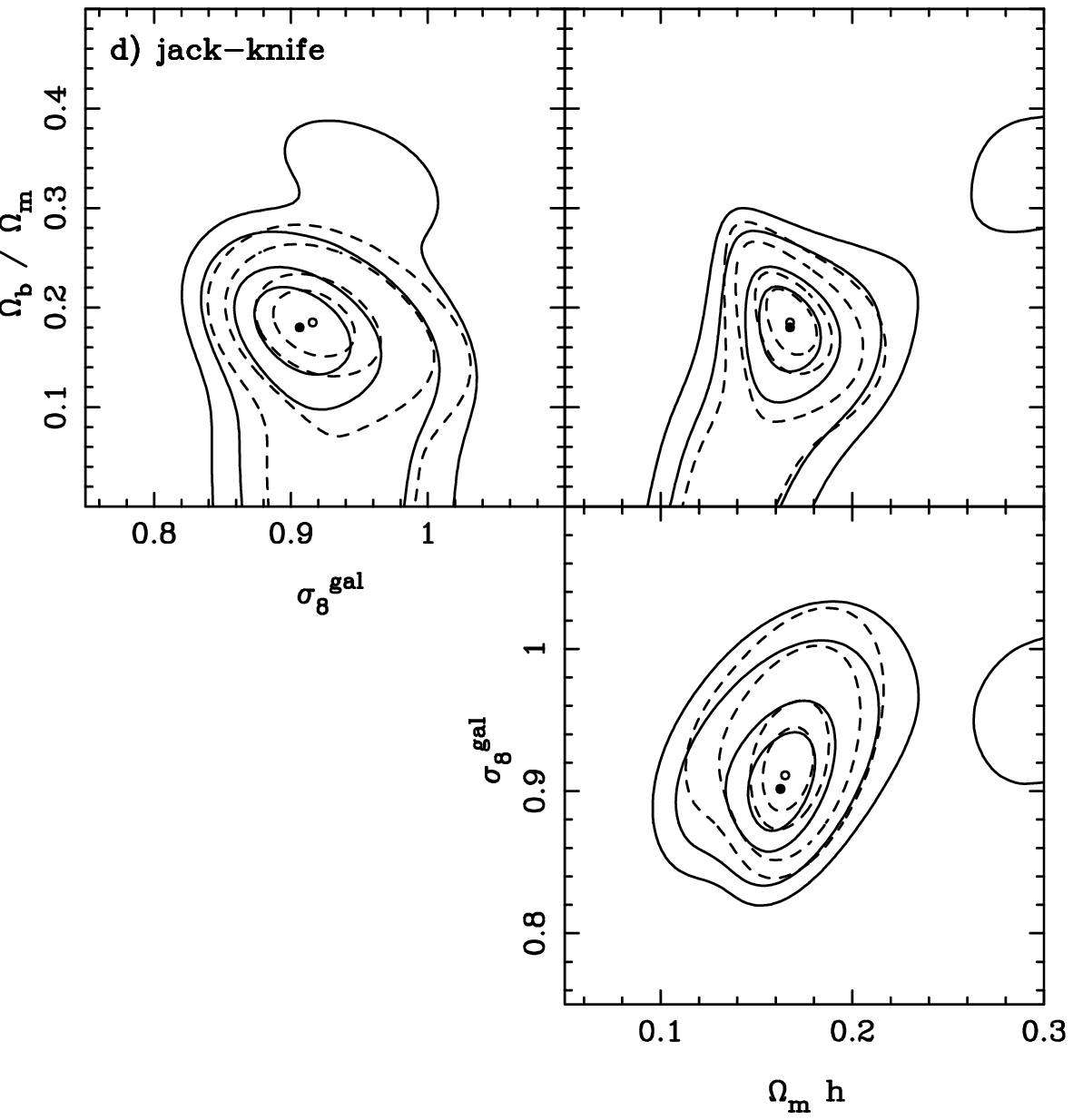}
\hspace{0.125\columnwidth}

\hspace{0.125\columnwidth}
\epsfxsize=0.75\columnwidth
  \epsfbox{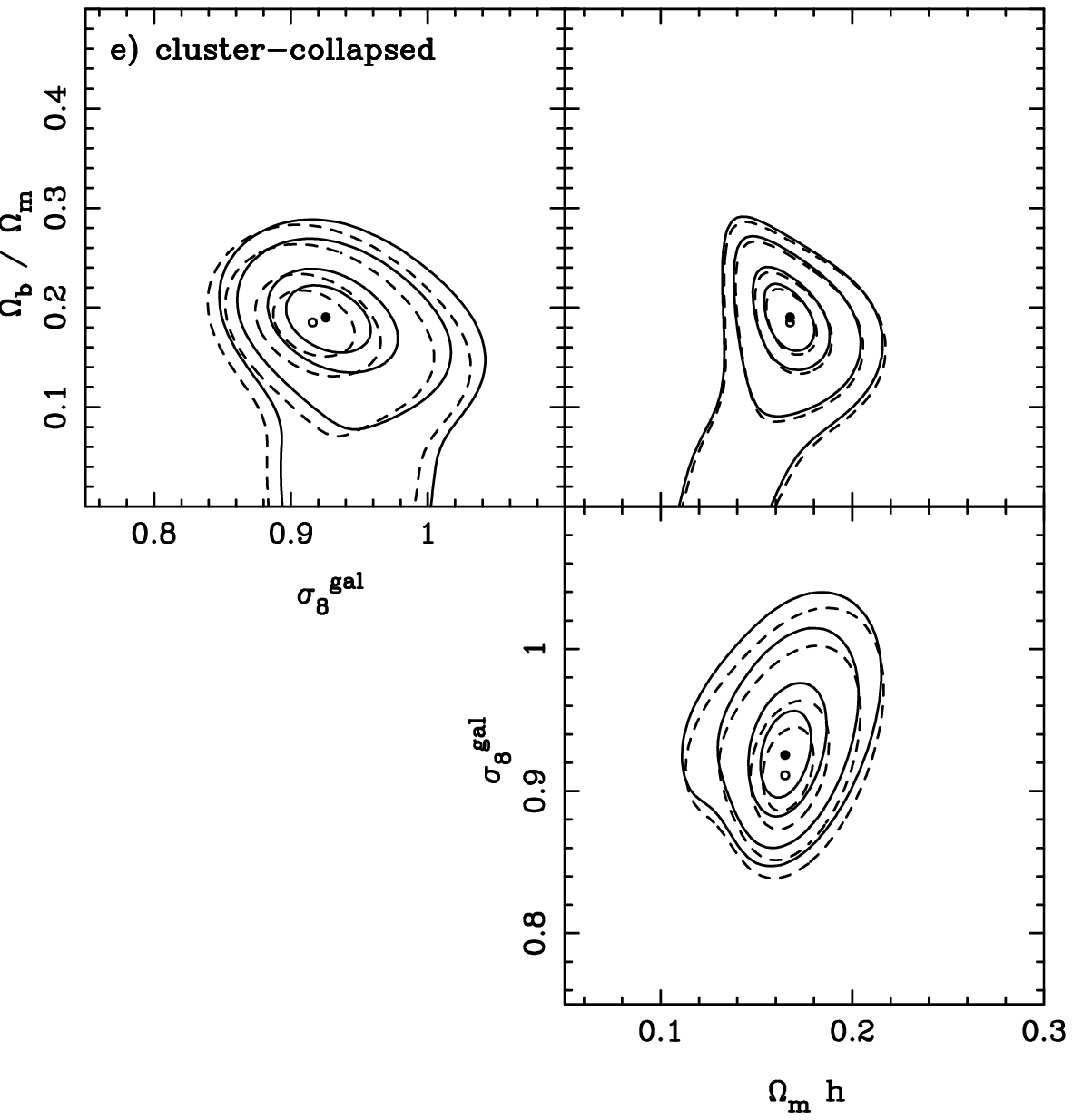}
\hfill
\epsfxsize=0.75\columnwidth
  \epsfbox{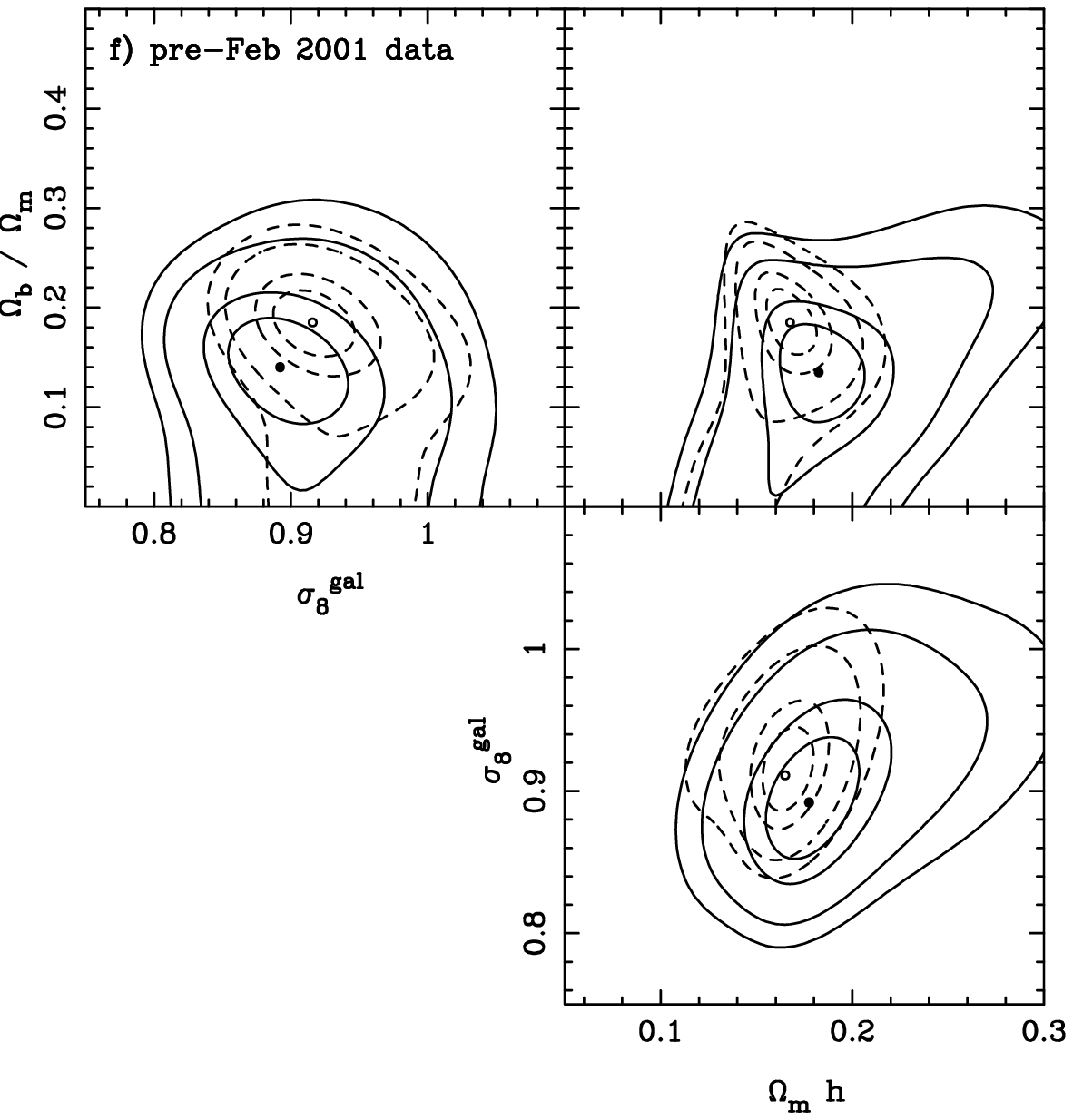}
\hspace{0.125\columnwidth}

\caption{Contour plots showing changes in the likelihood from the
maximum of $2\Delta\ln{\cal L}=1.0, 2.3, 6.0, 9.2$ for different
parameter combinations for the redshift-space 2dFGRS power spectrum,
assuming a $\Lambda$CDM cosmology with $h=0.72$ and $\ns=1.0$.
Dashed contours in all plots are as in Fig.~\ref{fig:2dF_like} and
were fitted for $0.02<k<0.20\hompc$, marginalizing over $Q$. The
open circle marks the maximum likelihood position for each 2d
likelihood surface. The solid contours show the likelihood surfaces
calculated with: a) a fixed covariance matrix calculated from
log-normal catalogues with model power spectrum matched to the
best-fit 2dFGRS value. b) $Q=0$ fixed, and fitting to a reduced $k$
range of $0.02<k<0.15\hompc$. c) covariance matrix calculated from
log-normal catalogues with parameters at the Hubble Volume values.
d) covariance matrix calculated from jack-knife 2dFGRS power
spectra. e) the cluster-collapsed 2dFGRS catalogue marginalizing
over $4<Q<12$ instead of $0<Q<8$. f) the pre-Feb 2001 dataset, as
used in P01, but reanalysed using the revised method. g) the red
subsample of galaxies. h) the blue subsample of galaxies. For each plot,
the solid circle marks the maximum likelihood position of these
revised likelihood surfaces. See text for further details of each
likelihood calculation. \label{fig:2dF_like_variations} }
\end{figure*}

\begin{figure*}
\hspace{0.125\columnwidth}
\epsfxsize=0.75\columnwidth
  \epsfbox{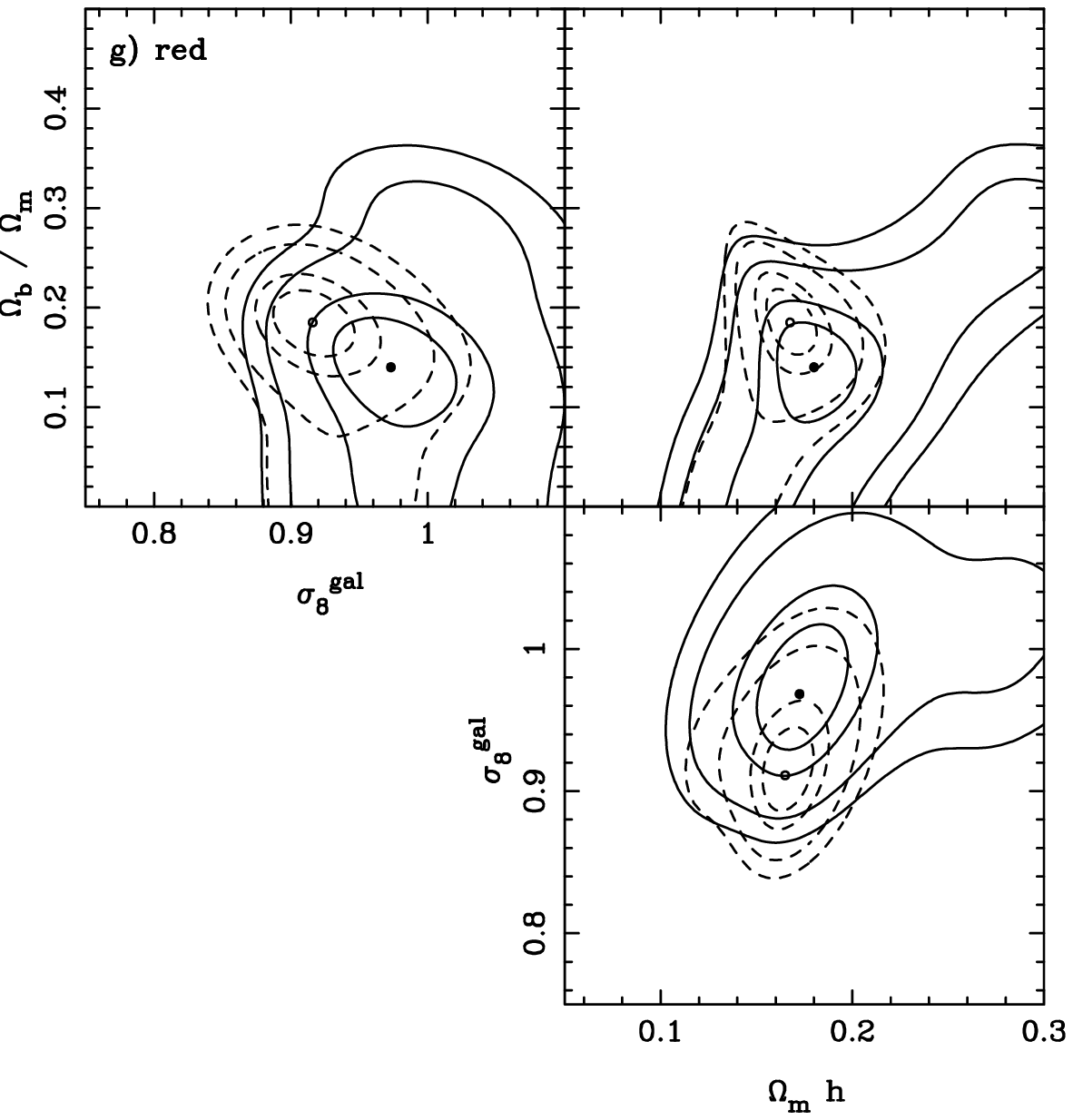}
\hfill
\epsfxsize=0.75\columnwidth
  \epsfbox{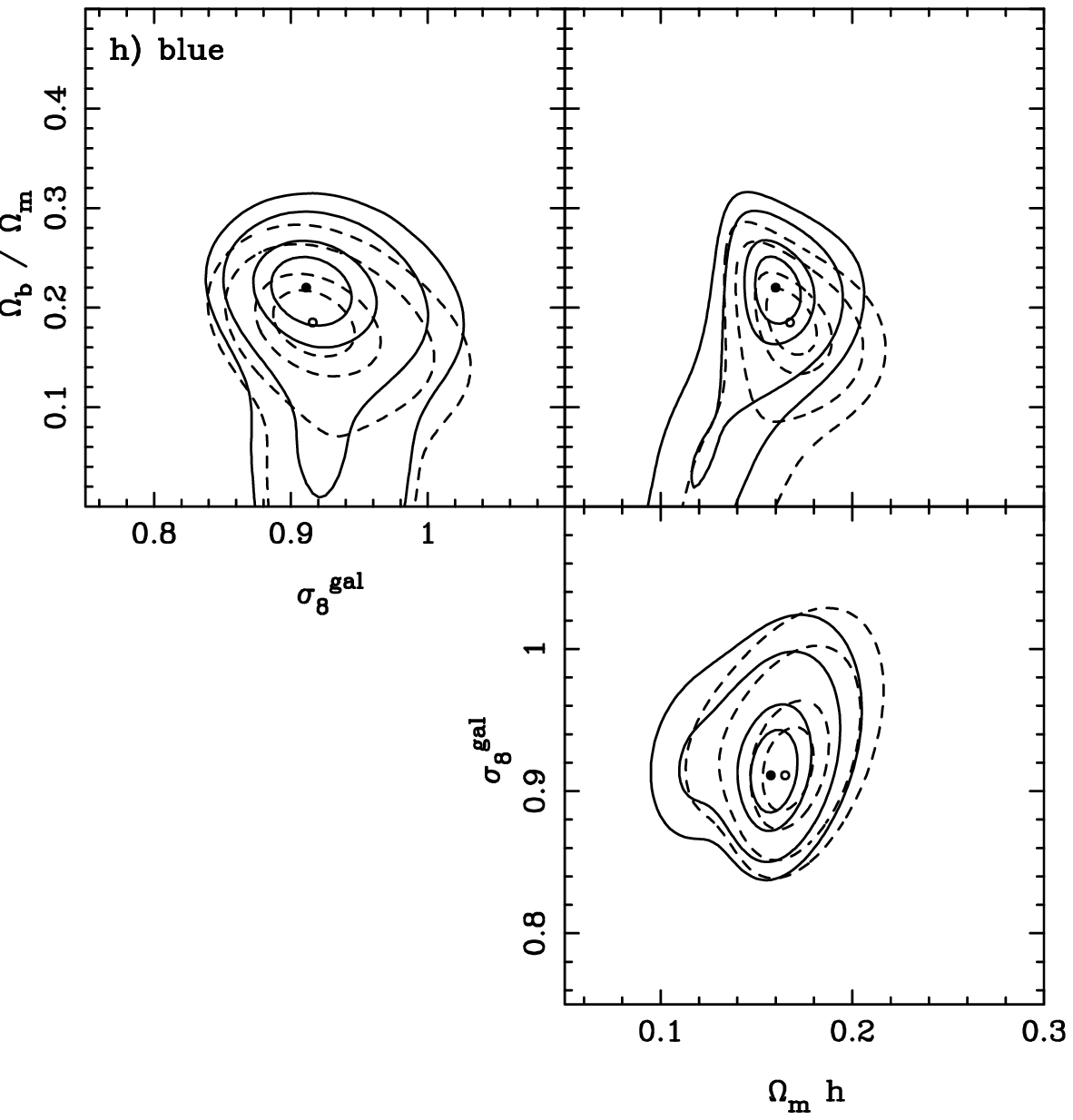}
\hspace{0.125\columnwidth}
\contcaption{}
\end{figure*}

In order to test the robustness of recovered parameters to the scales
probed, Fig.~\ref{fig:param_2dFGRS_kmax} shows marginalized values of
$\Om h$ and $\Ob /\Om $ as a function of $k_{\rm max}$ (i.e. fitting
to data with $0.02 <k<k_{\rm max}$), contrasting results
assuming that the observed galaxy power spectrum is directly
proportional to the linear matter power spectrum with results
involving marginalization over $Q$ and a large-scale correction as
described in Section~\ref{sec:nonlin}.  We also compare results from
the original redshift-space data to those calculated after the
clusters have been collapsed.  For the redshift-space data, we find
that including the $Q$ prescription makes very little difference to
the recovered parameters for $k_{\rm max}\ls0.15\hompc$, confirming
the premise of P01 (i.e. the solid and dotted lines in the left column
are similar for $k_{\rm max}\ls0.15\hompc$). However, this is not true
for smaller scales.

If we restrict ourselves to the assumption that the measured power
spectrum reflects linear theory exactly (dotted lines)
then there is a trend towards higher $\Om h$ and lower baryon fraction
with increasing $k_{\rm max}$.  This effect is especially marked for
the data where the clusters have been collapsed. However, if 
we apply our hybrid correction with $Q$ being allowed to float, these
variations disappear: the recovered parameters display no significant
change when $k_{\rm max}$ is increased from $0.15\hompc$ to
$0.3\hompc$.  This suggests that the $Q$ prescription is able to
capture the real distortions of the redshift-space power
spectrum with respect to linear theory.

On the other hand, it should be noted that the errors initially
fall with increasing $k_{\rm max}$, but beyond $k_{\rm max}\simeq0.2\hompc$ 
there is no further reduction in the
error -- there is little additional information in the small scale
data about the shape of the linear power spectrum.

\subsubsection{Dependence on other assumptions}

In Fig.~\ref{fig:2dF_like_variations} we compare the default likelihood
surface from Fig.~\ref{fig:2dF_like} (dashed lines), with
surfaces calculated using either different data, or with a revised
method.

(a) The three plots in the top-left of this figure show the likelihood
surface calculated using a fixed covariance matrix (solid lines). This
change in the method by which the likelihood is estimated is discussed
further in Section~\ref{sec:likefit}. The net effect here is very
small.

(b) In the three contour plots in the top-right of this figure, we
show likelihood surfaces fitting to $0.02<k<0.15\hompc$ fixing $Q=0$,
(i.e. not allowing for any correction for small-scale effects), but
still including the large-scale correction. The constraints on the
power spectrum normalization and $\Om h$ are consistent in the two
cases. $\Om h$ increases by about 2\%, and the baryon fraction
increases by about 10\%.

(c) The effect of calculating the log-normal catalogues with model
parameters other than the best-fit parameters is shown in
Fig.~\ref{fig:2dF_like_variations}c. Here the solid contours relate to
the parameters of the Hubble Volume simulation ($h=0.7$, $\Om =0.3$,
$\Omega_{\Lambda} =0.7$, $\Ob =0.04$ and $\sigma_8=0.9$). However,
this change in the assumed covariance matrix does not induce a
significant change in the recovered parameter values.

(d) Instead of directly using the log-normal catalogues to estimate
the covariance matrix, we have also considered using the jack-knife
resampling of the 2dFGRS data described in
Section~\ref{sec:system}. The jack-knife estimate of the covariance matrix
was unstable to direct inversion; as an alternative, we smoothed
the fractional difference between the jack-knife and log-normal
covariance matrices, and used this smoothed map to adjust the
log-normal covariance matrix.
This resulted in essentially negligible change in the likelihood contours.

(e) The bottom left part of Fig.~\ref{fig:2dF_like_variations} shows
likelihood surfaces calculated after collapsing the clusters in the
2dFGRS dataset. The scales fitted are the same in both cases, and
both surfaces were calculated after marginalizing over $Q$. The shapes
of the surfaces are in excellent agreement.

\begin{figure}
\epsfxsize=0.9\columnwidth \centering
  \epsfbox{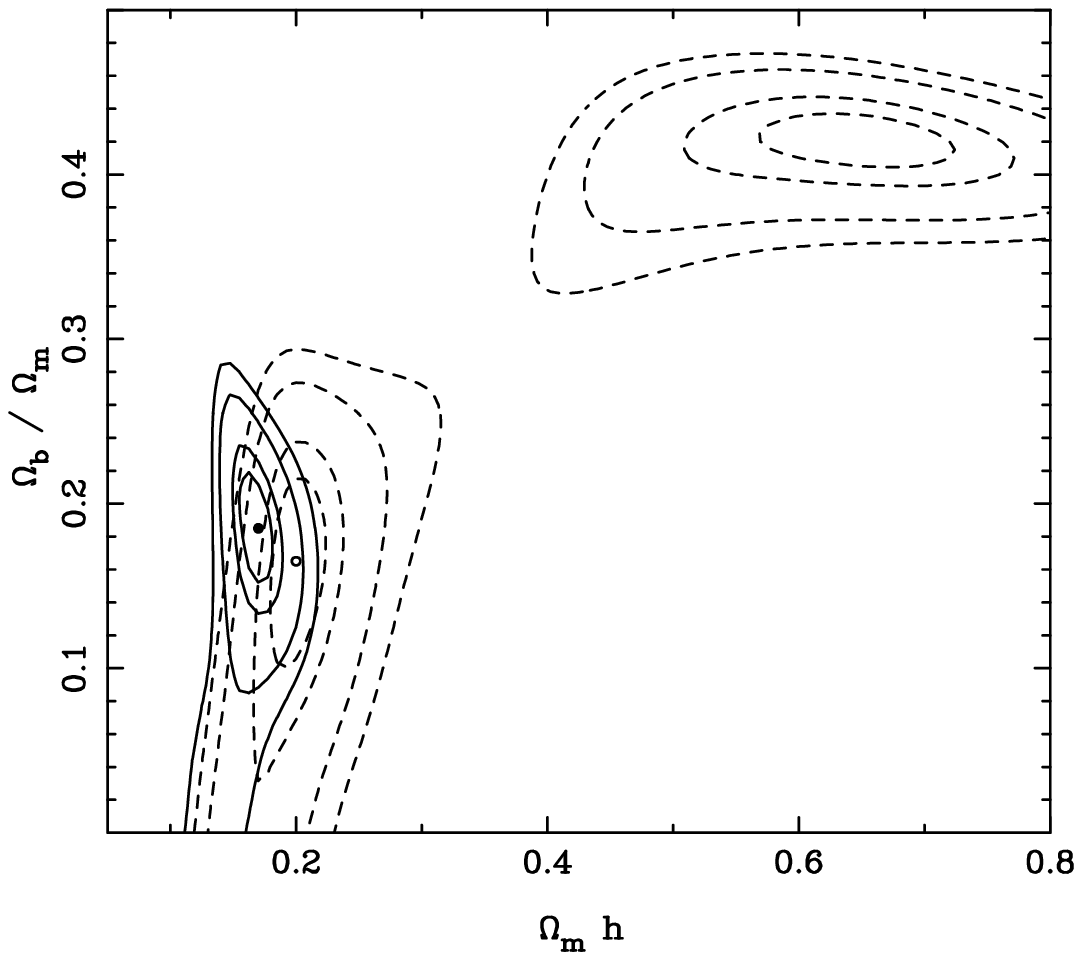}
\caption{Likelihood contours as in Fig.~\ref{fig:2dF_like}, but now
calculated using the data, covariance matrix and methodology of P01
(dashed lines). The cosmological model is as described in
Section~\ref{sec:model} (it differs from that of P01 because we fix
$h=0.72$). However, we have chosen to plot the contours for an
extended range of $\Om h$ to match the analysis of P01. For
comparison, the solid contours show our new default parameter
constraints. \label{fig:like_P01} }
\end{figure}

(f) In Fig.~\ref{fig:2dF_like_variations}f we compare the new likelihood
surface with that calculated using pre-Feb 2001 data. Rather than using
the P01 data and covariance matrix, we reanalyse
the pre-Feb 2001 data with our new method.
We see that most of the difference between the
result of P01 ($\Om h=0.20\pm0.03$, $\Ob/\Om =0.15\pm0.07$) 
and our current best-fit parameters
comes from the larger volume now probed: the parameter constraints in
this plot were calculated in the same way for both datasets.
For an alternative comparison, Fig.~\ref{fig:like_P01} compares our current
likelihood surface with that of P01 over an extended parameter range.
This shows that, in addition to the tightening of
the confidence interval on parameters, the high baryon fraction solution 
of P01 is now rejected at high confidence. 

(g) \& (h) In these panels we show likelihood surfaces for the two
samples defined by splitting the catalogue at a rest frame colour of
$(\bj-\rf)_{z=0}=1.07$. In contrast to the samples discussed in
Section~\ref{sec:redblue} and plotted in Fig.~\ref{fig:pk_redblue}, we
do not force the mean weight per unit redshift to be the
same for both samples. The samples therefore sample different
regions, and some of the difference will be caused by cosmic variance.
In both cases, a consistent $\Om h\simeq 0.17$ is derived.

\begin{figure*}

\hspace{0.05\columnwidth}
\epsfxsize=0.9\columnwidth
  \epsfbox{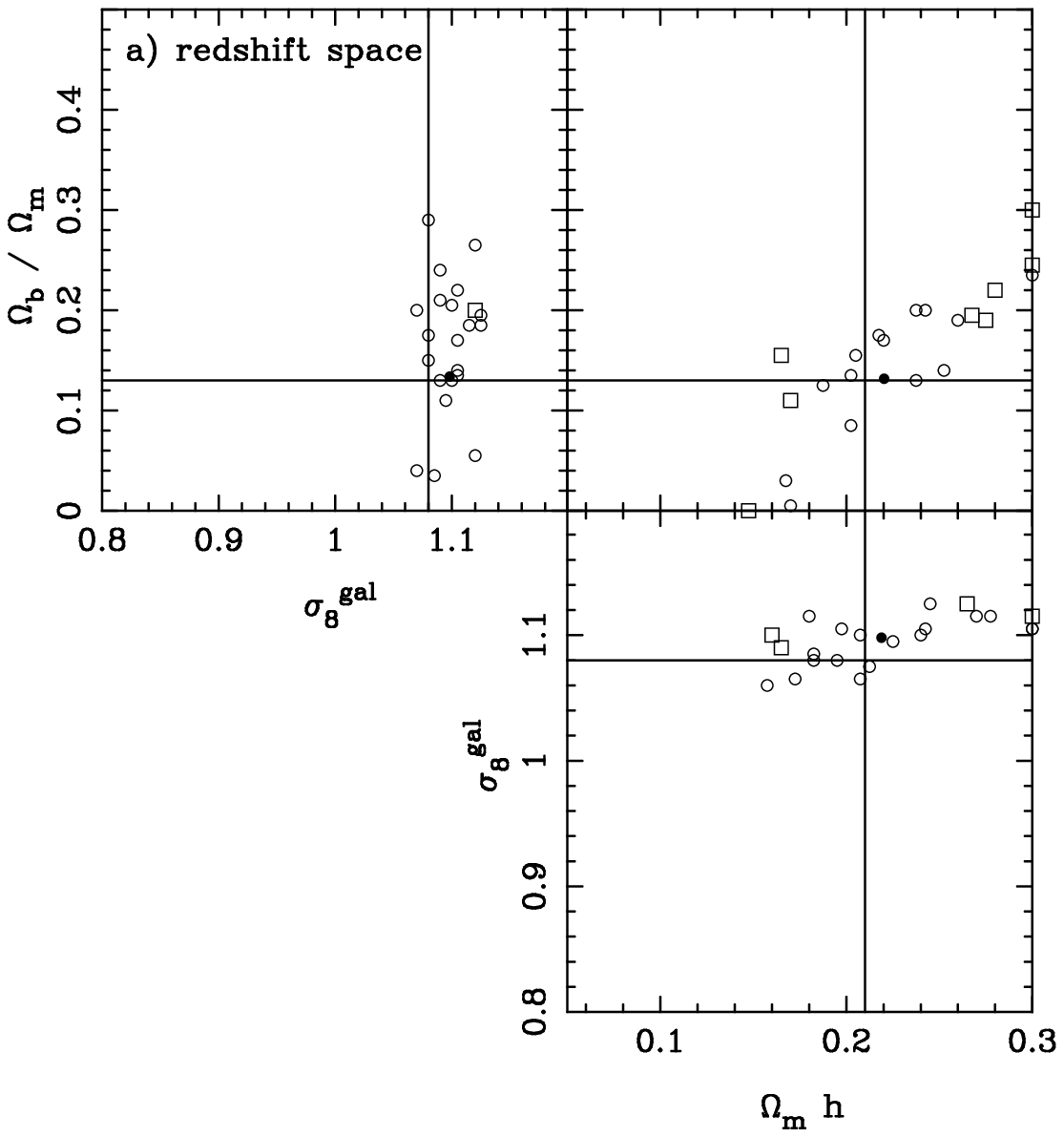}
\hfill
\epsfxsize=0.9\columnwidth
  \epsfbox{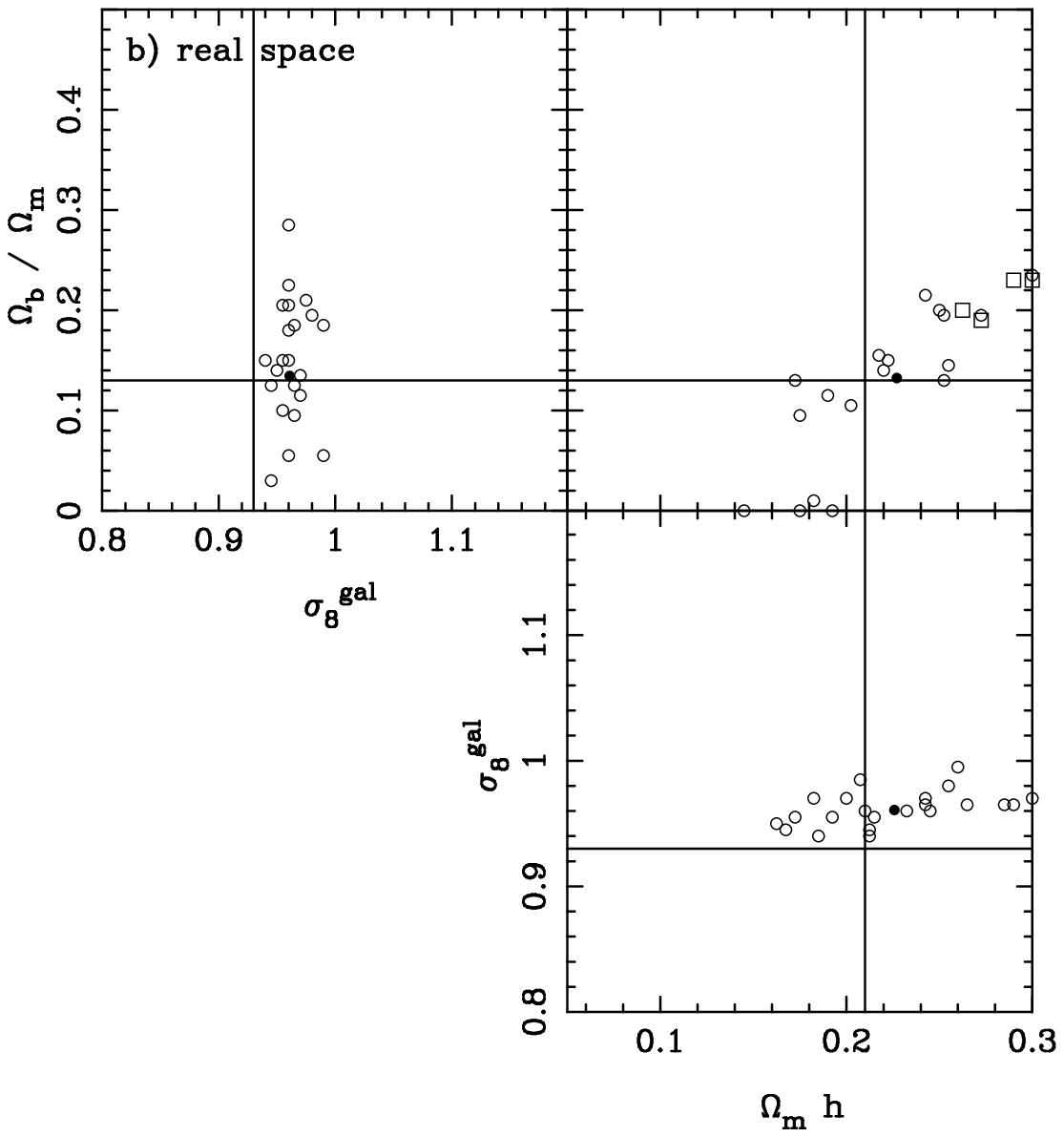}
\hspace{0.05\columnwidth}

\vspace{0.5cm}

\hspace{0.05\columnwidth}
\epsfxsize=0.9\columnwidth
  \epsfbox{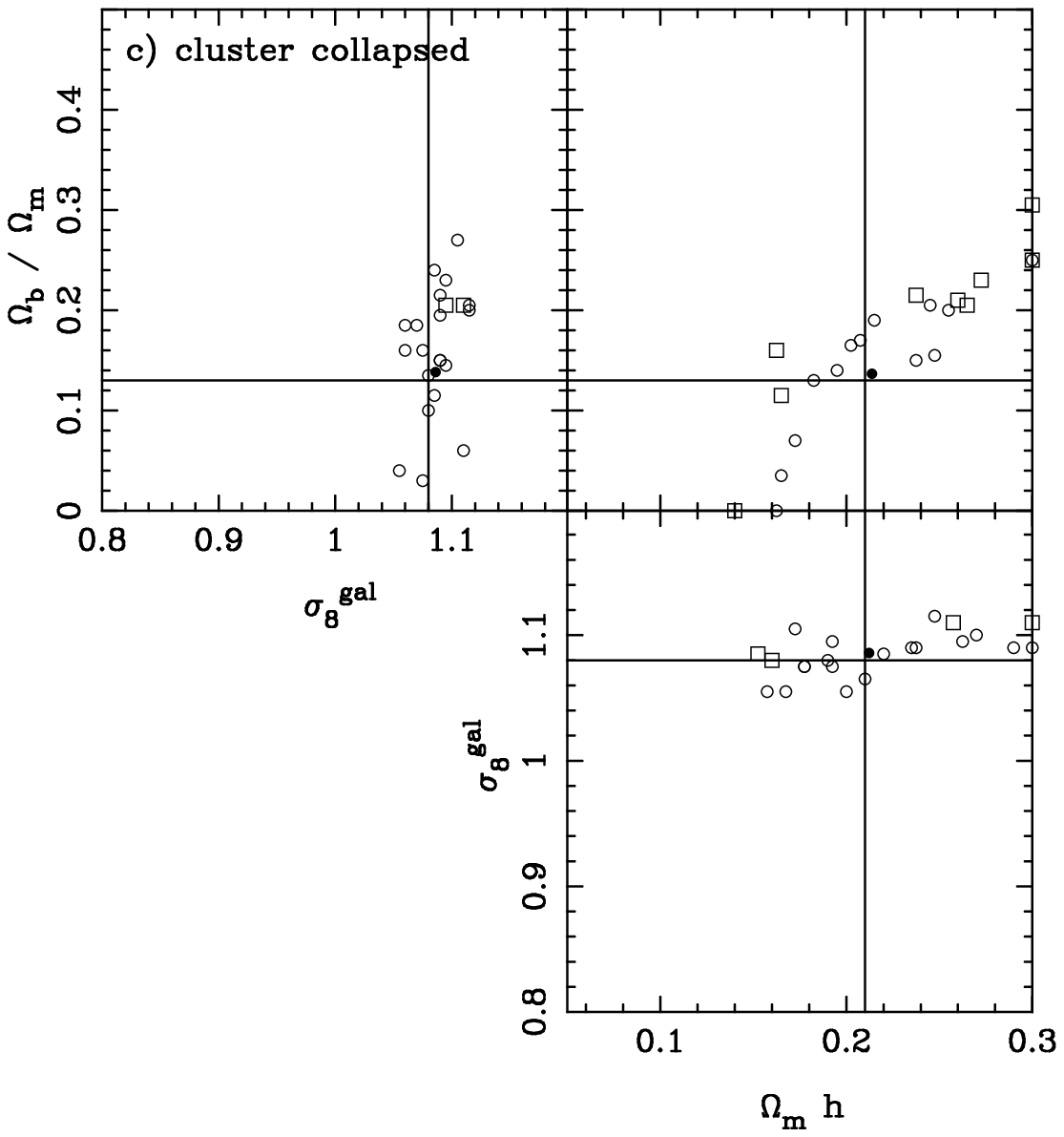}
\hfill
\epsfxsize=0.9\columnwidth
  \epsfbox{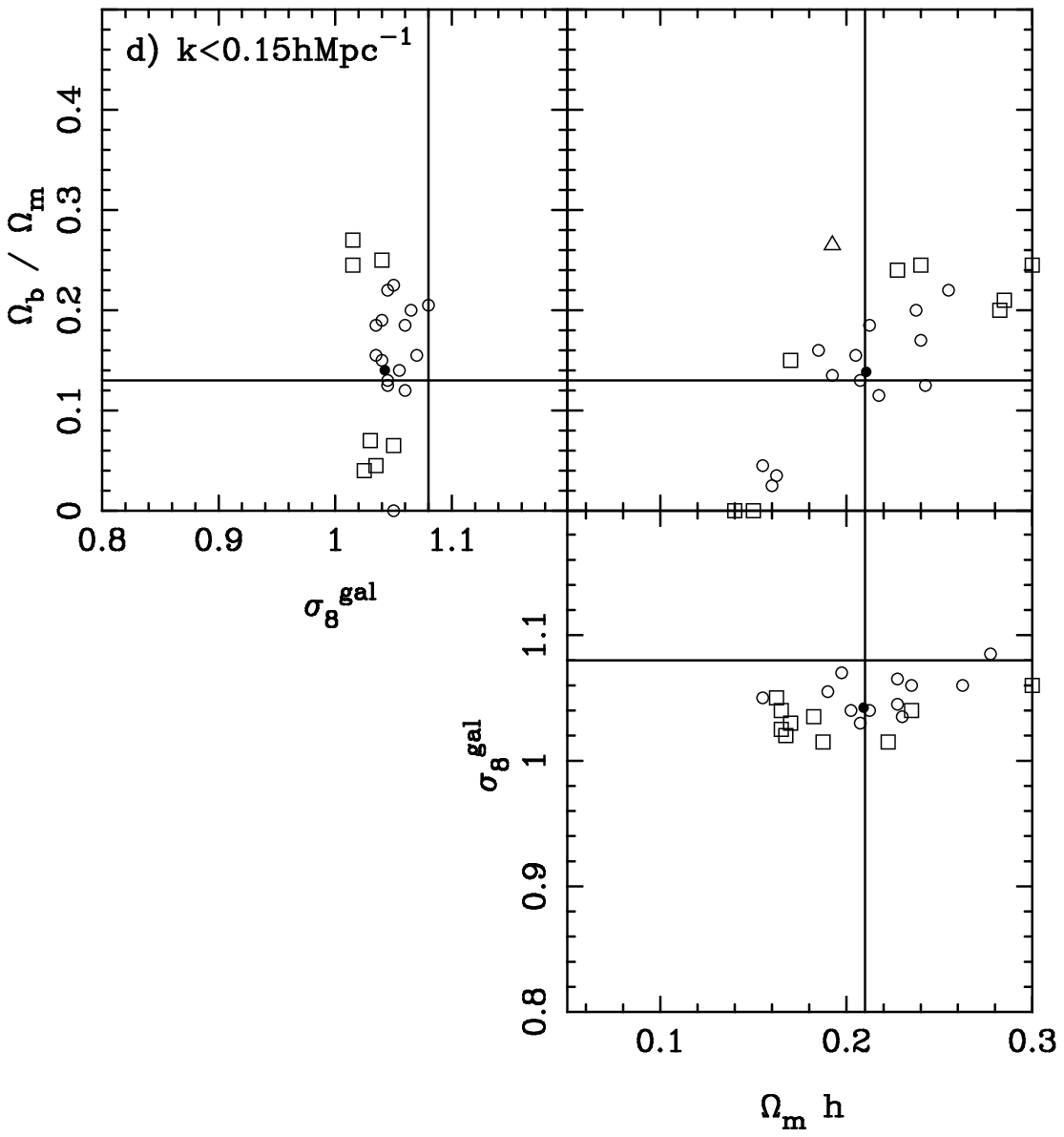}
\hspace{0.05\columnwidth}

\caption{ 
Recovered marginalized parameters from 22
Hubble Volume mock catalogues, demonstrating our ability to recover
the true input parameters from samples that accurately match the size 
and geometry of the 2dFGRS survey.
The solid lines mark the true cosmological parameters and 
normalization of the Hubble Volume simulation, 
calculated from a large realization of galaxies covering the full Hubble Volume
cube. Open circles mark the marginalized parameters recovered from
mocks with $2(\ln{\cal L}_{\rm max}-\ln{\cal L}_{\rm true})<2.3$
(corresponding to mocks with recovered parameters less than 1$\sigma$
from the true values), open squares $2.3<2(\ln{\cal L}_{\rm
max}-\ln{\cal L}_{\rm true})<6.0$ (1$\sigma$ to 2$\sigma$ from the
true values), and open triangles $2(\ln{\cal L}_{\rm max}-\ln{\cal
L}_{\rm true}>6.0$ ($>$2$\sigma$ from the true values). The solid
circle marks the average recovered parameters from all of the
mocks. a) for the redshift-space Hubble Volume mocks fitting to
$0.02<k<0.20\hompc$ marginalizing over $Q$. b) for the real-space
Hubble Volume mocks. c) for the cluster-collapsed redshift-space
mocks. d) for the redshift-space Hubble Volume mocks fitting to
$0.02<k<0.15\hompc$ with $Q=0$.  \label{fig:HV_like} }
\end{figure*}

\subsection{Fitting to the HV mocks}
\label{sec:hvfit}

As a final test, we apply the full fitting machinery to 
a set of 22 mock catalogues drawn
from the Hubble Volume simulation (see Section~\ref{sec:HV_mocks}).
As discussed above, the choice of a fixed covariance matrix has only
a minor effect on the results, so we use a single covariance matrix
to analyse all these mock surveys. This approach also
has the advantage that it is easier to test directly whether the
distribution of the recovered parameters from these
catalogues is consistent with the predicted confidence intervals.

In Fig.~\ref{fig:HV_like} we plot the recovered marginalized
parameters from different sets of 22 redshift-space, real-space and
cluster collapsed Hubble Volume mock catalogues. In general, the
distribution of $\Om h$ and $\Ob /\Om$ values follows the
general degeneracy of cosmological models which give parameter surfaces
with the same approximate shape as shown in
Fig.~\ref{fig:2dF_like}. There is no evidence for a strong bias in the
recovered parameters, and we find that the average recovered
parameters are close to the true values. 

Because the \HV\ mocks do not have luminosity dependent bias and we analyse
them with the FKP estimator, the normalisation we recover corresponds to
the typical galaxies, which have a bias that is approximately $1.26$ higher
that of $L_*$ galaxies. Also, the prescription for scale-dependent
bias given by equation (\ref{eq:Qdef})  
does not accurately match the artificial bias put into the
\HV\ mocks, and the recovered $\sigma_8$ values are seen to be slightly
offset from the expected numbers. This effect is not significant and
merely relates to the crude 
bias model (equation \ref{eq:HV_bias})  used for the \HV\ mocks.

There is some evidence that the average recovered value of
$\Om h$ is higher for the real-space catalogues than for
the redshift-space catalogues. This reflects the slight difference
in large-scale shape between real and redshift space power
spectra observed in Fig.~\ref{fig:pk_hub1}. Even so, this
deviation is smaller than the 1$\sigma$ errors on the
recovered parameters from an individual catalogue.

\section{Summary and discussion}
\label{sec:summary}

\subsection{Results from the complete 2dFGRS} 

This paper has been devoted to a detailed discussion of
the galaxy power spectrum as measured by the final 2dFGRS.
We have deduced improved versions of the masks that describe
the angular selection of the survey, and modelled the radial
selection via a new empirical treatment of evolutionary corrections.
We have carried out extensive checks of our methodology, 
varying assumptions in the treatment of the data and applying
our full analysis method to realistic mock catalogues.

Based on these investigations, we are confident that the 2dFGRS power
spectrum can be used to infer the matter content of the universe,
via fitting to a CDM model. Assuming a primordial $\ns=1$ spectrum
and $h=0.72$, the best fitting model has $\chi^2/{\rm d.f.}= 36/32$ and 
the preferred parameters are 
\begin{equation}
\Om  h = 0.168 \pm 0.016
\end{equation}
and a baryon fraction 
\begin{equation}
\Ob /\Om  = 0.185\pm 0.046.
\end{equation}
We have kept $\ns$ and $h$ fixed so that the
quoted errors reflect only the uncertainties that arises from the
uncertainty in the shape of power spectrum and not 
additional uncertainties due the choice of $\ns$ and $h$.
However the values and errors are insensitive to the choice of $h$.
Allowing 10\% Gaussian uncertainty gives
$\Om  h = 0.174 \pm 0.019$ and $\Ob /\Om  = 0.190\pm 0.053$.

These values represent in some respects an important change with
respect to P01, who found $\Om  h = 0.20 \pm 0.03$ and
$\Ob /\Om  = 0.15 \pm 0.07$. Statistically, the shift in
the preferred parameters is unremarkable. However, the precision
is greatly improved, by nearly a factor 2. This reflects a
substantial increase in the survey volume since P01, both because
the survey sky coverage is 50\% larger, and because our improved
understanding of the selection function enables us to work to
larger redshifts. 
In particular, the reduced error on the baryon fraction means that P01's
suggestion of a non-zero baryon content can now be regarded as a definite
measurement. Our figure of $\Ob/\Om=0.185\pm0.046$ appears at face value
to be a 4-$\sigma$ detection of baryon features, although this overstates
the significance. The difference in $\chi^2$ between the best zero-baryon
model and the best overall model is 6.3, so the likelihood ratio is
$L = \exp(-6.3/2)$. This might suggest a probability for no baryons of
$L/(1+L)=0.04$, but such a figure is too generous: for a Gaussian
distribution, this value of $L$ would be a 2.5-$\sigma$ effect, with
one-tailed probability of 0.006. It therefore seems fair to reject the
zero-baryon hypothesis at about the 1\% level.

It should be emphasised that the above statements depend on the
theoretical framework of the $\Lambda$CDM model. This is important not
only because the theory quantifies the relation between the baryon
fraction and any features in the power spectrum, but because it constrains
the allowed {\it form\/} of any baryon signature. What is impressive in
our data is not simply that the results suggest departures from a smooth
curve, but that these deviations occur in the locations expected from
theory. It is this prior knowledge that gives the extra statistical power
needed in order to reject a zero-baryon model with confidence.

Of course, proving that the universe contains baryons hardly ranks as a
great novelty. It is an inevitable prediction of the $\Lambda$CDM model
that the matter power spectrum should contain baryon features, and it has
recently been confirmed directly that these should survive in the galaxy
spectrum \citep{springel05}. The signature is much smaller than the
corresponding acoustic oscillations in the CMB, so this measurement in no
way competes with the CMB as a means of pinning down the baryon density.
Nevertheless, by
demonstrating a clearcut connection between the temperature fluctuations
in the CMB and the present-day galaxy distribution, the identification of
the baryon signal in the 2dFGRS provides an important verification of
our fundamental model of structure formation.

\subsection{Cosmological implications} \label{sec:implications}

The ability of the matter power spectrum to determine
cosmological parameters in isolation is limited owing to the inherent
physical degeneracies in the CDM model.  As is well known, these can
be overcome by combination with data on CMB anisotropies. The most
striking success of this method to date has been the combination of
the 2dFGRS results from P01 with the year-1 WMAP data (Spergel et
al. \citeyear{spergel03}), the results of which were subsequently 
confirmed using the SDSS galaxy power spectrum by \citet{tegmark04}.
It is of interest to see how our earlier
conclusions alter in the light of our new results. We have used the
Markov-Chain Monte-Carlo (MCMC; Lewis \& Bridle \citeyear{lewisbridle02}) method 
to fit cosmological models to our new
power-spectrum data combined with WMAP year 1 (Hinshaw et
al. \citeyear{hinshaw03}) CMB data. For the choice of model, we
adopted the philosophy of \citet{P02}, allowing $\Om$, $\Ob$, $h$,
$\ns$, $\tau$, $\sigma_8$ and $\sigma_8^{\rm gal}$ to vary while
assuming negligible neutrino contribution and a flat cosmology. The
results, ignoring the normalization of the model power spectra, are
as follows:
\begin{equation}
\eqalign{
\Om &= 0.231 \pm 0.021 \cr
\Ob &= 0.042 \pm 0.002 \cr
h   &= 0.766 \pm 0.032 \cr
\ns &= 1.027 \pm 0.050.\cr
}
\end{equation}
We see that using the new 2dFGRS result decreases $\Om$ by
approximately 15\% from the best-fit WMAP value of $\Om\simeq 0.27$. This
change is easily understood because our new best-fit $\Om h=0.168$ is
lower than that of P01. The CMB acoustic peak locations constrain
$\Om h^{3}$, so to fit the new data requires a lower value of
$\Om$ coupled with a higher value of $h$. 
Again, what is impressive is that the accuracy is significantly
improved, breaking the 10\% barrier on $\Om$. For comparison,
The WMAP analysis in Spergel et al. (2003) achieved 15\% accuracy on $\Om$.
As a result, we are able to achieve a firm rejection of the common
`concordance' $\Om=0.3$ in favour of a lower value ($0.19 < \Om < 0.27$ at
95\% confidence).
This result demonstrates
that large-scale structure measurements continue to play a crucial
role in determining the cosmological model.

\section*{Acknowledgements}
The data used here were obtained with the 2 degree field facility on the
3.9m Anglo-Australian Telescope (AAT). We thank all those involved in
the smooth running and continued success of the 2dF and the AAT.
We thank Valerie de Lapparent
for kindly making available the ESO-Sculptor photometry.
JAP and OL are grateful for the support of PPARC Senior Research Fellowships.
PN acknowledges the receipt of an ETH Zwicky Prize Fellowship.
We thank the anonymous referee for many useful comments.

\setlength{\bibhang}{2.0em}

\end{document}

%% file: defs.tex
\def\gs{\mathrel{\lower0.6ex\hbox{$\buildrel {\textstyle >}
 \over {\scriptstyle \sim}$}}}
\def\ls{\mathrel{\lower0.6ex\hbox{$\buildrel {\textstyle <}
 \over {\scriptstyle \sim}$}}}
\def\japsub{\rm\scriptscriptstyle}
\def\kms{{\,\rm km\,s^{-1}}}
\def\kmsmpc{{\,\rm km\,s^{-1}Mpc^{-1}}}
\def\hompc{{\,h\,\rm Mpc^{-1}}}
\def\mpcoh{{\,h^{-1}\,\rm Mpc}}
\def\hmpc{{\,h^{-1}\,\rm Mpc}}

\def\bj{b_{\japsub J}}
\def\bjsc{b_{\japsub J,SC}}
\def\rf{r_{\japsub F}}
\def\r{{\bf r}}
\def\k{{\bf k}}
\def\nrl{{n_{\rm L}^{\rm r}}}
\def\bnrl{{{\bar n}_{\rm L}^{\rm r}}}
\def\bngl{{{\bar n}_{\rm L}^{\rm g}}}
\def\bng{{{\bar n}^{\rm g}}}
\def\ngl{{n_{\rm L}^{\rm g}}}
\def\wa{w^{\rm A}}
\def\HV{Hubble Volume}
\def\LN{log-normal}

\def\m@th{\mathsurround=0pt }
\def\eqalign#1{\null\,\vcenter{\openup1\jot \m@th
 \ialign{\strut\hfil$\displaystyle{##}$&$\displaystyle{{}##}$\hfil
 \crcr#1\crcr}}\,}

\def\twodF{{2dFGRS}}

\def\eg{{e.g.\ }}

\def \etal {et al.\ }
\def \ie {{\rm i.e.\ } }

\def \bj {{\rm b_J}}

\def \logh {{\log_{10} h}} 
\def \date {May 2001}

\def \Om{\Omega_{\rm m}}
\def \Ob{\Omega_{\rm b}}
\def \ns{n_{\rm s}}